# Probabilistic Genotype-Phenotype Maps Reveal Mutational Robustness of RNA Folding, Spin Glasses, and Quantum Circuits

Anna Sappington[1, 2, 3, *] and Vaibhav Mohanty[4, 2, 3, *]

[1]*Department of Electrical Engineering and Computer Science, Massachusetts Institute of Technology, Cambridge, MA 02139*

[2]*Program for Health Sciences and Technology, Harvard Medical School, Boston, MA 02115 and Massachusetts Institute of Technology, Cambridge, MA 02139*

[3]*Harvard/MIT MD-PhD Program, Harvard Medical School, Boston, MA 02115 and Massachusetts Institute of Technology, Cambridge, MA 02139*

[4]*Department of Chemistry and Chemical Biology, Harvard University, Cambridge, MA 02138*

Recent studies of genotype-phenotype (GP) maps have reported universally enhanced phenotypic robustness to genotype mutations, a feature essential to evolution. Virtually all of these studies make a simplifying assumption that each genotype—represented as a sequence—maps deterministically to a single phenotype, such as a discrete structure. Here, we introduce probabilistic genotype-phenotype (PrGP) maps, where each genotype maps to a vector of phenotype probabilities, as a more realistic and universal language for investigating robustness in a variety of physical, biological, and computational systems. We study three model systems to show that PrGP maps offer a generalized framework which can handle uncertainty emerging from various physical sources: (1) thermal fluctuation in RNA folding, (2) external field disorder in spin glass ground state finding, and (3) superposition and entanglement in quantum circuits, which are realized experimentally on IBM quantum computers. In all three cases, we observe a novel biphasic robustness scaling which is enhanced relative to random expectation for more frequent phenotypes and approaches random expectation for less frequent phenotypes. We derive an analytical theory for the behavior of PrGP robustness, and we demonstrate that the theory is highly predictive of empirical robustness.

## I. INTRODUCTION

Systems which take a sequence as input and nontrivially produce a structure, function, or behavior as output are ubiquitous throughout the sciences and engineering. In biological systems such as RNA folding [1–11], lattice protein folding [4], protein self-assembly [12, 13], and gene regulatory networks [14, 15], the relationship between genotype (stored biological information) and phenotype (observable or functional properties) can be structured as genotype-phenotype (GP) maps, which have a rich history of computational and analytical investigation [1–34]. Systems from physics and computer science have also been analyzed as GP maps, including the spin glass ground state problem [30], linear genetic programming [26], and digital circuits [31].

Despite being completely disparate systems, all of the GP maps above share a number of common structural features, most notably an enhanced robustness of the phenotypes to genotype mutations. Phenotypic *robustness* $\rho_n$ of a phenotype $n$ is the average probability that a single character mutation of a genotype $g$ which maps to $n$ does not change the resultant phenotype $n$, averaged over all genotypes $g$ mapping to $n$. A completely random assignment of genotype to phenotype predicts that $\rho_n \approx f_n$ [4], where $f_n$ is the fraction of genotypes that map to phenotype $n$. However, the systems mentioned above display a surprising and substantially enhanced robustness, exhibiting the relationship $\rho_n \approx a + b \log f_n \gg f_n$ with system-dependent constants $a$ and $b$, meaning that even for rare phenotypes, a small changes to the genotype do not necessarily result in change of the phenotype. It has been shown that, in evolution, this enhanced robustness facilitates discovery of new phenotypes [11, 19, 20, 35] and is crucial for navigating fitness landscapes [5]. As a result, it is important to accurately quantify robustness and its relationship with phenotype frequency.

All GP map studies referenced above, spanning several decades of research, make the assumption that a genotype maps deterministically to a single phenotype. However, we argue that for most of the above systems, this is a major simplification. For instance, by mapping an RNA genotype to only the ground state energy structure, previous studies [1–11] make an implicit zero temperature approximation for the ensemble of molecules, even if the Gibbs free energy of an individual molecule itself is calculated within the folding software at finite temperature. Similarly, in studies of gene regulatory networks, spin glasses, linear genetic programs, and digital circuits, the systems investigated do not interact with external networks or variables. These investigations assume that the environmental effect on the GP mapping of the subsystem of interest is static. Probabilistic mappings from genotype to phenotype have certainly existed in many realms of science, such as probabilistic classifications of images by neural networks (i.e. sequences of pixel intensities mapping to probabilities of classes). However, the literature still lacks a unifying framework to analyze the single-character mutational robustness of these maps, among other properties, in the way that there already exists a universal language for the deterministic GP (DGP) maps mentioned above.

* The authors contributed equally to this work. Correspondence: asappington@hms.harvard.edu and mohanty@hms.harvard.edu.

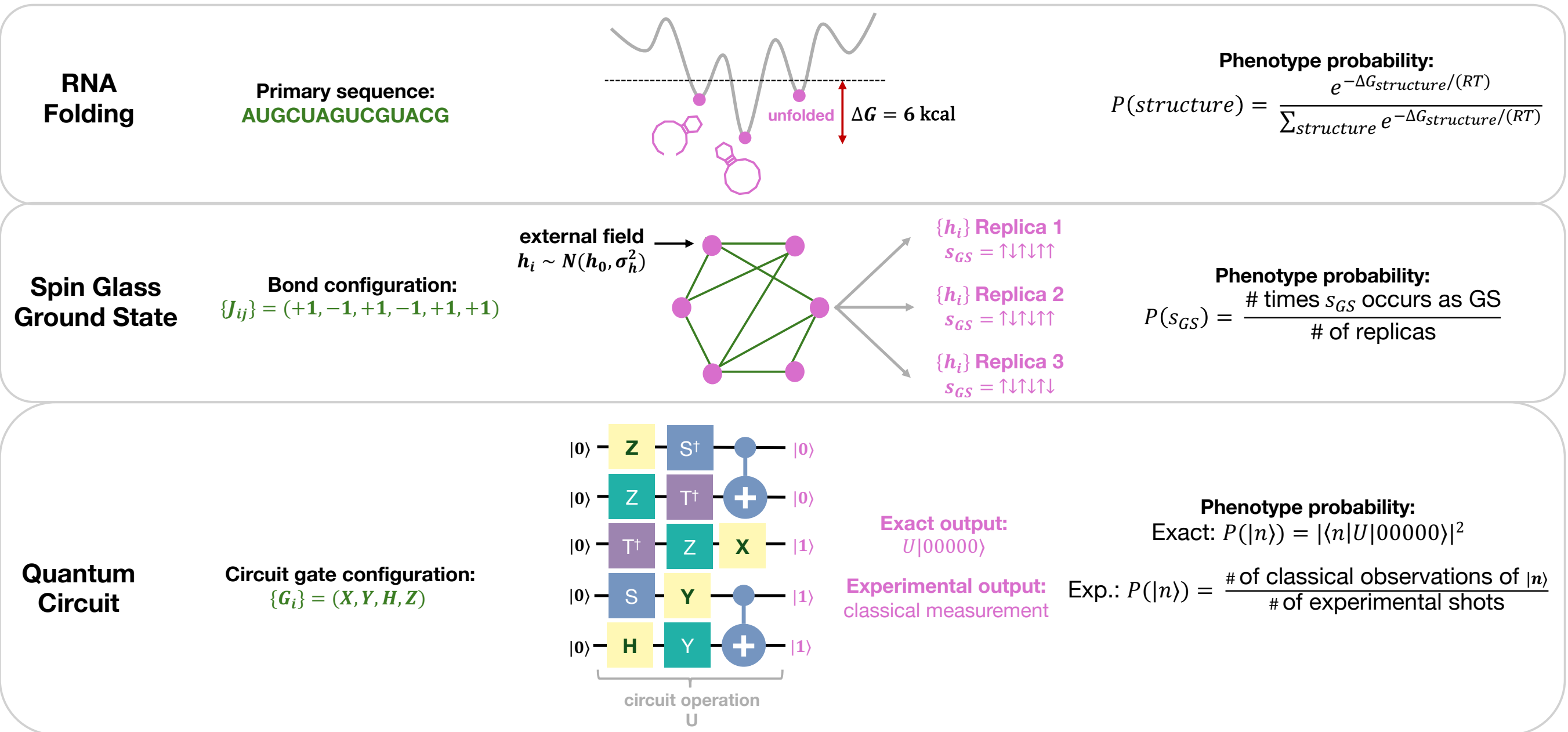

FIG. 1. Schematic representations of the PrGP model systems studied in this work. For each system, its respective genotype (green), a visualization of the system, its phenotypes (pink), and its method for calculating the phenotype probability vector are shown. For RNA folding, the genotype is a primary sequence of nucleotides and the phenotype is the folded dot-bracket structure. For spin glass ground states, the genotype is a bond configuration. Each spin $s_i$ (pink dots) are connected via this bond configuration $J_{ij}$ (green lines) and a disordered external field $h_i$ is applied. The phenotype is the fraction of replicas in which each ground state appears. For quantum circuits, the genotype consists of a subset of gates from a random circuit. The circuit is given a set input state, $|00\ldots0\rangle$, and the exact phenotype probability vector is the probability of classically measuring each basis state, $p_n(g) = |\langle n| U(g) |00\ldots0\rangle|^2$, where $U(g)$ is the circuit operation as a function of the genotype $g$. The experimental phenotype probability vector is computed from tallying classically measured states from 1000 experimental shots on a quantum computer.

| System | Genotype Alphabet | Alphabet size $k$ | Phenotype | Source of Uncertainty |
|---|---|---|---|---|
| RNA folding | {A, U, G, C} (or {G, C}) | 4 (or 2) | Folded dot-bracket structure | Thermal fluctuation, $T > 0$ |
| Spin glass ground state | {-1, +1} | 2 | Ground state spin configuration | Disordered external field |
| Quantum circuit | {Z, X, Y, H, S, S$^\dagger$, T, T$^\dagger$} | 8 | Classical measurement of circuit output | Superposition and entanglement |

TABLE I. Overview of the genotypes and phenotypes of each PrGP system, as well as their respective sources of uncertainty.

In this article, we introduce probabilistic genotype-phenotype (PrGP) maps—in contrast to the above systems which we call DGP maps—as a universal framework for analyzing the mutational robustness of sequence-to-discrete classifications. DGP maps thereby emerge as the limiting case of PrGP maps in each genotype or sequence maps to a single phenotype with probability 1 and all other phenotypes with probability 0. The definitions of phenotypic robustness and transition probabilities retain the same physical meaning in PrGP maps as in DGP maps, and we emphasize that PrGP maps can handle disorder and uncertainty emerging from a variety of sources.

To address the implicit zero temperature approximation in sequence-to-structure mappings (RNA, lattice protein folding, protein self-assembly), we study the folding of RNA primary sequences to a canonical ensemble of secondary structures corresponding to low-lying local free energy minima. To address external variable disorder with a known distribution, we study the zero temperature mapping of a spin glass bond configuration to its ground state with quenched external field disorder, building a phenotype probability vector using many replicas of the disordered field. This has implications for viral fitness landscape inference [36–40], where external fields, in part, model host immune pressure [39]. Lastly, to investigate inherent uncertainty in phenotypes, we introduce quantum circuit GP maps where uncertainty emerges from superposition and entanglement of classically measurable basis states. Our experimental realization of these quan-tum circuits on a 7-qubit IBM quantum computer also introduces measurement noise, which has a clear and unique effect on robustness. The PrGP map properties of the three model systems are summarized in Table I and visually in Figure 1. We observe empirically that PrGP maps exhibit a novel biphasic scaling of robustness versus phenotype frequency which, for higher frequency phenotypes, resembles the $\rho_n \propto \log f_n$ seen in DGP maps but is suppressed, and, for lower frequency phenotypes, settles closer to a linear relationship between $\rho_n$ and $f_n$. We then develop a set of approximations which yield an analytically solvable model of robustness which predicts empirical robustness well outside the approximation regime.

## II. THEORY

In this study, we focus on mappings of discrete genotypes, which can be written as sequences from a fixed alphabet, onto a discrete set of phenotypes (i.e. *discrete-to-discrete GP maps*).

Let $\Omega(g) = n$ represent the mapping of genotype $g$ to phenotype $n$, where $g$ is an element of $S_{\ell,k}$, the set of all $k^\ell$ sequences of length $\ell$ drawn from an alphabet of $k$ characters. A generalization of robustness is the *transition probability* $\phi_{mn}$, the average probability that a single character mutation of a genotype mapping to phenotype $n$ will change the phenotype to $m$, with the average taken over all genotypes mapping to $n$. For DGP maps, $\phi_{mn}$ is given by

$$\phi_{mn} = \frac{\sum_{\{g|\Omega(g)=n\}} |\{h \in \mathrm{nn}(g)|\Omega(h) = m\}|}{\ell(k-1)|\{g|\Omega(g) = n\}|}. \quad (1)$$

where $\mathrm{nn}(g)$ is the single character mutational neighborhood of sequence $g$. In this formula, the numerator is counting how many single-character mutational neighbors of some genotype $g$ (which maps to phenotype $m$) map to phenotype $n$. This means that the robustness can be written as the special case $m = n$:

$$\rho_n = \phi_{nn} = \frac{\sum_{\{g|\Omega(g)=n\}} |\{h \in \mathrm{nn}(g)|\Omega(h) = n\}|}{\ell(k-1)k^\ell f_n}. \quad (2)$$

For PrGP maps, we show in Appendix A that the transition probability formula becomes a modified version of eq. (1) in which we take a *weighted* sum in the numerator. In particular, we have

$$\phi_{mn} = \frac{\sum_{\{g,h\}\in\Delta_{\ell,k}} [\mathbf{p}(g) \otimes \mathbf{p}(h) + (\mathbf{p}(g) \otimes \mathbf{p}(h))^T]_{mn}}{\ell(k-1)k^\ell f_n}, \quad (3)$$

where $\mathbf{p}(g) = (p_0(g), p_1(g), \ldots)$ with $p_n(g) = \mathbb{P}[\Omega(g) = n]$, the probability that genotype $g$ maps to phenotype $n$. Again, the robustness $\rho_n = \phi_{nn}$. In the above equation, $\Delta_{\ell,k}$ is the set of all $k^\ell \ell(k-1)/2$ unordered pairs of sequences in $S_{\ell,k}$ which differ by exactly one character. The phenotype probability vector obeys the normalization conditions $k^\ell \mathbf{f} = \sum_{g\in S_{\ell,k}} \mathbf{p}(g)$ and $1 = \sum_{n\in\{\text{phenotypes}\}} p_n(g)$ for all $g \in S_{\ell,k}$, and phenotype robustnesses are given by the diagonal of the transition probability matrix, $\rho_n = \phi_{nn}$. We also are interested in the phenotype entropy $S(g) = -\sum_{n\in\{\text{phenotypes}\}} p_n(g) \log p_n(g)$, which quantifies the spread of a genotype's mappings onto multiple phenotypes, and the genotype entropy

$$S_n^\gamma = -\sum_{g\in\{\text{genotypes}\}} \frac{p_n(g)}{f_n k^\ell} \log \frac{p_n(g)}{f_n k^\ell}, \quad (4)$$

which quantifies the spread of a phenotype across all genotypes. In particular, we will show that the genotype entropy can be useful for estimating robustness.

In DGP maps, a random null model [4] for robustness can be built by randomly assigning genotype-phenotype pairings while keeping the frequencies $\mathbf{f}$ constant. As a result, the probability of a single character mutation leading to a change from phenotype $n$ to phenotype $m$ is approximately $\phi_{mn} \approx f_m$ for all $m$. For PrGP maps, a naive expectation can be built by letting all phenotype probability vectors equal the frequency vector, $\mathbf{p}(g) = \mathbf{f}$ for all genotypes $g$. From eq. (3), one finds that $\phi_{mn} = f_m$; thus, the two random expectations are the same, even though they physically represent different scenarios.

A fundamental difference between PrGP maps and DGP maps is that DGP maps can have no frequencies lower than $k^{-\ell}$, but PrGP phenotypes in principle could have arbitrarily small frequencies, suggesting that the PrGP robustness curve has a tail, representing very rare phenotypes, that is not necessarily predictable from existing DGP robustness theory [4, 12, 34]. In this work, we show that under two approximations, the robustness becomes analytically solvable in terms of the phenotype frequency $f_n$ and genotype entropy $S_n^\gamma$. Although these approximations make specific assumptions on the shape and distribution of the phenotype's probability over genotypes which do not necessarily coincide with empirical distributions, we demonstrate in the Results section that, amazingly, the resulting robustness formula below exhibits exceptional predictive performance well outside of the approximations made on the phenotype structure on all 3 systems empirically studied.

The two key approximations are as follows: (1) a phenotype $n$ with frequency $f_n$ has probability mass evenly spread across a fixed number of genotypes, and (2) that fixed number of genotypes would be a robustness-maximizing set in the DGP sense (i.e. maximizing eq. (1)). Two central results of this paper which follow from the above assumptions (see Appendix B for the derivation) are the approximate PrGP robustness as a function of the phenotype frequency $f_n$ and the genotype entropy $S_n^\gamma$:

$$\rho_n(f_n, S_n^\gamma) \approx \frac{k^\ell f_n S_n^\gamma e^{-S_n^\gamma}}{\ell \log k}, \quad (5)$$

and approximate upper bounds on the PrGP robustness given by the piecewise continuous function

$$\rho_n^{\text{PrGP upper}}(f_n) \approx \begin{cases} \dfrac{f_n k^{\ell-1}}{\ell} & f_n \leq k^{1-\ell} \\ 1 + \dfrac{\log f_n}{\ell \log k} & f_n \geq k^{1-\ell}. \end{cases} \tag{6}$$

The upper bound illustrates two distinct scaling laws—namely, a DGP-like $\rho_n \sim a + b \log f_n$ scaling for sufficiently large frequencies, and a null model-like linear scaling $\rho_n \sim f_n$ for small frequencies. Since empirical DGP robustness often scales like a "suppressed" down-scaling of the DGP maximum $\rho_n^{\text{DGP max}} \approx 1 + \frac{\log f_n}{\ell \log k}$, the biphasic scaling of the PrGP upper bound suggests that empirical PrGP robustness may also appear biphasic and suppressed relative to the upper bound.

The upper bounds here are approximate because we rely on the genotypes forming a robustness-maximizing set (in the DGP sense, meaning the genotypes tend to be clustered in the sequence space) before we optimize or approximate the spread of the phenotype's probability mass over those genotypes. Although this may very well be an exact and/or tight bound, we do not prove its tightness here. However, we discuss specific cases of how each of these upper bounds can be achieved in real phenotypes: for rare phenotypes in the "tail" of the robustness upper bound ($f_n \leq k^{1-\ell}$), we find that phenotypes which maximize the robustness are spread evenly over exactly $k$ genotypes whose sequences all differ at exactly one character in the sequence, with each of those genotypes having probability $p_n(g) = f_n k^{\ell-1}$ of mapping to the phenotype. For more common phenotypes with $f_n \geq k^{1-\ell}$, an instructive example appears when a phenotype frequency is $f_n = k^{r-\ell}$ for some integer $1 \leq r \leq \ell$. We consider $\ell - r$ of the $\ell$ sites in the sequence to be "constrained" (using terminology from ref. [12]), meaning that mutating any of those sites will lead to a change in phenotype. The remaining $r$ sites are "unconstrained," meaning that phenotypes at those sites will not lead to any change in phenotype. If the phenotype probability is $p_n(g) = 1$ at all $k^r$ of those genotypes, the robustness is exactly equal to the DGP robustness and is simply equal to the probability of mutating an unconstrained site, namely $\rho_n = r/\ell$, which attains the upper bound in eq. (6). For frequencies in which $f_n = k^{r-\ell}$ for some non-integer value of $r$, finding a configuration in which robustness maximized is nontrivial; for DGP maps, this problem was solved in ref. [34] and the maximal robustness was found to be given by a fractal curve though it asymptotically behaves like eq. (6) with small corrections. The exact upper bound for PrGP maps remains an open problem.

In the Results section, we show that eq. (5), which is highly successful at recapitulating empirical robustness in 3 systems (RNA, spin glasses, quantum circuits), is amenable to further analytical approximation given system-specific information about the genotype entropy $S_n^\gamma$, yielding such biphasic scaling in different frequency regimes.

## III. NUMERICAL METHODS

### A. RNA

In RNA folding DGP map studies [1–11], the global free energy minimum secondary structure (reported as a "dot-bracket" string indicating polymer connectivity) was calculated for every RNA sequence of fixed length drawn from the alphabet of the four canonical nucleotides $\{\mathrm{A}, \mathrm{C}, \mathrm{G}, \mathrm{U}\}$ (alphabet size $k = 4$). Here, we are interested in not only the global free energy minimum structures but also the low-lying local minima, and we additionally investigate the temperature-dependent behavior of the robustness. We use the `RNAsubopt` program from the ViennaRNA package (version 2.4.17) [41] to calculate the secondary structures and associated Gibbs free energies for the local free energy minima within 6 kcal/mol of the global free energy minimum (or all the nonpositive free energy local minima, if the global minimum is greater than $-6$ kcal/mol). Because of the increased computational time required to discover all the local minima within an energy range, we use a reduced alphabet of $\{\mathrm{C}, \mathrm{G}\}$ for our main simulations with sequence length $\ell = 20$. A validation study with $\ell = 12$ and the full $k = 4$ alphabet is reported in the Supplementary Material (SM) [42]. Simulations for the $\ell = 20$, $k = 2$ trials were conducted at 20 °C, 37 °C (human body temperature), and 70 °C. We take the low-lying local free energy minima structures to comprise a canonical ensemble at the simulation temperature, so the probability of RNA sequence $g$ mapping to secondary structure $n$ is determined from $p_n(g) = e^{-\Delta G_n/(RT)}/Z$, where $Z$ normalizes the vector.

### B. Spin Glasses

In a previous spin glass [43, 44] DGP map study [30], a zero temperature $\pm J$ spin glass on a random graph $\mathcal{G}(V, E)$ with Hamiltonian $H(\mathbf{s}; \mathbf{J}) = -\sum_{\{i,j\} \in E} J_{ij} s_i s_j - \sum_{i \in V} h_i s_i$ was considered. The genotype is the bond configuration where each $J_{ij} \in \{-1, +1\}$, and the phenotype is the ground state configuration where each $s_i \in \{-1, +1\}$. Degeneracies of the ground state were broken by the uniformly drawn, i.i.d. random external fields $h_i \in [-10^{-4}, 10^{-4}]$ which were fixed for each simulation. In our spin glass PrGP map, we use a similar setup, but we are interested in the effect of external field disorder on robustness. We therefore incorporate the effects of Gaussian-distributed external fields $h_i \sim \mathcal{N}(h_{0,i}, \sigma_h^2)$, where the uniformly distributed means $h_{0,i} \in [-0.1, 0.1]$ are fixed across all realizations of the disorder for each simulation. To obtain accurate robustness measurements, we exactly calculate every ground state for spin glasses with $|V| = 9$, and $|E| = 15$ by exhaustive enumeration. We examine the effect of external field disorder by simulating 450 replicas of $\{h_i\}$ with variances $\sigma_h^2 = 0.001$, 0.01, and 0.1 and fixed means $\{h_{0,i}\}$. Phenotype probability vectors for each genotype $g \equiv \mathbf{J}$ were constructed by tallying and normalizing the number of appearances of each ground state across each replica. Graph topology $\mathcal{G}(V, E)$ corresponding to data presented here, as well as validation trial data, are in the SM [42].

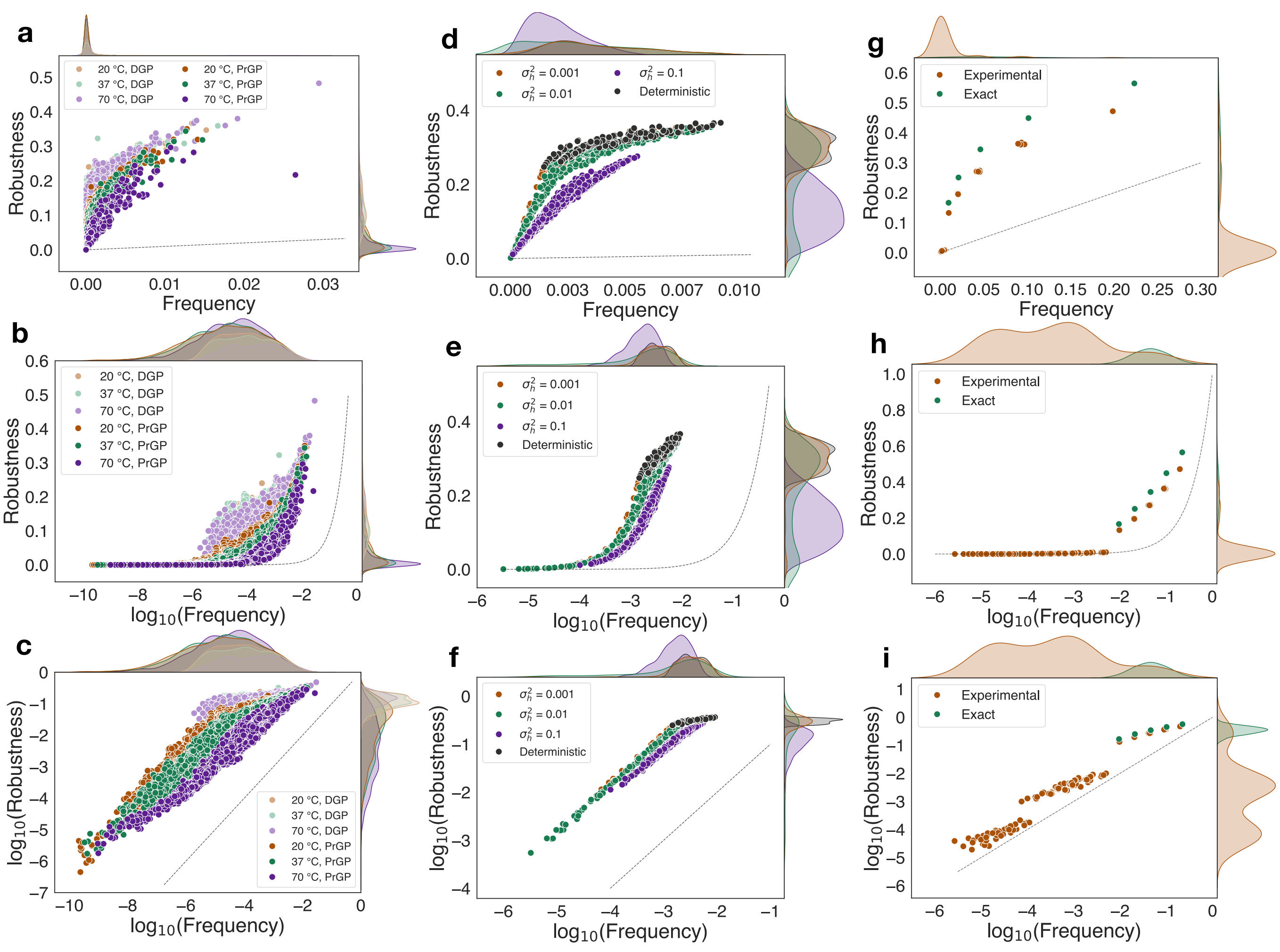

FIG. 2. Plots of (a, d, g) robustness versus frequency, (b, e, h) robustness versus $\log_{10}$(frequency), and (c, f, i) $\log_{10}$(robustness) versus $\log_{10}$(frequency) for (a, b, c) RNA folding, (d, e, f) spin glass ground state, and (g, h, i) quantum circuit PrGP maps. The dashed line is the random null expectation $\rho_n = f_n$.

### C. Quantum Circuits

Although methods to evolve quantum circuits have been suggested [45], to our knowledge this work is the first to analyze the structural properties of quantum circuit GP maps. We generate perform 7 trials in which we generate random quantum circuits (see SM for algorithm) with 7 qubits and 4 layers of gates; we also conduct an additional trial with 11 qubits and 4 layers of gates. Circuits are randomly seeded with *CNOT* gates which cannot participate in the genotype, and the remaining spaces are filled with single-qubit gates drawn from the alphabet $\{Z, X, Y, H, S, S^\dagger, T, T^\dagger\}$. We choose $\ell = 4$ ($\ell = 5$ for the 11 qubit trial) of these gates to be variable gates which comprise the genotype. The input to the circuit is the prepared state $|00\ldots0\rangle \equiv |0\rangle \otimes \cdots \otimes |0\rangle$, and the exact probability of classically measuring the basis state $|n\rangle = \bigotimes_{|q_i\rangle \in \{|0\rangle, |1\rangle\}} |q_i\rangle$ is $p_n(g) = |\langle n| U(g) |00\ldots0\rangle|^2$, where $|q_i\rangle$ is the $i$-th qubit, and $U(g)$ is the total circuit operation. We realize these quantum circuits on the *ibm_lagos* v1.2.0 quantum computer [42], one of the 7-qubit IBM Quantum Falcon r5.11H processors. Experimental phenotype probability vectors are constructed from tallying classical measurements from 1000 shots for each genotype. The 11-qubit trial is conducted on a Qiskit Aer simulator instead of an experimental quantum computer, using the *ibm_brisbane* noise profile to simulate noise. The circuits from our experimental trials are depicted in the SM [42].

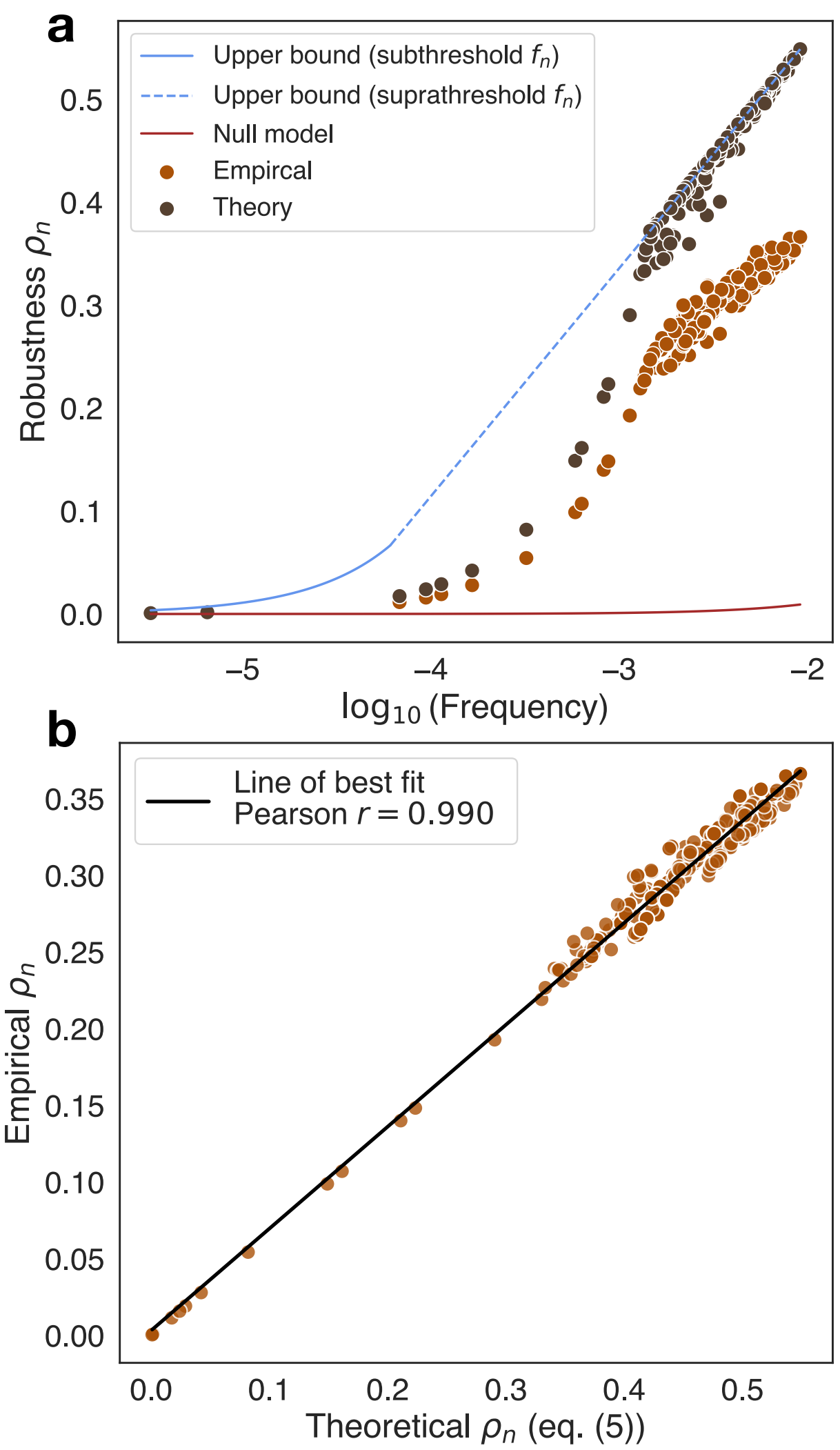

FIG. 3. (a) Plot of $\log_{10}$(frequency) versus robustness $\rho_n$, where $\rho_n$ has either been computed empirically from the experimental data or theoretically from eq. (5) for the spin glass system ($\sigma_h^2 = 0.001$). Includes upper bounds from eq. (6) and null model. (b) Scatter plot of theoretical $\rho_n$ versus empirical $\rho_n$ for the spin glass system ($\sigma_h^2 = 0.001$) with Pearson $r = 0.990$.

## IV. RESULTS

After running simulations to obtain the raw PrGP map data from the RNA, spin glass, and quantum circuit numerical experiments, we computed robustness, genotype entropy distributions, and phenotype distributions, which we plot in Figure 2, Appendix C, and Appendix E, respectively. Transition probabilities between different phenotypes and the RNA $k = 4, \ell = 12$ genotype entropy distribution are plotted in the SM [42]. As noted previously, validation trial data for spin glasses on a different random graph as well as multiple experimental quantum circuit trials' data are also provided in the SM [42]. The SM also contains a table in which we note the frequencies of the RNA "unfolded" phenotype; notably, in the $k = 2, \ell = 20$ cases, the unfolded phenotype frequency is less than 3% while for $k = 4, \ell = 12$ case, the unfolded phenotype frequency is more than 80% and there are much fewer phenotypes, as expected from the RNA12 DGP study [3].

In Figure 2 we plot robustness versus frequency, robustness versus log frequency, and log robustness versus log frequency for each of the 3 main systems studied (additional RNA, spin glass, and quantum circuit trials are in the SM [42]). Notable common features across all systems include robustness much higher than predicted by the null model for sufficiently large frequencies and a convergence toward the null model behavior for sufficiently small frequencies. The RNA PrGP maps, all show suppressed robustness relative to their DGP counterparts, and this scaling is further suppressed as temperature increases.

Similarly, in spin glasses, the DGP robustness is highest and closest to the linear-log relationship; the PrGP maps show increasingly suppressed scaling as the disorder variance is tuned higher. In quantum circuit PrGP maps, the trials with experimental or simulated noise show the appearance of a long tail of many new small-frequency phenotypes with, leading to the suppression of the robustness of the large-frequency phenotypes with a maintenance of the approximate $\log f_n$ scaling.

From the phenotype entropy distributions in Appendix E, we see that as disorder parameters are increased (temperature, field variance, measurement noise), phenotype entropy distributions widen, meaning a genotype is more likely to have a broader distribution of phenotypes to which it maps.

We now make predictions of robustness by directly plugging in measurements of $S_n^\gamma$ and $f_n$ into eq. (5). We show an example plot of the theoretical robustness, empirical robustness, null model, and upper bound for spin glasses with $\sigma_h^2 = 0.001$ in Figure 3(a). Not only does the the theoretical robustness, given only $S_n^\gamma$ and $f_n$, recapitulate the salient scaling behavior of the empirical robustness, as shown in Figure 3(b), but the Pearson correlation between the predicted and empirical robustness is $r = 0.990$; in Table II, we show that the Pearson correlations from robustness obtained from eq. (5) for all systems tabulated range from 0.947-0.99994 and outperformed the null model and DGP maximum robustness formulas across all systems, illustrating the success of eq. (5). While the Pearson correlations are high, the prediction from eq. (5) varies by additive or multiplicative constant factors likely due to violation of one or both assumptions mentioned in the Theory section. As disorder parameters increase, these violations become more prominent and eq. (5) and the null model's relative performance becomes better (see Table II), meaning that biphasic scaling starts to fade away in favor of null model-like linear scaling when there is too much disorder. Nonetheless, in all cases our analytical theory performs the best and remains highly predictive.

| System | Details | Theory vs. Real Pearson $r$ | Null vs. Real Pearson $r$ | DGP Max vs. Real Pearson $r$ |
|---|---|---|---|---|
| Spin Glass | $\sigma_h^2 = 0.001$ | **0.990** | 0.766 | 0.940 |
| Spin Glass | $\sigma_h^2 = 0.01$ | **0.994** | 0.874 | 0.924 |
| Spin Glass | $\sigma_h^2 = 0.1$ | **0.995** | 0.976 | 0.954 |
| RNA GC20 | 20 °C | **0.947** | 0.811 | 0.797 |
| RNA GC20 | 37 °C | **0.951** | 0.854 | 0.766 |
| RNA GC20 | 70 °C | **0.965** | 0.856 | 0.665 |
| RNA12 | 37 °C | **0.958** | 0.832 | 0.882 |
| Quantum Circuit | 11 qubit (exact) | **0.99994** | 0.901 | 0.997 |
| Quantum Circuit | 11 qubit (simulation) | **0.9991** | 0.865 | 0.576 |
| Quantum Circuit | 7 qubit, trial 1 (exact) | **0.99994** | 0.926 | 0.998 |
| Quantum Circuit | 7 qubit, trial 1 (exp.) | **0.9996** | 0.912 | 0.712 |

TABLE II. Pearson correlation coefficient $r$ between robustness predicted from eq. (5) versus empirically measured robustness. In general, the theory outperforms the null model and the DGP maximum approximations, and overall Pearson correlations are very high, close to 1, highlighting the success of eq. (5). Bold indicates the best-performing model.

We now combine empirical results for $S_n^\gamma$ versus $f_n$ with eq. (5) to develop a semi-empirical theory for understanding how robustness $\rho_n$ scales versus $f_n$. We observe in Appendix C that genotype entropy, which is exactly $S_n^\gamma = \ell \log k + \log f_n$ for DGP maps, empirically maintains similar scaling

$$S_n^\gamma \approx \alpha + \eta \log f_n \tag{7}$$

for some $\alpha$ and $\eta$ in PrGP maps, but is generally increased with respect to the DGP genotype entropy and generally with $0 \leq \eta \leq 1$. This means that a phenotype with fixed frequency is likely to be spread out over more genotypes in the PrGP case than in the DGP case, as expected. This relationship tends to hold over many orders of magnitude, for all 3 systems, though with slightly differing behavior. For instance, there are some cases where $\eta$ depends on $f_n$, but is constant for large stretches of frequencies. One example is spin glasses with $\sigma_h^2 = 0.001$, where $\eta \approx 1$ for most common phenotypes and then suddenly transitions to $\eta \approx 0$ for sufficiently small frequencies. Regardless, eq. (7) can still be combined with eq. (5) to understand how $\rho_n$ scales with $f_n$, and different values of $\eta$ can be used in limits of small or large $f_n$.

Substituting in eq. (7) into eq. (5), we have the robustness expression

$$\rho_n = \frac{k^\ell}{e^\alpha \ell \log k} \left[ f^{1-\eta} \left( \alpha + \eta \log f_n \right) \right], \tag{8}$$

Notably, when $\eta = 0$ (e.g. for sufficiently small frequencies in the spin glass $\sigma_h^2 = 0.001$), eq. (8) becomes $\rho_n \sim f_n$. In RNA, quantum circuits, and the spin glass $\sigma_h^2 = 0.1$ case, a slope $0 < \eta < 1$ is observed. In these cases, we can see from the formula above that $\rho_n \to -\infty$ when $f_n \to 0$. However, the appropriate limit should actually take into account the fact that the smallest phenotype frequency, $f_{\min}$, is finite. We show in Appendix D that when $f_n \geq f_{\min} \gg e^{-\alpha/\eta}$, which is the case for the empirical systems in which $0 < \eta < 1$, then a power law relationship $\log \rho_n \sim C + (1-\eta) \log f_n$ is expected for many orders of magnitude of frequency. Only after frequencies are so small that $f_{\min} \gg e^{-\alpha/\eta}$ is violated would a sharp divergence to $-\infty$ occur, but empirical frequencies observed in this study do not reach this regime. In Figure 2, we indeed observe a clear power law relationship for sufficiently small frequencies, which of course simplifies to the aforementioned linear relationship when $\eta = 0$. When $\eta$ is small but not 0, it may be difficult to distinguish a power law from a linear relationship from Figure 2.

Lastly, for sufficiently large frequencies, for $0 < \eta < 1$, we generally have a complex behavior $\rho \sim f^{1-\eta} \left( \alpha + \eta \log f_n \right)$ which can be expanded to leading order in $f_n$ or to leading order in $\log f_n$, depending on the variable choice. For example, substituting $x_n = \log f_n$ into eq. (8), the leading order behavior for small $x_n$ becomes $\rho_n \sim a + b \log f_n$, which is the expected large frequency behavior. It is important to note that, for examples such as the aforementioned spin glasses with $\sigma_h^2 = 0.001$ and even $\sigma_h^2 = 0.01$, $\eta \approx 1$ yields the "robust" logarithmic scaling seen in robust DGP maps and the PrGP upper bound computed here. Moreover, in the RNA systems, for sufficiently high frequencies the $S_n^\gamma$ versus $f_n$ plot points reach the exact DGP $S_n^\gamma$ curve, which has $\eta = 1$. In quantum circuits, the general trend of $S_n^\gamma$ is to lay parallel or almost parallel to the DGP $S_n^\gamma$ curve, which also suggests $\eta \approx 1$ or slightly less than 1. However, there are clusters of phenotypes with $0 < \eta < 1$, which would each have different estimated $\alpha$ values. This leads to a fragmented genotype entropy curve and robustness curve which may not be entirely explained by the monotonic behavior predicted by eq. (8), though combining empirical measurements in eq. (5) still yields excellent predictions. Nonetheless, different regimes of eq. (8) provide evidence for the complex behavior seen in PrGP map robustness.

## V. DISCUSSION

Compared to existing DGP maps, PrGP maps not only allow for the inclusion of realistic, physical sources of disorder like thermal fluctuation and external variables, but they also permit the analysis of new systems like quantum circuits with inherent uncertainty. We emphasize the broad applicability of this framework to a vast array of systems across biology, physics, and computer science, and other disciplines for the analysis of robustness and stability. The analytical theory introduced here, which functions well outside of the approximation regimes used to derive it, provides a link between a phenotype's frequency $f_n$, the genotype entropy $S_n^\gamma$, and the robustness $\rho_n$. Given the empirical observation of a logarithmic relationship between $S_n^\gamma$ and $f_n$, we can show that for high frequencies a complex $\rho \sim f^{1-\eta}(\alpha + \eta \log f_n)$ robustness relationship, which becomes linear-log (DGP-like) robustness when $\eta = 1$ is obtained, while for small frequencies linear or power law relationship is expected, depending on system specific information. Moreover, as disorder in a system is increased, phenotypes spread over a larger number of genotypes, leading to increasingly suppressed robustness and more null model-like behavior. Most notably, our theory in eq. (5) is highly successful, measured by Pearson correlation, in predicting empirical robustness across all systems.

The scaling we observe empirically and justify theoretically in this article is observed in all three studied systems, despite being disparate, hinting at its universality. How this robustness trend affects navigability of (probabilistic) fitness landscapes is an important direction for further investigation. We suggest that evolutionary dynamics on fitness landscapes where the genotype-to-phenotype mapping is probabilistic may display unique phenomena which are not present on fitness landscapes with purely deterministic GP mapping.

We also suggest that the mapping of genotypes to probability vectors instead of discrete phenotypes may facilitate the taking of gradients of, for instance, a negative loss-likelihood loss function in the process of learning PrGP or even DGP maps using statistical learning methods. Specifically, one might model a GP map using a graph neural network [46] and predict the phenotype or related properties of neighboring nodes. Such a model may ultimately aid in inferring fitness landscapes from limited initial GP data [47–49].

## VI. ACKNOWLEDGEMENTS

We acknowledge helpful discussions with Nora Martin, comments from anonymous referees, and the use of IBM Quantum services and the MIT Engaging Cluster for this work. This work was supported by awards T32GM007753 and T32GM144273 from the National Institute of General Medical Sciences, Hertz Foundation Fellowships (VM; AS), and a PD Soros Fellowship (VM). The content is solely the responsibility of the authors and does not necessarily represent the official views of the National Institute of General Medical Sciences, the National Institutes of Health, IBM, or the IBM Quantum Team. The authors declare no known conflict of interest.

[1] M. Weiß and S. E. Ahnert, Neutral components show a hierarchical community structure in the genotype–phenotype map of RNA secondary structure, Journal of The Royal Society Interface **17**, 20200608 (2020).

[2] M. Weiß and S. E. Ahnert, Using small samples to estimate neutral component size and robustness in the genotype–phenotype map of RNA secondary structure, Journal of The Royal Society Interface **17**, 20190784 (2020).

[3] J. Aguirre, J. M. Buldú, M. Stich, and S. C. Manrubia, Topological Structure of the Space of Phenotypes: The Case of RNA Neutral Networks, PLoS ONE **6**, e26324 (2011).

[4] S. F. Greenbury, S. Schaper, S. E. Ahnert, and A. A. Louis, Genetic Correlations Greatly Increase Mutational Robustness and Can Both Reduce and Enhance Evolvability, PLOS Computational Biology **12**, e1004773 (2016).

[5] S. F. Greenbury, A. A. Louis, and S. E. Ahnert, The structure of genotype-phenotype maps makes fitness landscapes navigable, Nature Ecology & Evolution **6**, 1742 (2022).

[6] K. Dingle, S. Schaper, and A. A. Louis, The structure of the genotype–phenotype map strongly constrains the evolution of non-coding RNA, Interface Focus **5**, 20150053 (2015).

[7] K. Dingle, C. Q. Camargo, and A. A. Louis, Input–output maps are strongly biased towards simple outputs, Nature Communications **9**, 761 (2018).

[8] K. Dingle, F. Ghaddar, P. Šulc, and A. A. Louis, Phenotype Bias Determines How Natural RNA Structures Occupy the Morphospace of All Possible Shapes, Molecular Biology and Evolution **39**, msab280 (2022).

[9] K. Dingle, G. V. Pérez, and A. A. Louis, Generic predictions of output probability based on complexities of inputs and outputs, Scientific Reports **10**, 4415 (2020).

[10] T. Jörg, O. C. Martin, and A. Wagner, Neutral network sizes of biological RNA molecules can be computed and are not atypically small, BMC Bioinformatics **9**, 464 (2008).

[11] A. Wagner, Robustness and evolvability: a paradox resolved, Proceedings of the Royal Society B: Biological Sciences **275**, 91 (2008).

[12] S. F. Greenbury, I. G. Johnston, A. A. Louis, and S. E. Ahnert, A tractable genotype–phenotype map modelling the self-assembly of protein quaternary structure, Journal of The Royal Society Interface **11**, 20140249 (2014).

[13] S. Tesoro and S. E. Ahnert, Non-deterministic genotype-phenotype maps of biological self-assembly, EPL (Europhysics Letters) **123**, 38002 (2018).

[14] C. Q. Camargo and A. A. Louis, Boolean Threshold Networks as Models of Genotype-Phenotype Maps, Complex Networks XI , 143 (2020).

[15] S. Kauffman, Homeostasis and Differentiation in Random Genetic Control Networks, Nature **224**, 177 (1969).

[16] A. Wagner, Distributed robustness versus redundancy as causes of mutational robustness, BioEssays **27**, 176 (2005).

[17] A. Wagner, *Robustness and evolvability in living systems*, 3rd ed., Princeton studies in complexity (Princeton Univ. Press, Princeton, NJ, 2007) oCLC: 845177181.

[18] J. L. Payne and A. Wagner, Constraint and Contingency in Multifunctional Gene Regulatory Circuits, PLoS Computational Biology **9**, e1003071 (2013).

[19] J. L. Payne, J. H. Moore, and A. Wagner, Robustness, evolvability, and the logic of genetic regulation, Artificial Life **20**, 111 (2014).

[20] J. L. Payne and A. Wagner, The Robustness and Evolvability of Transcription Factor Binding Sites, Science **343**, 875 (2014).

[21] S. Schaper and A. A. Louis, The Arrival of the Frequent: How Bias in Genotype-Phenotype Maps Can Steer Populations to Local Optima, PLoS ONE **9**, e86635 (2014).

[22] S. F. Greenbury and S. E. Ahnert, The organization of biological sequences into constrained and unconstrained parts determines fundamental properties of genotype–phenotype maps, Journal of The Royal Society Interface **12**, 20150724 (2015).

[23] S. E. Ahnert, Structural properties of genotype–phenotype maps, Journal of The Royal Society Interface **14**, 20170275 (2017).

[24] M. Weiß and S. E. Ahnert, Phenotypes can be robust and evolvable if mutations have non-local effects on sequence constraints, Journal of The Royal Society Interface **15**, 20170618 (2018).

[25] D. Nichol, M. Robertson-Tessi, A. R. A. Anderson, and P. Jeavons, Model genotype–phenotype mappings and the algorithmic structure of evolution, Journal of The Royal Society Interface **16**, 20190332 (2019).

[26] T. Hu, M. Tomassini, and W. Banzhaf, A network perspective on genotype–phenotype mapping in genetic programming, Genetic Programming and Evolvable Machines 10.1007/s10710-020-09379-0 (2020).

[27] S. Manrubia, J. A. Cuesta, J. Aguirre, S. E. Ahnert, L. Altenberg, A. V. Cano, P. Catalán, R. Diaz-Uriarte, S. F. Elena, J. A. García-Martín, P. Hogeweg, B. S. Khatri, J. Krug, A. A. Louis, N. S. Martin, J. L. Payne, M. J. Tarnowski, and M. Weiß, From genotypes to organisms: State-of-the-art and perspectives of a cornerstone in evolutionary dynamics, Physics of Life Reviews **38**, 55 (2021).

[28] J. L. Payne and A. Wagner, The causes of evolvability and their evolution, Nature Reviews Genetics **20**, 24 (2019).

[29] S. Schaper, I. G. Johnston, and A. A. Louis, Epistasis can lead to fragmented neutral spaces and contingency in evolution, Proceedings of the Royal Society B: Biological Sciences **279**, 1777 (2012).

[30] V. Mohanty and A. A. Louis, Robustness and stability of spin-glass ground states to perturbed interactions, Physical Review E **107**, 014126 (2023), publisher: American Physical Society.

[31] A. H. Wright and C. L. Laue, Evolving Complexity is Hard (2022), arXiv:2209.13013 [cs].

[32] I. G. Johnston, K. Dingle, S. F. Greenbury, C. Q. Camargo, J. P. K. Doye, S. E. Ahnert, and A. A. Louis, Symmetry and simplicity spontaneously emerge from the algorithmic nature of evolution, Proceedings of the National Academy of Sciences **119**, e2113883119 (2022).

[33] V. Mohanty, *Robustness of evolutionary and glassy systems*, Ph.D. thesis, University of Oxford (2021).

[34] V. Mohanty, S. F. Greenbury, T. Sarkany, S. Narayanan, K. Dingle, S. E. Ahnert, and A. A. Louis, Maximum mutational robustness in genotype–phenotype maps follows a self-similar blancmange-like curve, Journal of The Royal Society Interface **20**, 20230169 (2023), publisher: Royal Society.

[35] J. A. Draghi, T. L. Parsons, G. P. Wagner, and J. B. Plotkin, Mutational robustness can facilitate adaptation, Nature **463**, 353 (2010), number: 7279 Publisher: Nature Publishing Group.

[36] R. H. Y. Louie, K. J. Kaczorowski, J. P. Barton, A. K. Chakraborty, and M. R. McKay, Fitness landscape of the human immunodeficiency virus envelope protein that is targeted by antibodies, Proceedings of the National Academy of Sciences **115**, E564 (2018).

[37] T. C. Butler, J. P. Barton, M. Kardar, and A. K. Chakraborty, Identification of drug resistance mutations in HIV from constraints on natural evolution, Physical Review E **93**, 022412 (2016).

[38] J. P. Barton, N. Goonetilleke, T. C. Butler, B. D. Walker, A. J. McMichael, and A. K. Chakraborty, Relative rate and location of intra-host HIV evolution to evade cellular immunity are predictable, Nature Communications **7**, 11660 (2016).

[39] K. Shekhar, C. F. Ruberman, A. L. Ferguson, J. P. Barton, M. Kardar, and A. K. Chakraborty, Spin models inferred from patient-derived viral sequence data faithfully describe HIV fitness landscapes, Physical Review E **88**, 062705 (2013).

[40] T. A. Hopf, J. B. Ingraham, F. J. Poelwijk, C. P. I. Schärfe, M. Springer, C. Sander, and D. S. Marks, Mutation effects predicted from sequence co-variation, Nature Biotechnology **35**, 128 (2017).

[41] R. Lorenz, S. H. Bernhart, C. Höner zu Siederdissen, H. Tafer, C. Flamm, P. F. Stadler, and I. L. Hofacker, ViennaRNA Package 2.0, Algorithms for Molecular Biology **6**, 26 (2011).

[42] See Supplemental Material for this work.

[43] S. F. Edwards and P. W. Anderson, Theory of spin glasses, Journal of Physics F: Metal Physics **5**, 965 (1975).

[44] D. Sherrington and S. Kirkpatrick, Solvable Model of a Spin-Glass, Physical Review Letters **35**, 1792 (1975).

[45] D. Tandeitnik and T. Guerreiro, Evolving Quantum Circuits (2022), arXiv:2210.05058 [quant-ph].

[46] T. N. Kipf and M. Welling, enSemi-Supervised Classification with Graph Convolutional Networks (2017), arXiv:1609.02907 [cs, stat].

[47] R. G. Shaw and C. J. Geyer, Inferring Fitness Landscapes, Evolution **64**, 2510 (2010), publisher: [Society for the Study of Evolution, Wiley].

[48] T. Nozoe, E. Kussell, and Y. Wakamoto, enInferring fitness landscapes and selection on phenotypic states from single-cell genealogical data, PLOS Genetics **13**,

e1006653 (2017), publisher: Public Library of Science.
[49] S. D'Costa, E. C. Hinds, C. R. Freschlin, H. Song, and P. A. Romero, Inferring protein fitness landscapes from laboratory evolution experiments, PLOS Computational Biology **19**, e1010956 (2023).
[50] O. E. Galkin and S. Y. Galkina, Global extrema of the Delange function, bounds for digital sums and concave functions, Sbornik: Mathematics **211**, 336 (2020).

### Appendix A: Definitions of DGP and PrGP Map Transition Probability and Robustness

Let a genotype-phenotype map $\Omega$ map a genotype $g$ to phenotype $n$ be denoted by $\Omega(g) = n$. The set of all genotypes $g$ which map to phenotype $n$ is called a *neutral set* and is given by $\{g|\Omega(g) = n\}$. As stated in the main text, the transition probability $\phi_{mn}$ is the average probability that a single character mutation of a genotype mapping to phenotype $n$ will change the phenotype to $m$, with the average taken over all genotypes mapping to $n$. This is exactly

$$\phi_{mn} = \frac{\sum_{\{g|\Omega(g)=n\}} |\{h \in \mathrm{nn}(g)|\Omega(h) = m\}|}{\ell(k-1)|\{g|\Omega(g) = n\}|}. \tag{A1}$$

The numerator is the number of neighbors of some genotype $g$ which map to a phenotype $m$, averaged over all genotypes $g$ which map to a phenotype $n$. The *robustness* of a phenotype $n$ is the probability that a neighboring genotype maps back onto the same phenotype $n$:

$$\rho_n = \phi_{nn}. \tag{A2}$$

We note that we can write

$$|\{h \in \mathrm{nn}(g)|\Omega(h) = m\}| = \sum_{h\in \mathrm{nn}(g)} \mathbb{I}[\Omega(h) = m], \tag{A3}$$

where $\mathbb{I}[\cdot]$ is an indicator function, and similarly

$$|\{g|\Omega(g) = n\}| = \sum_{g\in S_{\ell,k}} \mathbb{I}[\Omega(g) = n], \tag{A4}$$

where $S_{\ell,k}$ is the set of all sequences of length $\ell$ drawn from an alphabet of $k$ letters. We can also write the summation

$$\sum_{\{g|\Omega(g)=n\}} y(g) = \sum_{g\in S_{\ell,k}} \mathbb{I}[\Omega(g) = n]y(g). \tag{A5}$$

Putting together all three substitutions, we have

$$\phi_{mn} = \frac{\sum_{g\in S_{\ell,k}} \mathbb{I}[\Omega(g) = n] \sum_{h\in \mathrm{nn}(g)} \mathbb{I}[\Omega(h) = m]}{\ell(k-1)\sum_{g\in S_{\ell,k}} \mathbb{I}[\Omega(g) = n]}. \tag{A6}$$

Now, we relax the indicator functions to probabilities $\mathbb{I}[\Omega(g) = n] \mapsto \mathbb{P}[\Omega(g) = n]$. Note that this preserves the meaning of the transition probability—namely, that of an average probability of a change from phenotype $m$ to $n$—as we now are weighting the contribution of every genotype by the probability that that genotype maps to phenotype $m$ or $n$. So, we have for PrGP maps,

$$\phi_{mn} = \frac{\sum_{g\in S_{\ell,k}} \mathbb{P}[\Omega(g) = n] \sum_{h\in \mathrm{nn}(g)} \mathbb{P}[\Omega(h) = m]}{\ell(k-1)\sum_{g\in S_{\ell,k}} \mathbb{P}[\Omega(g) = n]}. \tag{A7}$$

Now, writing $p_n(g) = \mathbb{P}[\Omega(g) = n]$, and rearranging the numerator, we have

$$\phi_{mn} = \frac{\sum_{g\in S_{\ell,k}} \sum_{h\in \mathrm{nn}(g)} p_n(g)p_m(h)}{\ell(k-1)\sum_{g\in S_{\ell,k}} p_n(g)}. \tag{A8}$$

We recognize that

$$\sum_{g\in S_{\ell,k}} p_n(g) = f_n k^\ell \tag{A9}$$

because the total probability of encountering phenotype $n$ is still $f_n$. We also recognize that the sums

$$\sum_{g\in S_{\ell,k}} \sum_{h\in \mathrm{nn}(g)} y(g,h) = \sum_{\{g,h\}\in\Delta_{\ell,k}} [y(g,h)+y(h,g)], \tag{A10}$$

where the left hand summations represent a sum over all genotypes and each genotype's neighbors, whereas the right hand summations are a sum over all pairs of neighboring genotypes. We now have

$$\phi_{mn} = \frac{\sum_{\{g,h\}\in\Delta_{\ell,k}} [\mathbf{p}(g)\otimes\mathbf{p}(h) + (\mathbf{p}(g)\otimes\mathbf{p}(h))^T]_{mn}}{\ell(k-1)k^\ell f_n}, \tag{A11}$$

where we use the vector of phenotype probabilities $\mathbf{p}(g) = (p_0(g), p_1(g), \ldots)$ at a particular genotype for convenience.

## Appendix B: Analytical Derivation of Biphasic Robustness Curve for PrGP maps

Here we show an analytical derivation of the biphasic robustness curve which supports the empirical observations that we report in this paper that for larger frequencies $\rho_n \propto \log f_n$ as in the DGP case and $\rho_n \propto f_n$.

### 1. Review of Recent Analytical Results for DGP Robustness

We first review recent exact analytical results on the upper bound of robustness in DGP maps, shown recently by one of the authors [34]. In DGP maps, the maximum mutational robustness for a phenotype $n$ which has frequency $f_n = m_n k^\ell$ is given by

$$\rho_n^{\mathrm{DGP\ max}}(m_n) = \frac{2S_k(m_n)}{m_n\ell(k-1)} = \frac{\log_k m_n}{\ell} - \frac{g_k(k^{\{log_k m_n\}})}{\ell(k-1)}, \tag{B1}$$

where $\{x\}$ indicates the fractional part of $x$, $S_k(m) = \sum_{i=0}^{m-1} s_k(m)$, with $s_k(m)$ being the sum of all the digits in the base-$k$ representation of integer $m = fk^\ell$, where $f$ is the frequency of the phenotype, and

$$g_k(x) = (k-1)\log_k x + \frac{D_k(x)}{x}. \tag{B2}$$

Here, $D_k(x)$ is the modified Delange function defined by

$$D_k(x) = \sum_{n=0}^{\infty} \frac{D_{k,0}(k^x)}{k^n}, \quad D_{k,0}(x) = \int_0^x \mathrm{d}t\,(2k[t] - 2[kt] + k - 1), \tag{B3}$$

where $[x]$ is the integer part of $x$. The function above is a rescaling and shifting of the self-similar, continuous-everywhere, but differentiable-nowhere Takagi function or blancmange curve. The maximum mutational robustness curve is tightly bounded [34]

$$\frac{\log_k m_n}{\ell} + \frac{2A_k}{(k-1)\ell} \leq \rho_n^{\mathrm{DGP\ max}} \leq \frac{\log_k m_n}{\ell}, \tag{B4}$$

where [50] has defined

$$A_k = \frac{k}{2}\left[1 - \frac{\log\log k}{\log k} + \mathcal{O}\left(\frac{1}{\log k}\right)\right]. \tag{B5}$$

$A_k$ is typically very small relative to the value of the robustness, so the maximum robustness is well-approximated by

$$\rho_n^{\text{DGP max}} \approx \frac{\log_k m_n}{\ell} = 1 + \frac{\log f_n}{\ell \log k}, \quad \frac{1}{k^\ell} \leq f_n \leq 1 \tag{B6}$$

We find from empirical studies [4, 30] that empirical robustness in many DGP systems is, **due to noise in the** GP map, a suppression of the maximum robustness curve, *i.e.*

$$\rho_n^{\text{DGP emp}} \approx a + b \log f_n. \tag{B7}$$

generally with $a < 1$ and for various $b$. This motivates our search for an equivalent robustness bound which, when suppressed, matches the empirical observations we have shown in RNA, spin glasses, and quantum circuits.

### 2. Derivation of the Biphasic PrGP Robustness and Upper Bound

To justify why biphasic robustness is necessary/expected, we first point out a critical difference between PrGP maps and DGP maps. DGP maps have a fundamental limit on the lower bound of the frequency of a phenotype—namely, in DGP maps, since a single genotype maps to a single phenotype, a phenotype with nonzero frequency must be mapped to by at least one genotype. Therefore, the lower bound on the phenotype frequency for DGP maps is $f_n = 1/k^\ell$, and for this frequency the DGP robustness is always 0 since $S_k(0) = 0$; intuitively, a single genotype (node) will have zero neighbors which map to the same phenotype, so this phenotype must necessarily have zero robustness. But, PrGP maps have no such restriction on the lower bound of the frequency of a phenotype. In fact, we find many phenotypes which occur at frequencies $f_n < 1/k^\ell$. Therefore, we must expect a different scaling below $f_n < 1/k^\ell$, and perhaps even for some frequencies $f_n \geq 1/k^\ell$. We will now show this.

For sufficiently large frequencies $f_n$, we expect the PrGP robustness to obey at least

$$\rho_n^{\text{PrGP}} \lesssim \rho_n^{\text{DGP max}} \approx 1 + \frac{\log f_n}{\ell \log k} \tag{B8}$$

We will later be able to derive a condition for "sufficiently large." Now, consider a phenotype with small frequency $f_n$; for our current purposes, "small" means $f_n < 1/k^\ell$, but we will actually later find that this holds for some frequencies $f_n \geq 1/k^\ell$. Placing all of the probability mass $f_n$ of this phenotype onto a single genotype ensures that the robustness is zero. If we were to instead map one genotype to phenotype $n$ with probability $f_n/2$ and a neighboring genotype to the same phenotype $n$ also with probability $f_n/2$, we'd find that the robustness is larger than zero. So, the question is, over how many nodes $\xi_n \times k^\ell \geq 1$ should the probability mass $f_n$ be spread out so that the robustness is maximized? Now, $\xi_n$ is normalized to be between $1/k^\ell$ and 1.

#### a. *PrGP Robustness Approximations Enable Analytical Solution*

We now take a graph theoretic approach. We consider the space of all genotypes to be the Hamming graph $H_{\ell,k}$, a graph in which each of the $k^\ell$ vertices represents a genotype, and an edge exists between two vertices if and only if those two corresponding genotype sequences differ by exactly one character. Let $G(\xi_n)$ be the subgraph containing $|V(G(\xi_n))| = \xi_n k^\ell$ number of vertices over which the $n$-th phenotype's probability mass is spread. It can readily be shown from the definition of robustness given in the main text (or, for example, in [33, 34]) that DGP robustness in graph theoretic terms is

$$\rho(\xi_n) = \frac{2}{\ell(k-1)} \frac{|E(G(\xi_n))|}{|V(G(\xi_n))|}, \tag{B9}$$

where $E(G(\xi_n))$ is the set of edges of $G(\xi_n)$, and $V(G(\xi_n))$ is the set of vertices of $G$. We now readily substitute $|V(G(\xi_n))| = \xi_n k^\ell$ to find that

$$|E(G(\xi_n))| = \frac{k^\ell \ell(k-1)}{2} \xi_n \rho(\xi_n) \tag{B10}$$

$$\rho_n(\xi_n) \approx 1 + \frac{\log \xi_n}{\ell \log k}. \tag{B11}$$

Strictly, the equation above should be an inequality, but assuming that $G$ is maximally robust means that eq. (C3) applies (see previous section). It now follows that

$$|E(G(\xi_n))| \approx \frac{k^\ell \ell(k-1)}{2}\xi_n \left[1 + \frac{\log \xi_n}{\ell \log k}\right]. \tag{B12}$$

Now, we can use the main text definition of transition probability to compute the PrGP robustness for a phenotype with frequency $f_n$ whose probability mass is spread optimally (in the sense of robustness-maximizing) over $\xi_n k^\ell$ nodes. From the main text, we have the PrGP transition probability

$$\phi_{mn} = \frac{\sum_{\{g,h\}\in\Delta_{\ell,k}} [\mathbf{p}(g)\otimes\mathbf{p}(h) + (\mathbf{p}(g)\otimes\mathbf{p}(h))^T]_{mn}}{\ell(k-1)k^\ell f_n}, \tag{B13}$$

where we can identify $\Delta_{\ell,k} \equiv E(H_{\ell,k})$ as the edge set of the Hamming graph $H_{\ell,k}$ (genotype space). For small $f_n$, the maximum PrGP robustness is thus

$$\begin{aligned}
\rho_n(f_n) = \phi_{nn} &= \frac{\sum_{\{g,h\}\in\Delta_{\ell,k}} [\mathbf{p}(g)\otimes\mathbf{p}(h) + (\mathbf{p}(g)\otimes\mathbf{p}(h))^T]_{nn}}{\ell(k-1)k^\ell f_n} \\
&= \frac{\sum_{\{g,h\}\in\Delta_{\ell,k}} p_n(g)p_n(h) + p_n(h)p_n(g)}{\ell(k-1)k^\ell f_n} \\
&= \frac{2\sum_{\{g,h\}\in\Delta_{\ell,k}} p_n(g)p_n(h)}{\ell(k-1)k^\ell f_n}.
\end{aligned} \tag{B14}$$

We now note that, within our approximation scheme, for $g \in G(\xi_n)$, $p_n(g) = f_n/\xi_n$, and for $g \notin G(\xi_n)$, $p_n(g) = 0$. Therefore,

$$\begin{aligned}
\sum_{\{g,h\}\in\Delta_{\ell,k}} p_n(g)p_n(h) &\approx \left(\frac{f_n}{\xi_n}\right)^2 \sum_{\{g,h\}\in E(G(\xi_n))} 1 \\
&= \left(\frac{f_n}{\xi_n}\right)^2 |E(G(\xi_n))| \\
&\approx \frac{k^\ell \ell(k-1)}{2}\frac{f_n^2}{\xi_n}\left[1 + \frac{\log \xi_n}{\ell \log k}\right].
\end{aligned} \tag{B15}$$

Finally, we can write that for small $f_n$, the maximum PrGP robustness is

$$\rho_n(f_n) \approx \frac{f_n}{\xi_n}\left[1 + \frac{\log \xi_n}{\ell \log k}\right], \tag{B16}$$

which is eq. (5) in the main text. This an approximate upper bound due to the assumption that the probability mass is equally spread over the $\xi_n k^\ell$ nodes, though we expect it to be greater than most, if not all, real PrGP robustness values because the network of $\xi_n k^\ell$ nodes is optimally robust.

It is apparent that measurement of the cluster size $\xi_n k^\ell$ may depend substantially on the measurement resolution at a given genotype and/or on some arbitrary threshold value at which a phenotype is "detected" at a genotype. That is, a phenotype which has extremely low probability at a particular genotype may lead to undercounting or overcounting (depending on measurement resolution and detection threshold) of $\xi_n k^\ell$. To address this issue, we can use the relationship eq. (C6):

$$S_n^{\gamma,\text{PrGP approx}} = \ell \log k + \log \xi_n, \tag{B17}$$

later derived in Appendix C, as a better measurement of the phenotype's "spread" or effective number of genotypes occupied by the phenotype's distribution over all genotypes. Substituting this into eq. (B16), we can instead write

$$\rho_n(f_n) \approx \frac{k^\ell f_n S_n^\gamma e^{-S_n^\gamma}}{\ell \log k}. \tag{B18}$$

Within the approximation scheme presented in this paper, eq. (B16) and eq. (B18) provide the same result, but empirically measuring and plugging in $S_n^\gamma$ or $\xi_n$ may lead to differences in the final computed robustness. In this paper, we treat eq. (B18) as the primary analytical result since $S_n^\gamma$ is more rigorously measurable without defining a "detection" threshold.

For the purposes of our paper, we continue to use eq. (B16) as an upper bound and further optimize it in the next section and suggest that some rescaling of eq. (B16), such as

$$\rho_n(f_n) \approx c + \frac{f_n}{\xi_n}\left[a + b\frac{\log \xi_n}{\ell \log k}\right], \tag{B19}$$

is what we expect to empirically observe in real PrGP systems, in analogy with how empirical DGP robustness $a + b\log f_n$ is a suppression of the DGP maximum in eq. (C3). Since our upper bound is approximate, and also the empirical PrGP does not strictly to be a "suppression." Since eq. (B19) contains additional fitting parameters, we directly use eq. (B16) or eq. (B18) in most of this paper to test the theory. Equation (B16) and eq. (B18) constitute a central result of the paper.

#### b. *Upper Bound on PrGP Robustness for Small Frequencies (PrGP Robustness Tail)*

The first approach we take to derive the PrGP robustness tail (for small frequencies) is direct optimization of eq. (5). We will show that this result has some special properties, but is not the tightest bound.

*a. Direct Optimization of eq.* (5). We proceed by determining function $\xi_n$ will maximize the robustness for small frequencies. By taking the functional derivative of $\rho_n(f_n)$ with respect to $\xi_n$ and setting it to zero

$$\frac{\delta \rho_n(f_n)}{\delta \xi_n} = -\frac{f_n}{\xi_n^2}\left[1 + \frac{\log \xi_n}{\ell \log k}\right] + \frac{f_n}{\xi_n}\left[\frac{1}{\xi_n \ell \log k}\right] = 0, \tag{B20}$$

we have that

$$-1 - \frac{\log \xi_n}{\ell \log k} + \frac{1}{\ell \log k} = 0, \tag{B21}$$

from which it follows that

$$\log \xi_n = 1 - \ell \log k, \tag{B22}$$

so

$$\xi_n = \frac{e}{k^\ell}. \tag{B23}$$

We find that $\xi_n$ is independent of $f_n$. Plugging eq. (B23) back into eq. (5), we find that the maximum robustness (for small $f_n$) is

$$\rho_n^{\text{PrGP max, "tail"}} = \frac{f_n k^\ell}{e\ell \log k}, \quad f_n < \frac{1}{k^\ell}. \tag{B24}$$

We now need to establish exactly what is the condition for "small $f_n$." As we mentioned previously, it is indeed possible that eq. (B24) holds as the maximum PrGP robustness for some $f_n \geq 1/k^\ell$. To check this, we compute the intersection between eq. (B24) and eq. (C3):

$$\begin{aligned} \rho_n^{\text{PrGP max, "tail"}} &= \rho_n^{\text{DGP max}} \\ \Rightarrow \quad \frac{f_n k^\ell}{e\ell \log k} &= 1 + \frac{\log f_n}{\ell \log k} \\ \Rightarrow \quad \frac{1}{e} &= \frac{\log\left(f_n k^\ell\right)}{f_n k^\ell}. \end{aligned} \tag{B25}$$

Letting $x = -\log\left(f_n k^\ell\right)$, we can rewrite the above equation as

$$x e^x = -\frac{1}{e}, \tag{B26}$$

whose solution is

$$x = W_0\left(-\frac{1}{e}\right) = W_{-1}\left(-\frac{1}{e}\right) = -1, \tag{B27}$$

where $W_0$ and $W_{-1}$ are the two branches of the Lambert $W$ function, and $-1/e$ is the point at which the two branches meet and produce the same value. It thus follows that $x = -1 = -\log\left(f_n k^\ell\right)$, so the point of intersection of the two curves is

$$f_n^{\text{intersection}} = \frac{e}{k^\ell}. \tag{B28}$$

It is quick to verify that for $f_n > \frac{e}{k^\ell}$, the DGP maximum eq. (C3) has higher robustness than the tail function eq. (B24), and for $f_n < \frac{e}{k^\ell}$, the DGP maximum eq. (C3) has lower robustness than the tail function eq. (B24).

We can now bound the PrGP robustness from above with the following piecewise smooth function:

$$\rho_n^{\text{PrGP upper}}(f_n) = \begin{cases} \dfrac{f_n k^\ell}{e\ell \log k} & f_n \leq \dfrac{e}{k^\ell} \\ 1 + \dfrac{\log f_n}{\ell \log k} & f_n \geq \dfrac{e}{k^\ell}. \end{cases} \tag{B29}$$

Although this is not the tightest possible upper bound on the robustness tail (shown in the next section), the above has some nice properties. First, both parts of the piecewise function are closely related to or derived from eq. (5), and eq. (B29) can be equivalently written as

$$\xi_n = \begin{cases} \dfrac{e}{k^\ell} & f_n \leq \dfrac{e}{k^\ell} \\ f_n & f_n \geq \dfrac{e}{k^\ell}. \end{cases} \tag{B30}$$

Next, we can show that eq. (B29) is once differentiable everywhere. We only need to show that the derivatives match at $f_n = e/k^\ell$. From the left,

$$\frac{\partial \rho_n^{\text{PrGP upper}}}{\partial f_n} = \frac{k^\ell}{e\ell \log k}, \quad f_n \leq \frac{e}{k^\ell}, \tag{B31}$$

so

$$\left.\frac{\partial \rho_n^{\text{PrGP upper}}}{\partial f_n}\right|_{f_n \to \frac{e}{k^\ell}^-} = \frac{k^\ell}{e\ell \log k}. \tag{B32}$$

and from the right

$$\frac{\partial \rho_n^{\text{PrGP upper}}}{\partial f_n} = \frac{1}{f_n \ell \log k}, \quad f_n \geq \frac{e}{k^\ell}, \tag{B33}$$

so

$$\left.\frac{\partial \rho_n^{\text{PrGP upper}}}{\partial f_n}\right|_{f_n \to \frac{e}{k^\ell}^+} = \frac{k^\ell}{e\ell \log k}. \tag{B34}$$

It is clear that eq. (B29) is an upper bound and once differentiable everywhere.

To interpret the above derivation, we consider a phenotype of frequency $f_n \leq 1/k^\ell$. The derivation in this section shows that by leaving the $f_n$ probability mass on a single genotype, the robustness is less than optimal (it is zero). It suggests that robustness could be optimized by spreading out the probability mass $f_n$ over $e$ genotypes. (Since the number of genotypes can be only integer valued, this means, that the phenotype would need to be spread over 2 or 3 nodes.) However, below we show that spreading the phenotype over $e$ nodes puts the actual robustness far enough below the bound eq. (B29) such that higher-robustness configurations can be constructed which are still below eq. (B29). Therefore, eq. (B29) is not the tightest possible bound, and we derive a tighter one below.

*b. Tighter Bound on PrGP Robustness Tail.* Consider a phenotype with small frequency $f_n \leq k/k^\ell = k^{1-\ell}$. Now, suppose that this frequency is spread over exactly $N$ genotypes, with $1 \leq N \leq k$. Since $N \leq k$, it is possible to form a complete graph with these $N$ nodes, which will have exactly $N(N-1)/2$ edges. Below, we will show that the PrGP robustness will be maximized by spreading probability mass evenly over $N$ nodes arranged in a complete graph, which can only occur if $N \leq k$.

**Theorem B.1.** *A phenotype with $f_n \leq k/k^\ell = k^{1-\ell}$ spread over a complete graph with $N$ nodes, with $1 \leq N \leq k$, will have maximum PrGP robustness when the phenotype probability is equal across all nodes and $N = k$. This maximum robustness is*

$$\rho = \frac{f_n k^{\ell-1}}{\ell}, \quad 0 < f_n \leq k^{1-\ell}. \tag{B35}$$

*Proof.* We can write down the adjacency matrix of the complete graph as $A = 11^T - \mathbb{I}$, where $\mathbb{I}$ is the $N \times N$ identity matrix, and 1 is the length $N$ vector of all ones. Letting $p_g \equiv p_n(g)$, the PrGP robustness is exactly

$$\begin{aligned} \rho(\mathbf{p}) &= \frac{2}{\ell(k-1)k^\ell f_n}\left[\frac{1}{2}\sum_{g,h} p_g A_{gh} p_h\right] \\ &= \frac{1}{\ell(k-1)k^\ell f_n}\left[\left(\sum_g p_g\right)^2 - \sum_g p_g^2\right]. \end{aligned} \tag{B36}$$

Using a Lagrange multiplier $\lambda$, we impose a normalization constraint on the probability vector with the Lagrangian

$$\mathcal{L}[\mathbf{p}] = \rho(\mathbf{p}) - \lambda\left(\sum_g p_g - f_n k^\ell\right). \tag{B37}$$

We optimize the Lagrangian

$$\begin{aligned} 0 = \frac{\delta \mathcal{L}}{\delta p_h} &= \frac{1}{\ell(k-1)k^\ell f_n}\left[2\left(\sum_g p_g\right)\sum_g \delta_{gh} - 2\sum_g p_g \delta_{gh}\right] - \lambda\left(\sum_g \delta_{gh}\right) \\ &= \frac{2}{\ell(k-1)k^\ell f_n}\left[\sum_g p_g - p_h\right] - \lambda, \end{aligned} \tag{B38}$$

and

$$\begin{aligned} 0 = \frac{\delta \mathcal{L}}{\delta \lambda} &= -\left(\sum_g p_g - f_n k^\ell\right) \\ \Rightarrow \quad \sum_g p_g &= f_n k^\ell. \end{aligned} \tag{B39}$$

Summing eq. (B38) over $h$ and plugging in the normalization constraint, we have

$$\begin{aligned} 0 &= \frac{2}{\ell(k-1)k^\ell f_n}\left(N f_n k^\ell - \sum_h p_h\right) - N\lambda \\ &= \frac{2}{\ell(k-1)k^\ell f_n}\left(N f_n k^\ell - f_n k^\ell\right) - N\lambda, \end{aligned} \tag{B40}$$

so

$$\lambda = \frac{2(N-1)}{N\ell(k-1)}. \tag{B41}$$

Plugging this into eq. (B38), we have

$$0 = \frac{2}{\ell(k-1)k^\ell f_n}\left[f_n k^\ell - p_h\right] - \frac{2(N-1)}{N\ell(k-1)}, \tag{B42}$$

from which it follows that

$$p_h = \frac{f_n k^\ell}{N} \tag{B43}$$

for any genotype $h$. This means that for a complete graph with $N$ nodes, PrGP robustness is optimized by spreading the phenotype probability equally over all $N$ nodes.

We can now plug in this result into eq. (B36) to find the robustness in terms of $N$:

$$\begin{aligned}\rho(N) &= \frac{1}{\ell(k-1)k^\ell f_n}\left[f_n^2 k^{2\ell} - \sum_g \frac{f_n^2 k^{2\ell}}{N^2}\right] \\ &= \frac{f_n k^\ell}{\ell(k-1)}\left[1 - \frac{1}{N}\right].\end{aligned} \tag{B44}$$

Since $1 \leq N \leq k$ and $N$ is an integer, it is clear that $\rho(N)$ is maximized when $N = k$, so the maximum robustness is

$$\rho(N = k) = \frac{f_n k^\ell}{\ell(k-1)}\left[1 - \frac{1}{k}\right] = \frac{f_n k^{\ell-1}}{\ell}. \tag{B45}$$

This completes the proof. □

For $f_n \leq k^{1-\ell}$, we have now provided a tight upper bound on robustness, and we point out the critical point that the tail in Theorem B.1 still scales as $\rho_n \propto f_n$, just like in eq. (B29). It can easily be verified that the bound from Theorem B.1 is strictly lower/tighter than eq. (B29) by comparing the slopes; that is, $k^{\ell-1}\ell \geq k^\ell/(e\ell \log k)$, with equality when $k = e$. In terms of interpretation, we also now understand that spreading the phenotype over $e$ nodes, as suggested by eq. (B29), does not produce a robustness that is as high as spreading the phenotype over $k$ nodes.

We now compute the intersection between the robustness from Theorem B.1 and the approximation to the DGP maximum robustness in eq. (C3):

$$\begin{aligned}\frac{f_n k^{\ell-1}}{\ell} &= 1 + \frac{\log f_n}{\ell \log k} \\ \Rightarrow \quad \frac{(f_n k^\ell)}{k\ell} &= \frac{\log\left(f_n k^\ell\right)}{\ell \log k}.\end{aligned} \tag{B46}$$

By inspection we can see that $f_n k^\ell = k$ (i.e. $f_n = k^{1-\ell}$) is a solution to the above equation and provides the point of intersection between the two robustness curves. This leads to a tighter upper bound on robustness:

$$\rho_n^{\text{PrGP upper}}(f_n) = \begin{cases} \dfrac{f_n k^{\ell-1}}{\ell} & f_n \leq k^{1-\ell} \\ \\ 1 + \dfrac{\log f_n}{\ell \log k} & f_n \geq k^{1-\ell}. \end{cases} \tag{B47}$$

This form is not, in general, differentiable at $f_n = k^{1-\ell}$ , unlike eq. (B29). The derivative from the left is

$$\frac{\partial \rho_n^{\text{PrGP upper}}}{\partial f_n} = \frac{k^{\ell-1}}{\ell}, \quad f_n \leq k^{1-\ell}, \tag{B48}$$

so

$$\left.\frac{\partial \rho_n^{\text{PrGP upper}}}{\partial f_n}\right|_{f_n \to k^{1-\ell -}} = \frac{k^{\ell-1}}{\ell}. \tag{B49}$$

and from the right

$$\frac{\partial \rho_n^{\text{PrGP upper}}}{\partial f_n} = \frac{1}{f_n \ell \log k}, \quad f_n \geq k^{1-\ell}, \tag{B50}$$

so

$$\left.\frac{\partial \rho_n^{\text{PrGP upper}}}{\partial f_n}\right|_{f_n \to k^{1-\ell +}} = \frac{k^{\ell-1}}{\ell \log k}. \tag{B51}$$

Equating the two derivatives, using similar analysis as eq. (B26), we find that the first derivatives only match when $k = e$. Thus, since $k$ only takes on integer values $\geq 2$, in general the robustness upper bound is not differentiable at $f_n = k^{1-\ell}$. $=$

## Appendix C: Genotype Entropy Distributions for Various Systems

In the main text, we define the genotype entropy for a phenotype $n$ as

$$S_n^\gamma = -\sum_{g \in \{\text{genotypes}\}} \frac{p_n(g)}{f_n k^\ell} \log \frac{p_n(g)}{f_n k^\ell}. \tag{C1}$$

where the factor $f_n k^\ell$ in the probabilities ensures that $\frac{p_n(g)}{f_n k^\ell}$ can be treated as a probability mass since

$$\sum_{g \in \{\text{genotypes}\}} \frac{p_n(g)}{f_n k^\ell} = 1. \tag{C2}$$

We now derive a few key analytical results.

### 1. DGP Genotype Entropy

First, the relationship between genotype entropy $S_n^\gamma$ and frequency $f_n$ for DGP systems can be analytically derived, since $p_n(g)$ takes on values of 0 or 1 for each genotype $g$. Letting $p_n(g) = 1$ means that

$$S_n^{\gamma,\text{DGP}} = -\sum_{g \in \{\text{genotypes}\}} \frac{1}{f_n k^\ell} \log \frac{1}{f_n k^\ell} = -\log \frac{1}{f_n k^\ell} = \ell \log k + \log f_n. \tag{C3}$$

Therefore, for DGP maps, the scaled genotype entropy $S_n^{\gamma,\text{DGP}}/(\ell \log k)$ is equal to the asymptotic maximum robustness. All DGP maps will have genotype entropy which falls along this curve.

### 2. PrGP Genotype Entropy within Approximation Scheme

Second, we obtain the genotype entropy within the approximation scheme used to derive the PrGP robustness in eq. (5): namely, we assume that the phenotype has frequency $f_n$, is spread over $\xi_n k^\ell$ genotypes, and has equal probability across all of these genotypes. From the normalization rule, we must have that

$$\xi_n k^\ell p_n(g) = f_n k^\ell, \tag{C4}$$

so the probability on any node will be

$$p_n(g) = \frac{f_n}{\xi_n}. \tag{C5}$$

It follows that

$$S_n^{\gamma,\text{PrGP approx}} = -\sum_{g\in\{\text{genotypes}\}} \frac{1}{\xi_n k^\ell} \log \frac{1}{\xi_n k^\ell} = \ell \log k + \log \xi_n. \tag{C6}$$

### 3. Maximum PrGP Genotype Entropy

Third, we can now find the maximum PrGP robustness. Since there are no constraints other than normalization, it would be straightforward to show that entropy should be maximized when probability mass is equally spread over all genotypes that map at all to phenotype $n$. This is the same as maximizing eq. (C6) with respect to $\xi_n k^\ell$ over the domain of integers $1, \ldots, k^\ell$. It is clear that for $\xi_n k^\ell = k^\ell$ (or $\xi_n = 1$), we have that

$$S_n^{\gamma,\text{PrGP max}} = \ell \log k, \tag{C7}$$

regardless of the phenotype frequency $f_n$.

### 4. Empirical Results for Genotype Entropy

In Figure 4, we plot the genotype entropy $S_n^\gamma$ for all phenotypes $n$ for each of the three RNA GC20 PrGP map temperatures. For each of the three temperatures, the plots on the left show $\log_{10}$(frequency) versus scaled genotype entropy $\frac{S_n^\gamma}{\ell \log k}$ for each phenotype. In these plots, the DGP entropy line is also shown, as well as the maximum PrGP entropy line. The plots on the right show scaled genotype entropy $\frac{S_n^\gamma}{\ell \log k}$ versus robustness $\rho_n$ for each phenotype. The line $\rho_n = \frac{S_n^\gamma}{\ell \log k}$ is also plotted. The PrGP genotype entropy tends to be lower than the corresponding DGP genotype entropy, but generally maintains the overall relationship between $S_n^\gamma \sim a + b \log f_n$. In the case of RNA GC20 at 70 °C, we find that PrGP genotype entropy can exceed the DGP entropy but remains far from the maximum value. We also plot robustness versus the scaled genotype entropy and generally find a positive correlation, though with PrGP maps typically having lower genotype entropy for the same robustness. At different temperatures (hence, disorder), we see behavior that is consistent with theory. Phenotype probability mass spreads spread over more genotypes as temperature increases, thus acquiring higher genotype entropy and straying farther away from the maximally robust PrGP spread as derived in eq. (C7). In the SM [42], we similarly plot the genotype entropy $S_n^\gamma$ for all phentoypes $n$ for the RNA12 PrGP map.

In Figure 5 we plot the genotype entropy $S_n^\gamma$ for all phenotypes $n$ for each of the three spin glass PrGP map external field disorder settings for the main text spin glass trial. For each of the three external field disorder settings, the plots on the left show $\log_{10}$(frequency) versus scaled genotype entropy $\frac{S_n^\gamma}{\ell \log k}$ for each phenotype. In these plots, the DGP entropy line is also shown, as well as the maximum PrGP entropy line. The plots on the right show scaled genotype entropy $\frac{S_n^\gamma}{\ell \log k}$ versus robustness $\rho_n$ for each phenotype. The line $\rho_n = \frac{S_n^\gamma}{\ell \log k}$ is also plotted. The PrGP genotype entropy tends to be lower than the corresponding DGP genotype entropy, but generally maintains the overall relationship between $S_n^\gamma \sim a + b \log f_n$. We also plot robustness versus the scaled genotype entropy and generally find a positive correlation, though with PrGP maps typically having lower genotype entropy for the same robustness. At different external field disorder settings, we see behavior that is consistent with theory. Phenotype probability mass spreads spread over more genotypes as external field disorder increases, thus acquiring higher genotype entropy and straying farther away from the maximally robust PrGP spread as derived in eq. (C7).

In Figure 6 we plot the genotype entropy $S_n^\gamma$ for all phenotypes $n$ for an 11-qubit quantum circuit PrGP map validation trial. We observe several groups of phenotypes behave linearly between log frequency and scaled genotype entropy. It is interesting to note that the genotype entropy from the noiseless quantum circuit fall exactly along the DGP genotype entropy line, while the noisy simulations show groups of phenotypes which each cross the same line, each with higher slopes than in eq. (C6).

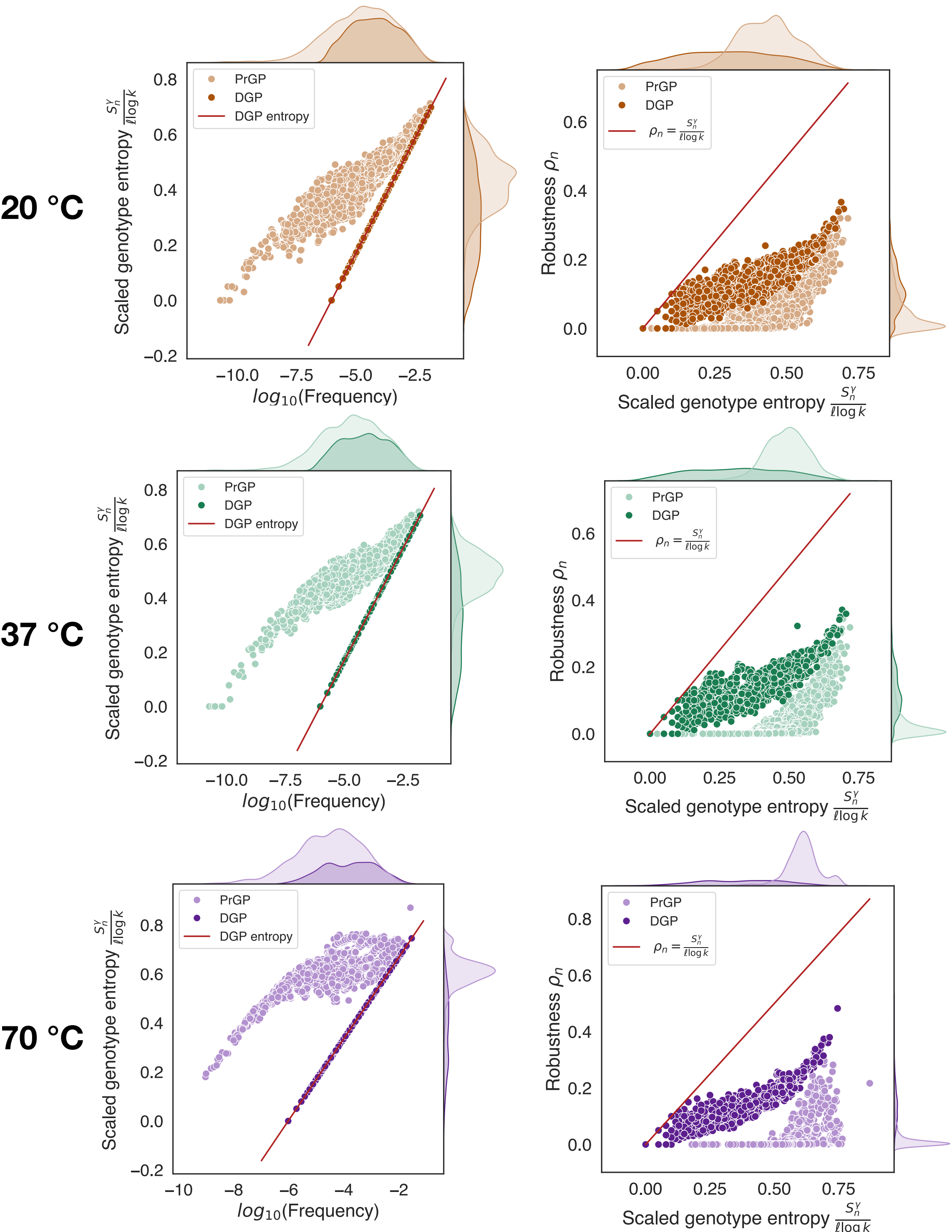

FIG. 4. Plots of (left) $\log_{10}$(frequency) versus scaled genotype entropy $\frac{S_n^\gamma}{\ell \log k}$ and (right) scaled genotype entropy $\frac{S_n^\gamma}{\ell \log k}$ versus robustness $\rho_n$ for RNA folding PrGP maps across three temperatures with $k = 2, \ell = 20$.

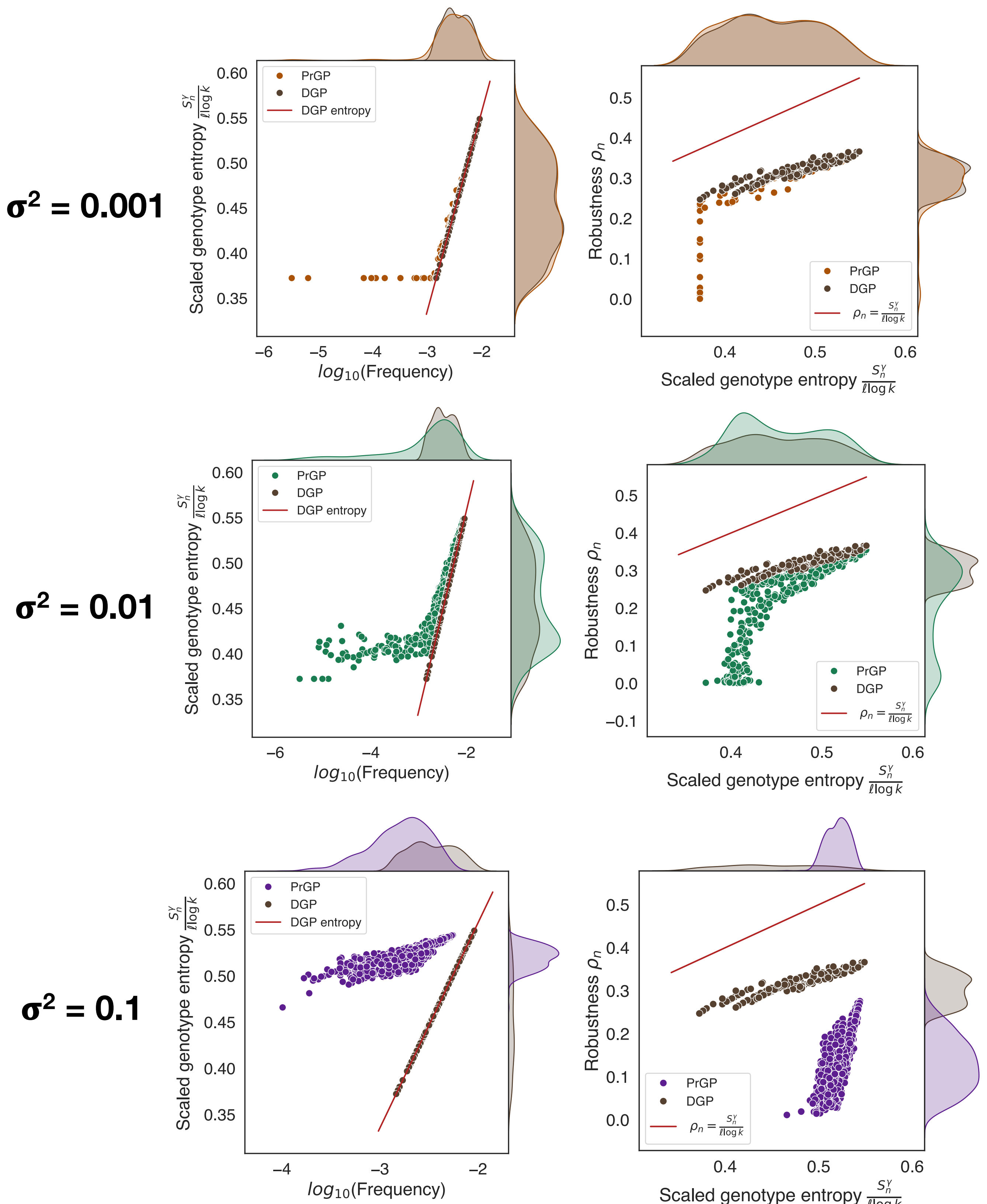

FIG. 5. Plots of (left) $\log_{10}$(frequency) versus scaled genotype entropy $\frac{S_n^\gamma}{\ell \log k}$ and (right) scaled genotype entropy $\frac{S_n^\gamma}{\ell \log k}$ versus robustness $\rho_n$ for each spin glass ground state at three different external field variances for the spin glass PrGP maps.

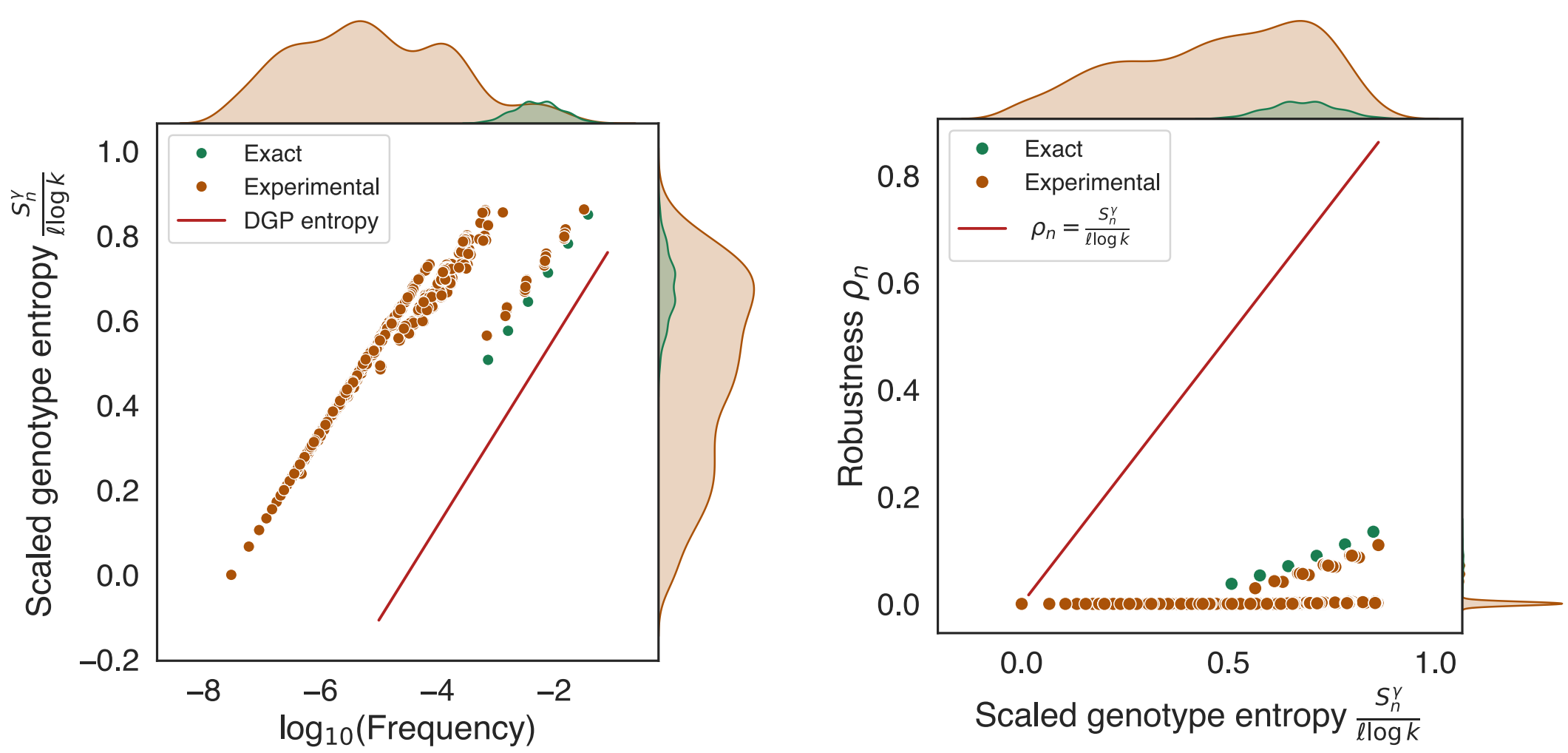

FIG. 6. Plots of (left) $\log_{10}$(frequency) versus scaled genotype entropy $\frac{S_n^\gamma}{\ell \log k}$ and (right) scaled genotype entropy $\frac{S_n^\gamma}{\ell \log k}$ versus robustness $\rho_n$ for the simulated 11-qubit quantum circuit trial.

### 5. Semi-Empirical Robustness Formula and Biphasic Behavior

We return to a discussion of eq. (B18), and see how the genotype entropy empirical results allow for further simplification of eq. (B18) into a semi-empirical equation in which robustness $\rho_n$ only depends explicitly on frequency $f_n$, not both $f_n$ and genotype entropy $S_n^\gamma$.

In the systems studied here, we found that, in general

$$S_n^\gamma \approx \alpha + \eta \log f_n, \tag{C8}$$

where $\alpha$ and $\eta$ are some constants. There are examples, such as the spin glass $\sigma_h^2 = 0.001$ case, where $\eta$ does change with $f_n$, but it remains constant across large ranges of frequencies and changes relatively sharply. Therefore, we can still plug in eq. (C8) into eq. (B18) to better understanding limiting behaviors of robustness. Within our PrGP approximation from Appendix B, eq. (C8) is the equivalent of saying that $\xi_n$ behaves as a power law with respect to $f_n$.

Substituting eq. (C8) into eq. (B18), we have

$$\rho_n \approx \frac{k^\ell}{e^\alpha \ell \log k} \left[ f^{1-\eta} \left( \alpha + \eta \log f_n \right) \right], \tag{C9}$$

which is eq. (8) in the main text. We first note that when $\eta = 0$, we simply have

$$\rho_n \approx \frac{k^\ell \alpha f_n}{e^\alpha \ell \log k} \sim f_n, \tag{C10}$$

which is linear, and represents null-model like behavior for all frequencies. For the spin glass $\sigma_h^2 = 0.001$, we find $\eta = 0$ for sufficiently small frequencies and $\eta \approx 0$ for $\sigma_h^2 = 0.01$. In both of these cases, the linear, null model-like behavior of robustness for the smallest frequencies is prominent in the log-log plot in main text Figure 2. We note that in both of these cases, there is a transition for larger frequencies to $\eta \approx 1$, so the logarithmic behavior

$$\rho_n \approx \frac{k^\ell}{e^\alpha \ell \log k} \left[ \alpha + \log f_n \right] \sim a + b \log f_n \tag{C11}$$

is apparent for the most common frequencies, as seen in the linear-log plot in main text Figure 2. This captures the idealized biphasic robustness behavior shown in the PrGP robustness upper bound derived previously in Appendix B.

For various systems, we also find $0 < \eta < 1$. This occurs in the RNA systems and spin glasses with $\sigma_h^2 = 0.1$. Here, it would appear that in the limit $f_n \to 0$, we would see a divergence $\rho_n \to -\infty$. However, the correct limit must take into consideration that the smallest phenotype frequency, $f_{\text{min}}$, is finite. If consider the log robustness as given by eq. (C8), we have

$$\log \rho_n \approx \log \frac{k^\ell}{e^\alpha \ell \log k} + (1 - \eta) \log f_n + \log \left( \alpha + \eta \log f_n \right). \tag{C12}$$

Notably, if

$$f_n \geq f_{\text{min}} \gg e^{-\alpha/\eta}, \tag{C13}$$

we would have

$$\log \left( \alpha + \eta \log f_n \right) \approx \log \alpha, \tag{C14}$$

which means that

$$\log \rho_n \approx \log \frac{k^\ell \alpha}{e^\alpha \ell \log k} + (1 - \eta) \log f_n \sim C + (1 - \eta) \log f_n, \tag{C15}$$

which is a power law relationship. Of course, when $\eta$ is 0 or very small, this becomes a linear relationship, as discussed above. The question remains, how much greater does $f_{\text{min}}$ really need to be compared to $e^{-\alpha/\eta}$? We plot eq. (C12) in Figure 7 for the RNA $k = 2, \ell = 20$ main text PrGP maps at all 3 temperatures, using $\alpha$ and $\eta$ values derived from linear regression of $S_n^\gamma$ versus $f_n$. For 20° C, we have $\alpha = 10.4862$ and $\eta = 0.4076$, for 37° C, we have $\alpha = 10.5819$ and $\eta = 0.34994$, and for 70° C, we have $\alpha = 11.037673$ and $\eta = 0.2497868$. We note that in Figure 7, as frequencies get smaller there is a clear, sharp dropoff of $\log \rho_n$, though an approximate power law relationship with $f_n$ remains for several orders of magnitude. As we see in main text Figure 2, $f_{\min}$ for these RNA systems tends to be around $10^{-10}$, so sharp divergence is avoided even by the lowest frequency phenotypes, and the approximate power law relationship holds since it appears that $f_n \geq f_{\min} \gg e^{-\alpha/\eta}$ holds.

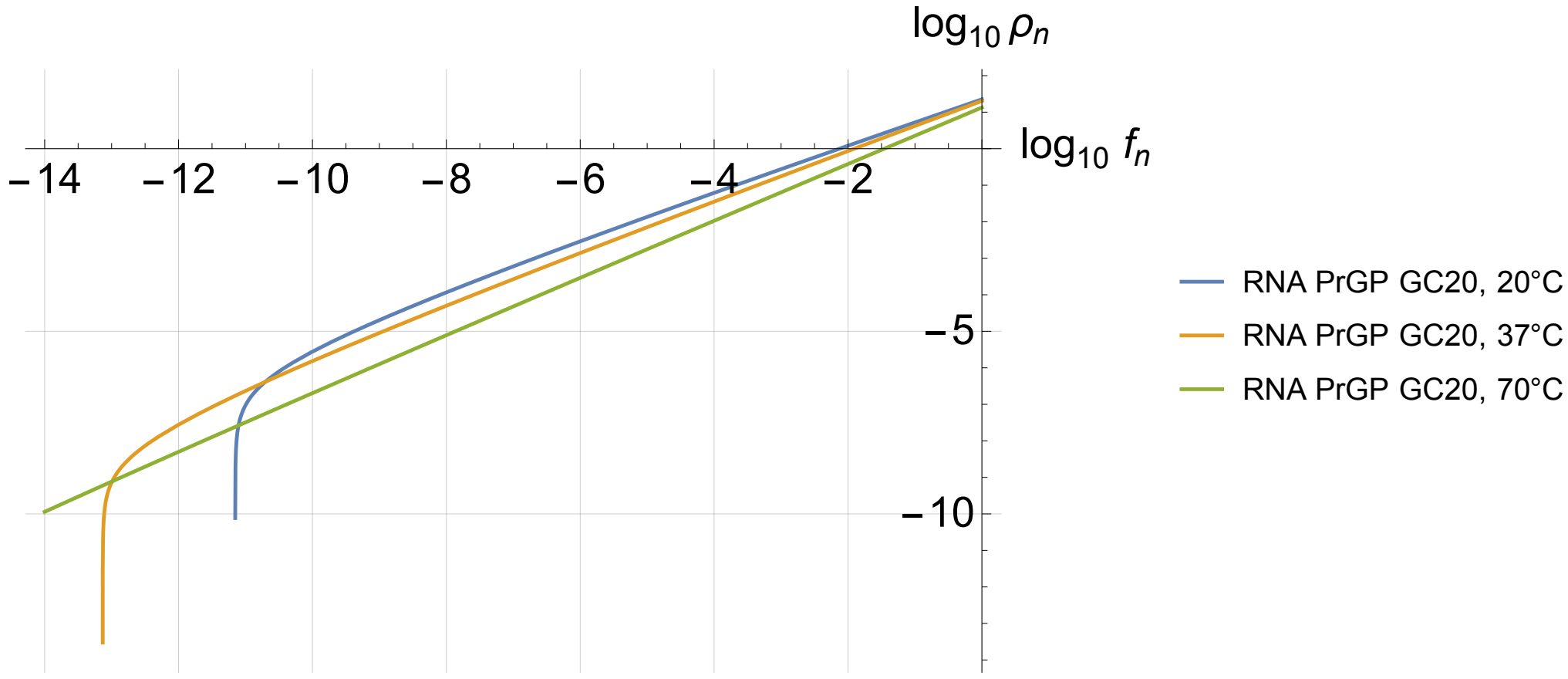

FIG. 7. Plot of eq. (C12) for RNA $k = 2, \ell = 20$ main text PrGP maps at all 3 temperatures using $\alpha$ and $\eta$ values obtained from linear regression of $S_n^\gamma$ versus $f_n$.

Lastly, for the high frequency case. In general, for $0 < \eta < 1$, robustness has a more complex behavior with

$$\rho_n \approx \frac{k^\ell}{e^\alpha \ell \log k} \left[ f^{1-\eta} \left( \alpha + \eta \log f_n \right) \right] \sim f^{1-\eta} \left( \alpha + \eta \log f_n \right), \tag{C16}$$

which can be expanded around $f_n = 1$ to leading order in $f_n$ or to leading order in $\log f_n$. We find for many systems, such as spin glasses with $\sigma_h^2 = 0.001$ and $\sigma_h^2 = 0.01$, that $\eta \approx 1$ for sufficiently large frequencies, as mentioned previously, which leads to logarithmic "DGP robust" behavior. We also see that for RNA, the PrGP genotype entropies tend to reach the DGP curve for sufficiently large frequencies, and the DGP $S_n^\gamma$ curve has $\eta = 1$ as well. In quantum circuits, we see an overall trend of $S_n^\gamma$ versus $f_n$ to lay parallel or almost parallel to the DGP curve, suggesting $\eta \approx 1$, or at least is very close. However, there are also clusters of phenotypes with $0 < \eta < 1$ which could be separated into different regressions with various $\alpha$ values. This leads to a fragmented genotype entropy curve and robustness curve which may not be entirely explained by the monotonic behavior predicted by fig. 7. Regardless, the different regimes of eq. (C12) we discuss here provide insight into PrGP map robustness.

## Appendix D: Validation of the PrGP Robustness Theory

We show that theoretical robustness, eq. (5) in the main text, recapitulates empirical robustness with very high Pearson correlation across all of the system studied.

### 1. Spin Glass PrGP Theoretical and Empirical Robustness

The spin glass PrGP maps across all settings of $\sigma_h^2$ show excellent agreement with the theory. In Figure 8 below, we plot theoretical and empirically observed $\log_{10}(f_n)$ versus $\rho_n$ for the spin glass systems, alongside the upper bound derived in eq. (6) and the robustness under the null model. We also present scatter plots displaying the strong correlation between our theoretically-derived $\rho_n$ and the empirically observed $\rho_n$.

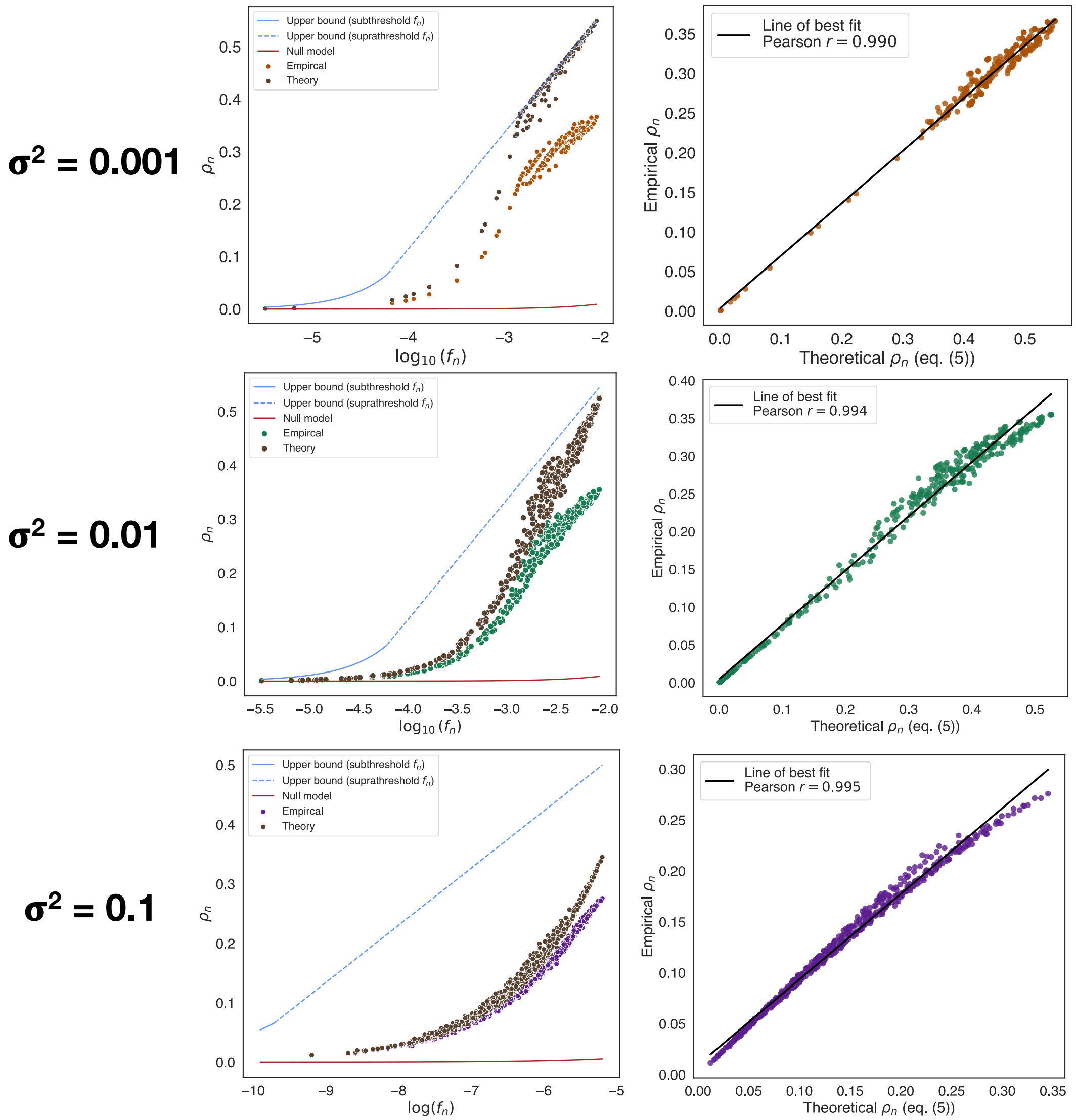

FIG. 8. Plots of (left) $\log_{10}(f_n)$ versus robustness $\rho_n$ and (right) theoretical $\rho_n$ versus empirical $\rho_n$ for various settings of $\sigma_h^2$ for the spin glass PrGP map system.

## 2. RNA PrGP Theoretical and Empirical Robustness

The RNA PrGP maps across all temperature settings show excellent agreement with the theory. In Figure 9 below, we plot theoretical and empirically observed $\log_{10}(f_n)$ versus $\rho_n$ for the RNA $k = 2, \ell = 4$ systems, alongside the upper bound derived in eq. (6) and the robustness under the null model. We also present scatter plots displaying the strong correlation between our theoretically-derived $\rho_n$ and the empirically observed $\rho_n$.

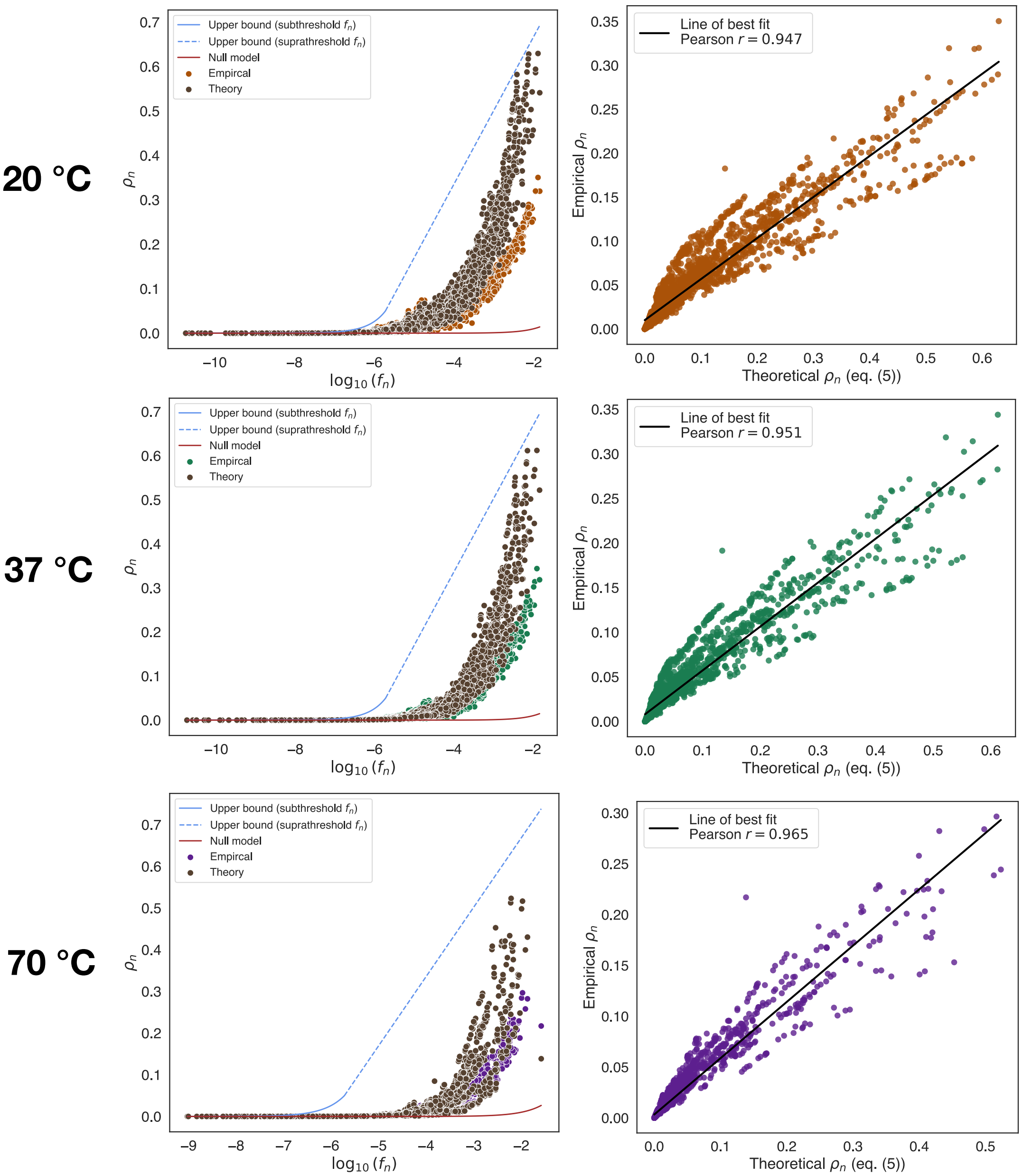

FIG. 9. Plots of (left) $\log_{10}(f_n)$ versus robustness $\rho_n$ and (right) theoretical $\rho_n$ versus empirical $\rho_n$ for various settings of $\sigma_h^2$ for the RNA PrGP map system.

### 3. Quantum Circuit PrGP Theoretical and Empirical Robustness

The quantum circuit PrGP maps show excellent agreement with the theory. In Figure 10 below, we plot theoretical and empirically observed $\log_{10}(f_n)$ versus $\rho_n$ for the 11-qubit quantum circuit system, exact and experimental, alongside the upper bound derived in eq. (6) and the robustness under the null model. We also present scatter plots displaying the strong correlation between our theoretically-derived $\rho_n$ and the empirically observed $\rho_n$.

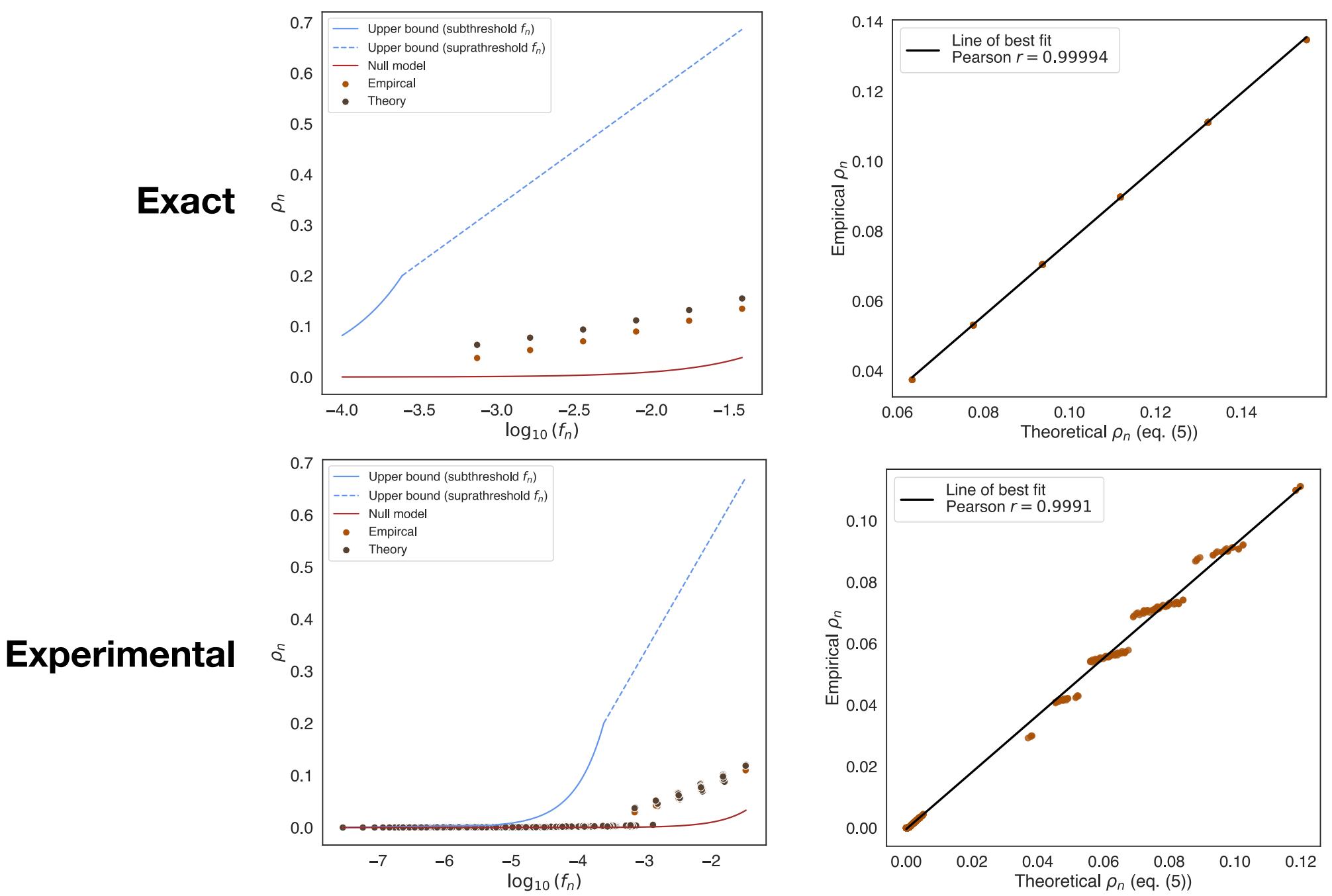

FIG. 10. Plots of (left) $\log_{10}(f_n)$ versus robustness $\rho_n$ and (right) theoretical $\rho_n$ versus empirical $\rho_n$ for various settings of $\sigma_h^2$ for the 11-qubit quantum circuit PrGP map system.

### Appendix E: Phenotype Entropy Distributions for Main Text Systems

In the main text, we presented robustness versus frequency plots for RNA folding, spin glass ground state, and quantum circuit PrGP maps. For the spin glass ground state and quantum circuit PrGP maps, data for a single representative realization were presented in the main text. In Figure 11, we plot the distribution of phenotype entropy $S(g)$ across all genotypes $g$ for each of these PrGP maps. Recall that phenotype entropy is defined in the main text as

$$S(g) = -\sum_{n \in \{\text{phenotypes}\}} p_n(g) \log p_n(g). \tag{E1}$$

For RNA folding and spin glasses, we observe that the phenotype entropy distributions shift rightward as the disorder parameter increases. For RNA, this corresponds to increasing temperature and for spin glasses, this corresponds to increasing external field variance $\sigma_h^2$. For the quantum circuits, we plot both exact and experimental results (from the 7-qubit IBM quantum computer); the experimental phenotype entropy distribution is shifted rightward relative to the exact result, due to measurement noise as well as a finite number of experimental trials.

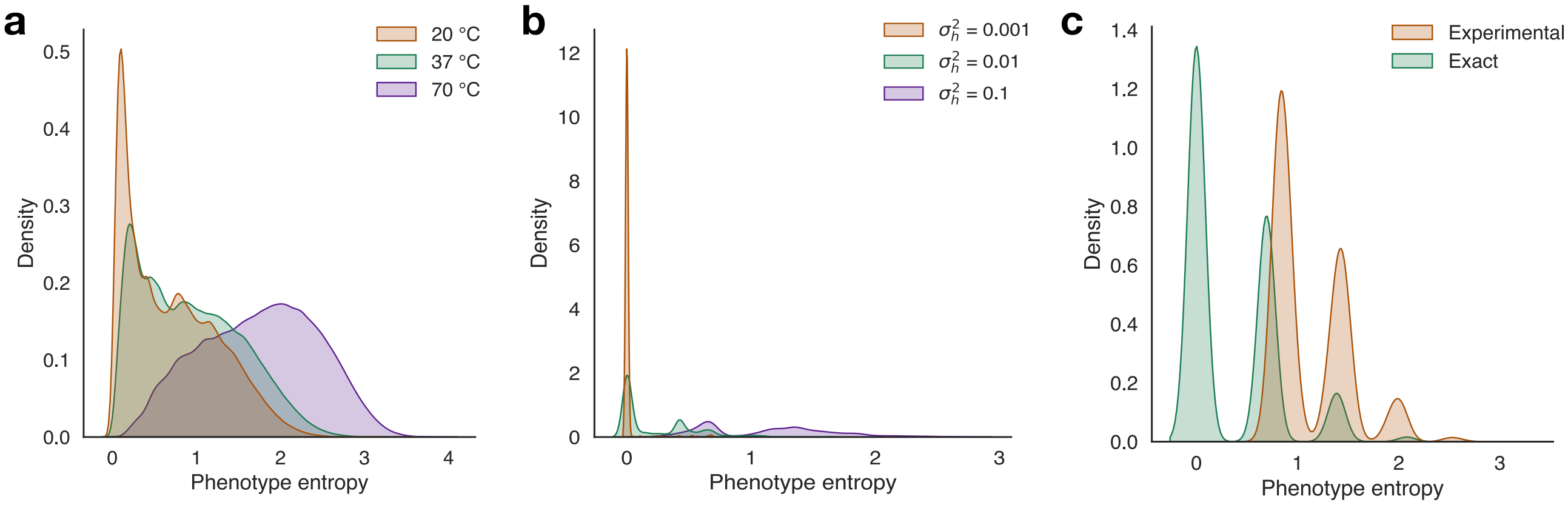

FIG. 11. Phenotype entropy distributions for the (a) RNA folding, (b) spin glass ground state, and (c) quantum circuit PrGP maps whose robustness plots are presented in the main text. As disorder parameters increase in (a) and (b) due to increased temperature and increased external field variance, respectively, the entropies shift rightward. The same occurs due to measurement noise in (c).

# Supplemental Materials for "Probabilistic Genotype-Phenotype Maps Reveal Mutational Robustness of RNA Folding, Spin Glasses, and Quantum Circuits"

Anna Sappington[1, 2, *] and Vaibhav Mohanty[3, 2, *]

[1] *Department of Electrical Engineering and Computer Science, Massachusetts Institute of Technology, Cambridge, MA 02139*

[2] *Harvard-MIT Health Sciences and Technology, Harvard Medical School, Boston, MA 02115 and Massachusetts Institute of Technology, Cambridge, MA 02139*

[3] *Department of Chemistry and Chemical Biology, Harvard University, Cambridge, MA 02138*

* The authors contributed equally to this work. Correspondence: asappington@hms.harvard.edu and mohanty@hms.harvard.edu.

## I. EXTENDED DATA FOR MAIN TEXT RNA FOLDING PrGP MAP, GC ALPHABET, $\ell = 20, k = 2$

In the main text, we presented robustness versus frequency plots with linear and log scaling for RNA folding PrGP and DGP maps at three temperatures. For clarity, we have included robustness versus frequency, robustness versus $\log_{10}$(frequency), and $\log_{10}$(robustness) versus $\log_{10}$(frequency) plots separately for PrGP and DGP maps in Figure S1. First, we see that the DGP map results reproduce the expected $\rho_n \propto \log f_n$ relationship for most phenotypes, with significant elevation above the random null model expectation. We also note that there is little temperature dependence in DGP robustness calculations, which suggests the effect of temperature does little to alter the exact ground state phenotype. In contrast, our PrGP map results showcase a different robustness behavior in which as simulation temperature increases, there is a gradual but clear suppression of the robustness versus frequency relationship; see main text for discussion of these features. In the PrGP map results we also note a biphasic behavior in which for high frequency phenotypes, the PrGP map robustness, similar to the DGP map robustness, is substantially elevated above the random null expectation and for lower frequencies, the robustness behaves more like the random model.

In Table S1, we include the Pearson correlation coefficient $r$ and Spearman rank correlation coefficient $\rho$ for each map (PrGP, DGP), temperature (20 °C, 37 °C, 70 °C), and axis transformation presented in Figure 2(a-b) and Figure S1. The primary feature we point out is the relative decrease of the PrGP Pearson $r$ coefficients in robustness versus $\log_{10}$(frequency) plots as compared to the DGP plots; this suggests a deviation from the empirical $\rho_n \propto \log f_n$ trend observed in DGP studies.

In the GP map literature, *phenotype bias*, the finding that phenotype frequencies can vary over many orders of magnitude with a small number of phenotypes being the targets of a large number of genotypes, has been shown for many systems [1–4]. In Figure S2, we present plots of $\log_{10}$(frequency) versus normalized rank and $\log_{10}$(frequency) versus $\log_{10}$(normalized rank) for each temperature and map pairing which show phenotype bias for this RNA folding system. Notably, the $\log_{10}$(frequency) versus $\log_{10}$(normalized rank) plot suggests a deviation from Zipf's law.

Figure S3 presents transition probabilities $\phi_{mn}$ for the most frequently occurring phenotype $n$ to the other phenotypes $m$ due to a single nucleotide mutation for both PrGP and DGP maps at three different temperatures. For each respective map, a plot including and excluding the most robust transition (*i.e.* from phenotype $n \to n$) is shown for added clarity. This figure demonstrates that the off-diagonal transition probabilities for PrGP maps maintained an approximate relationship $\phi_{mn} \propto f_m$ for $m \neq n$ in concordance with DGP maps, and in concordance with the random null expectation for PrGP maps (see main text). A proportionality constant not equal to 1 for $\phi_{mn} \propto f_m$ with $m \neq n$ is likely due to transition probability mass that is acquired by the diagonal element $\phi_{nn}$. It is also apparent that the most robust transition is much more likely than the transition to any other phenotype, in support of our claim that PrGP maps, like DGP maps, exhibit enhanced robustness.

| System | Alphabet, Length | Map | Temperature | Axes | Pearson $r$ | Spearman $\rho$ |
|---|---|---|---|---|---|---|
| RNA | GC, 20 | PrGP | 20 °C | Robust v. Freq | 0.811 | 0.962 |
| RNA | GC, 20 | PrGP | 20 °C | Robust v. $\log_{10}$(Freq) | 0.813 | 0.962 |
| RNA | GC, 20 | PrGP | 20 °C | $\log_{10}$(Robust) v. $\log_{10}$(Freq) | 0.951 | 0.962 |
| RNA | GC, 20 | PrGP | 37 °C | Robust v. Freq | 0.854 | 0.974 |
| RNA | GC, 20 | PrGP | 37 °C | Robust v. $\log_{10}$(Freq) | 0.784 | 0.974 |
| RNA | GC, 20 | PrGP | 37 °C | $\log_{10}$(Robust) v. $\log_{10}$(Freq) | 0.970 | 0.974 |
| RNA | GC, 20 | PrGP | 70 °C | Robust v. Freq | 0.856 | 0.982 |
| RNA | GC, 20 | PrGP | 70 °C | Robust v. $\log_{10}$(Freq) | 0.665 | 0.982 |
| RNA | GC, 20 | PrGP | 70 °C | $\log_{10}$(Robust) v. $\log_{10}$(Freq) | 0.982 | 0.982 |
| RNA | GC, 20 | DGP | 20 °C | Robust v. Freq | 0.721 | 0.860 |
| RNA | GC, 20 | DGP | 20 °C | Robust v. $\log_{10}$(Freq) | 0.868 | 0.860 |
| RNA | GC, 20 | DGP | 20 °C | $\log_{10}$(Robust) v. $\log_{10}$(Freq) | 0.839 | 0.860 |
| RNA | GC, 20 | DGP | 37 °C | Robust v. Freq | 0.717 | 0.856 |
| RNA | GC, 20 | DGP | 37 °C | Robust v. $\log_{10}$(Freq) | 0.859 | 0.856 |
| RNA | GC, 20 | DGP | 37 °C | $\log_{10}$(Robust) v. $\log_{10}$(Freq) | 0.836 | 0.856 |
| RNA | GC, 20 | DGP | 70 °C | Robust v. Freq | 0.759 | 0.914 |
| RNA | GC, 20 | DGP | 70 °C | Robust v. $\log_{10}$(Freq) | 0.903 | 0.914 |
| RNA | GC, 20 | DGP | 70 °C | $\log_{10}$(Robust) v. $\log_{10}$(Freq) | 0.884 | 0.914 |

TABLE S1. Pearson and Spearman correlation coefficients for all robustness versus frequency plots in main text/Supplemental Material for RNA $k = 2, \ell = 20$ simulations with reduced alphabet, for each simulation temperature.

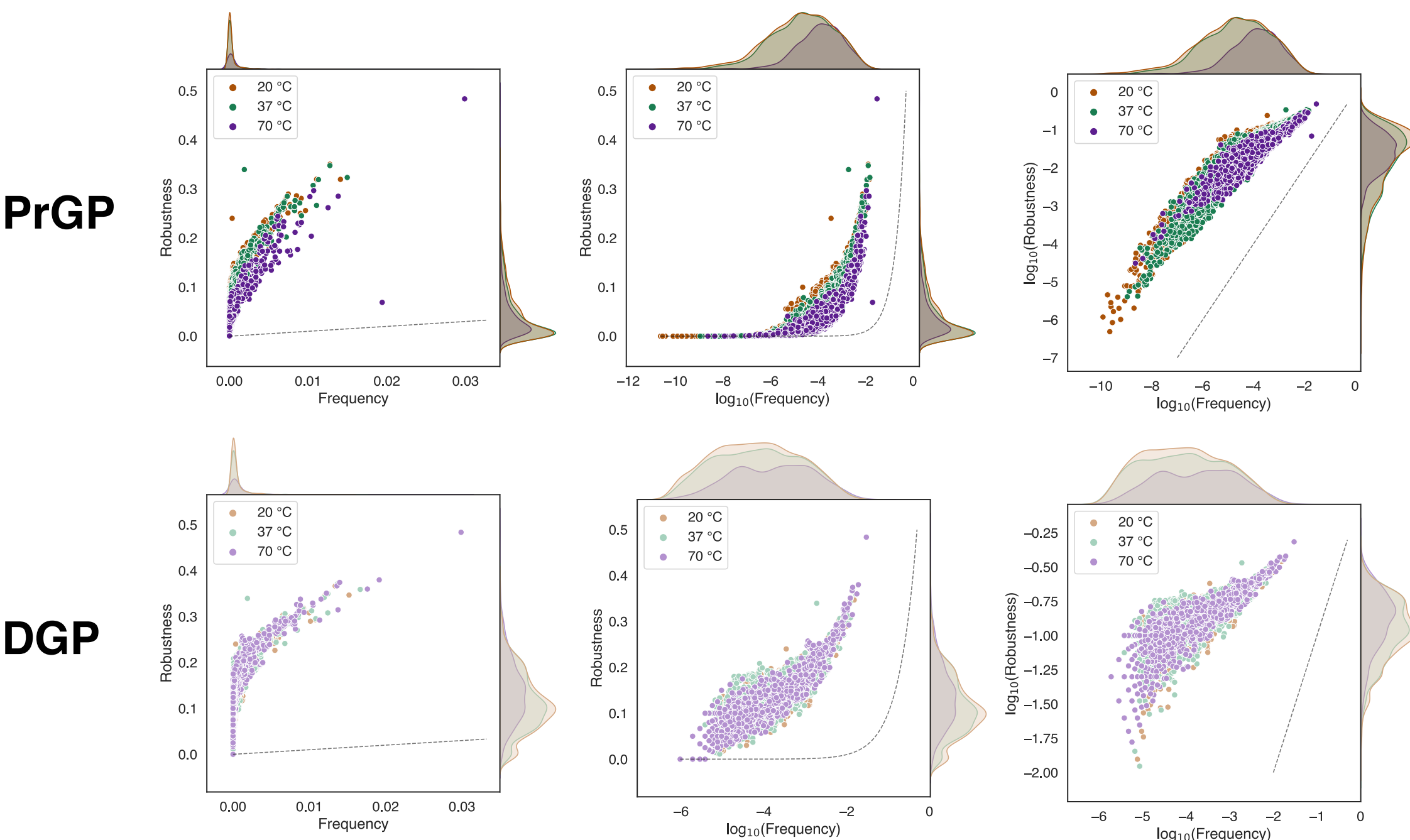

FIG. S1. Plots of (left) robustness versus frequency, (middle) robustness versus $\log_{10}$(frequency), and (right) $\log_{10}$(robustness) versus $\log_{10}$(frequency) for RNA folding (top row) PrGP maps and (bottom row) DGP maps for three temperatures. These data are the same results from main text Figure 2, with axis scaling adjusted and with PrGP and DGP data shown separately for clarity. The dashed line is the random null expectation for both PrGP and DGP maps given by $\phi_{mn} = f_m$ for all $m$ and $n$.

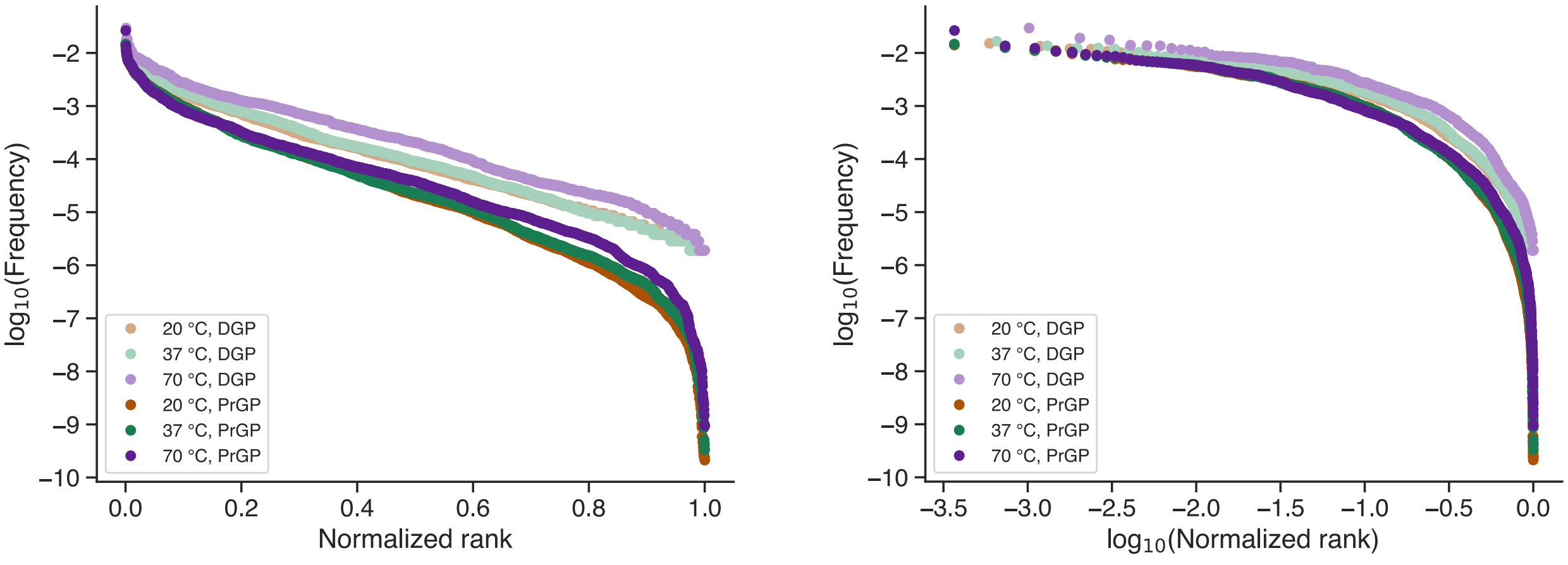

FIG. S2. Plots of (left) $\log_{10}$(frequency) versus normalized rank and (right) $\log_{10}$(frequency) versus $\log_{10}$(normalized rank) for RNA folding PrGP and DGP maps for three temperatures. When computing ranks, ties were broken arbitrarily.

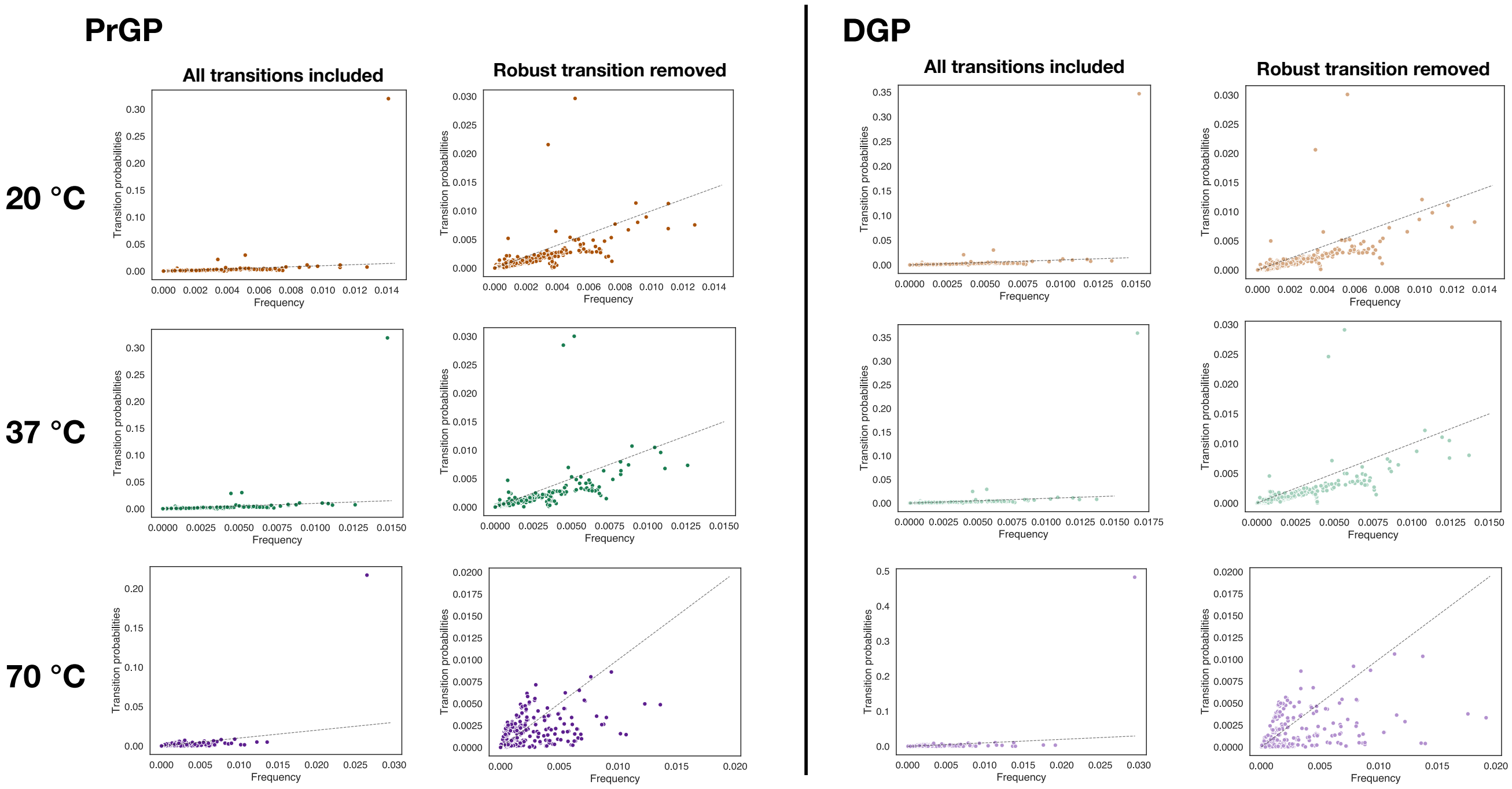

FIG. S3. Plots of transition probabilities versus frequency for RNA folding (left) PrGP maps and (right) DGP maps for three temperatures, (top) 20° C, (middle) 37° C, and (bottom) 70° C. For each respective map, plots include either (left) all transitions or (right) have the most robust transition removed. The dashed line is the random null expectation for both PrGP and DGP maps given by $\phi_{mn} = f_m$ for all $m$ and $n$.

## II. VALIDATION TRIAL FOR RNA FOLDING PrGP MAP, FULL ALPHABET, $\ell = 12, k = 4$

Here, we present results of a validation trial for RNA folding PrGP maps for sequences of length $\ell = 12$ utilizing the full alphabet of size $k = 4$, $\{A, C, G, U\}$. In Figure S4, we present robustness versus frequency, robustness versus $\log_{10}$(frequency), and $\log_{10}$(robustness) versus $\log_{10}$(frequency) plots for RNA folding for PrGP and DGP maps. As with the reduced alphabet case, we see both PrGP and DGP map results show significant elevation above the random null model expectation, with PrGP map results demonstrating a gradual but clear suppression of the robustness versus frequency relationship compared to DGP map results. The expected $\rho_n \propto \log f_n$ relationship for phenotypes in the DGP map results as well as the biphasic behavior of the PrGP map results is present but less clear in this case, likely due to a small size effect from the limited number of phenotypes present in this complete alphabet ($k = 4$, $\ell = 12$) system compared to the reduced alphabet system ($k = 2$), which contains sequences of longer length ($\ell = 20$). Also in Figure S4, we plot the distribution of phenotype entropy $S(g)$ across all genotypes $g$; most phenotype entropies are zero due to their being deterministic because for the RNA folding, $k = 4$, $\ell = 12$ system most genotypes do not fold.

In Table S2, we include the Pearson correlation coefficient $r$ and Spearman rank correlation coefficient $\rho$ for each map (PrGP, DGP) and axis transformation presented in Figure S4. In Figure S5, we present plots of $\log_{10}$(frequency) versus normalized rank and $\log_{10}$(frequency) versus $\log_{10}$(normalized rank) for each temperature and map pairing which show phenotype bias for this RNA folding system. Notably, the $\log_{10}$(frequency) versus $\log_{10}$(normalized rank) plot suggests a deviation from Zipf's law.

Figure S6 presents transition probabilities $\phi_{mn}$ for the most frequently occurring phenotype $n$ to the other phenotypes $m$ due to a single nucleotide mutation for both PrGP and DGP maps. For each respective map, a plot including and excluding the most robust transition is shown for added clarity. This figure demonstrates that the off-diagonal transition probabilities for PrGP maps maintained an approximate relationship $\phi_{mn} \propto f_m$ for $m \neq n$ in concordance with DGP maps, and in concordance with the random null expectation for PrGP maps (see main text). A proportionality constant not equal to 1 for $\phi_{mn} \propto f_m$ with $m \neq n$ is likely due to transition probability mass that is acquired by the diagonal element $\phi_{nn}$. It is also apparent that the most robust transition is much more likely than the transition to any other phenotype, in support of our claim that PrGP maps, like DGP maps, exhibit enhanced robustness.

| **System** | **Alphabet, Length** | **Map** | **Axes** | **Pearson $r$** | **Spearman $\rho$** |
|---|---|---|---|---|---|
| RNA | AUCG, 12 | PrGP | Robust v. Freq | 0.832 | 0.927 |
| RNA | AUCG, 12 | PrGP | Robust v. $\log_{10}$(Freq) | 0.882 | 0.927 |
| RNA | AUCG, 12 | PrGP | $\log_{10}$(Robust) v. $\log_{10}$(Freq) | 0.877 | 0.927 |
| RNA | AUCG, 12 | DGP | Robust v. Freq | 0.788 | 0.244 |
| RNA | AUCG, 12 | DGP | Robust v. $\log_{10}$(Freq) | 0.814 | 0.244 |
| RNA | AUCG, 12 | DGP | $\log_{10}$(Robust) v. $\log_{10}$(Freq) | 0.855 | 0.244 |

TABLE S2. Pearson and Spearman correlation coefficients for all robustness versus frequency plots for RNA $k = 4, \ell = 12$ validation trial with reduced alphabet. Simulations were conducted at 37 °C.

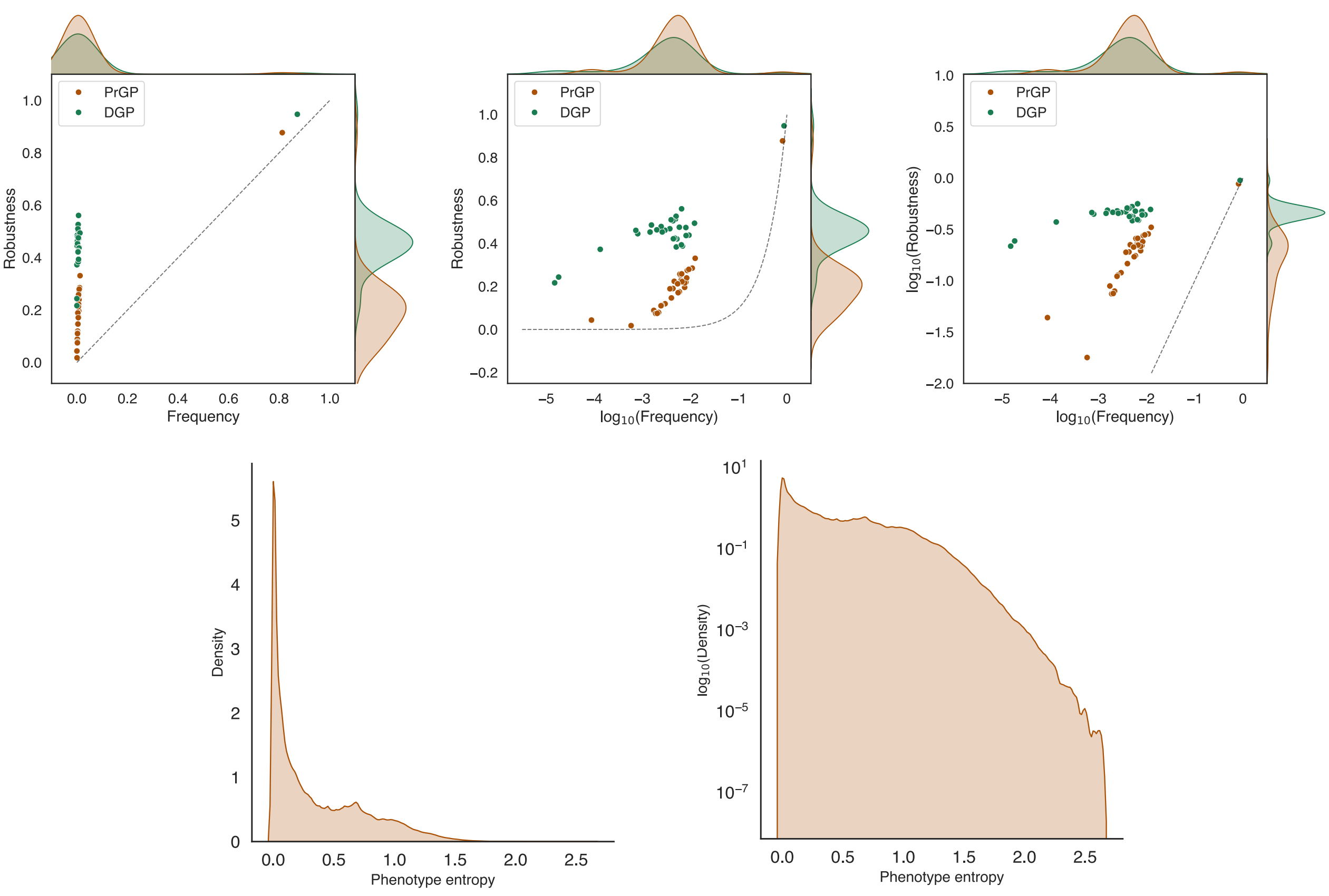

FIG. S4. Plots of (top left) robustness versus frequency, (top middle) robustness versus $\log_{10}$(frequency), and (top right) $\log_{10}$(robustness) versus $\log_{10}$(frequency) for RNA folding PrGP and DGP maps. Plots of (bottom left) density versus phenotype entropy and (bottom right) $\log_{10}$(density) versus phenotype entropy. The dashed line is the random null expectation for both PrGP and DGP maps given by $\phi_{mn} = f_m$ for all $m$ and $n$.

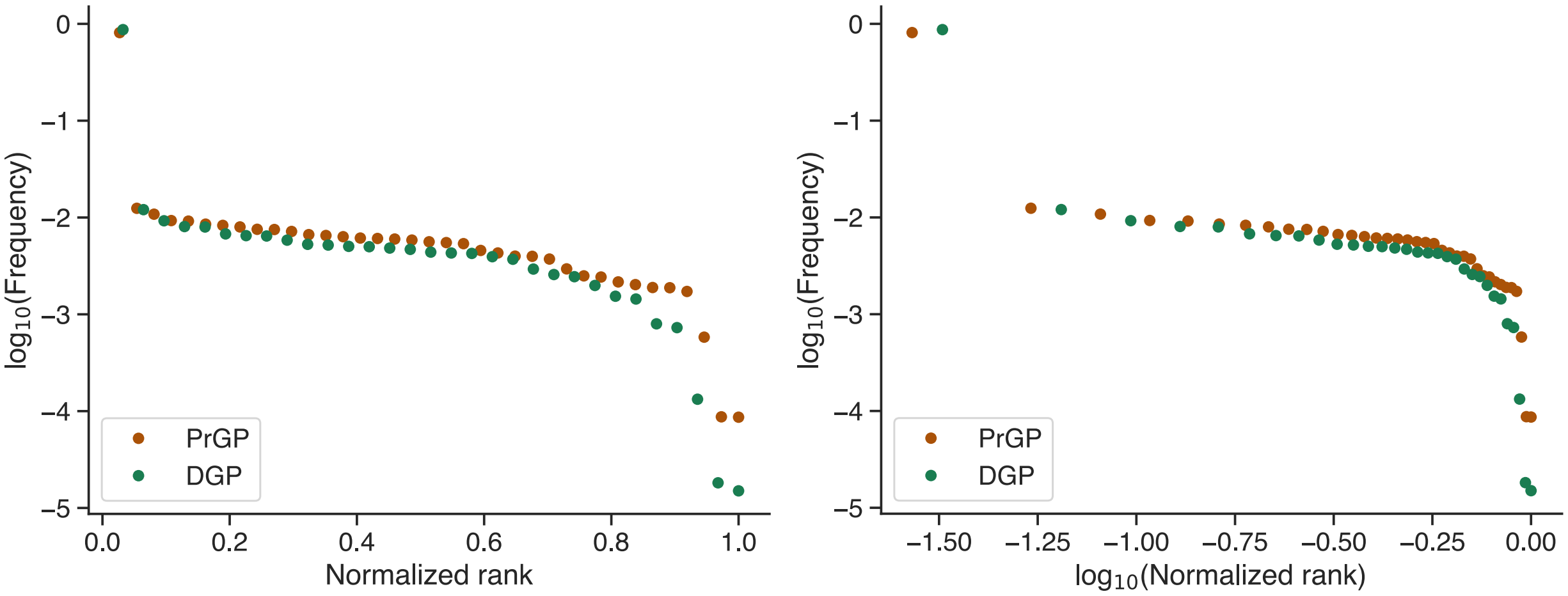

FIG. S5. Plots of (left) $\log_{10}$(frequency) versus normalized rank and (right) $\log_{10}$(frequency) versus $\log_{10}$(normalized frequency) for RNA folding PrGP and DGP maps. When computing ranks, ties were broken arbitrarily.

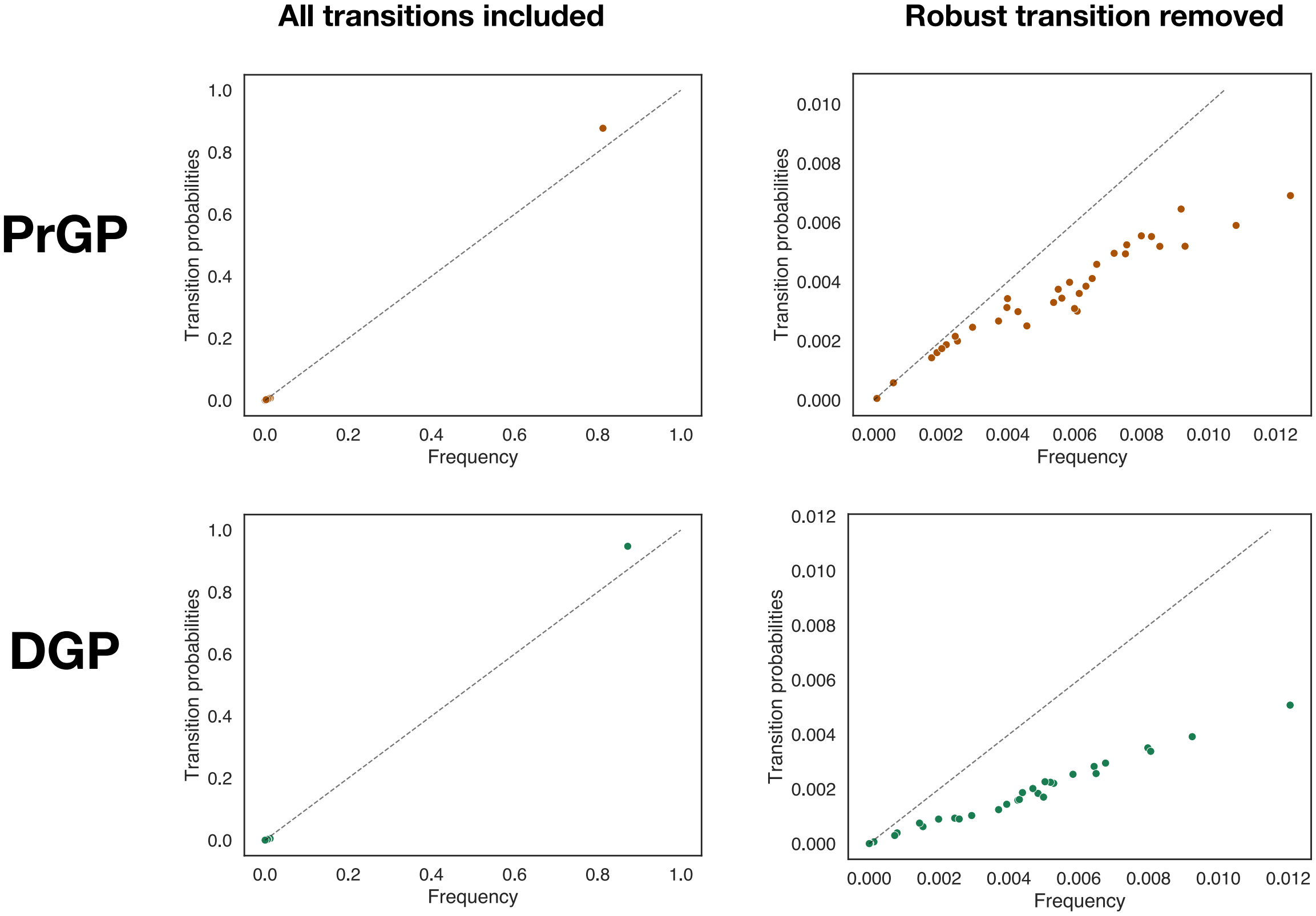

FIG. S6. Plots of transition probabilities versus frequency RNA folding (top) PrGP maps and (bottom) DGP maps. For each respective map, plots include either (left) all transitions or (right) have the most robust transition removed. The dashed line is the random null expectation for both PrGP and DGP maps given by $\phi_{mn} = f_m$ for all $m$ and $n$.

### A. Genotype Entropy for $k = 4, \ell = 12$ Case

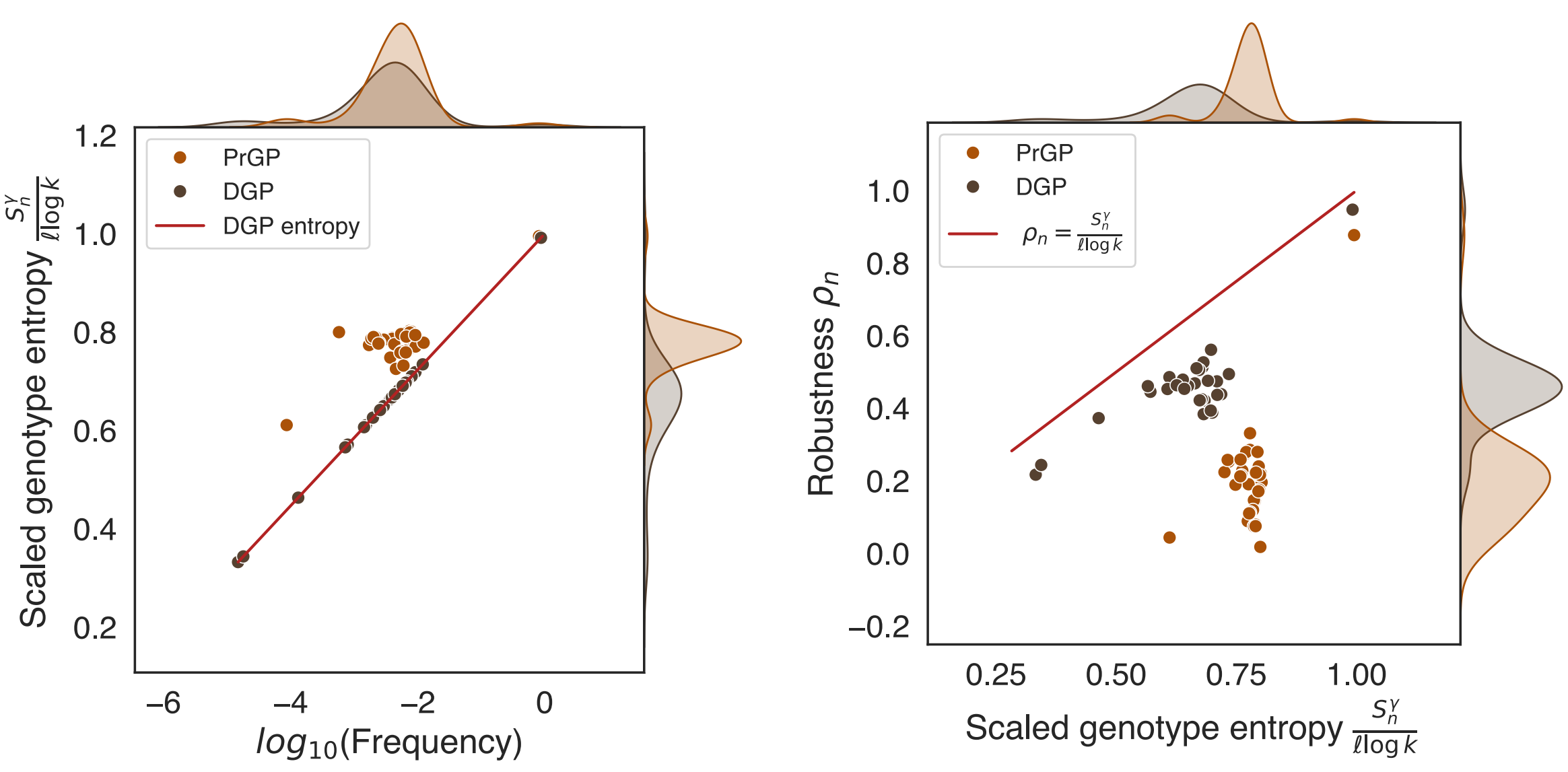

FIG. S7. Plots of (left) $\log_{10}$(frequency) versus scaled genotype entropy $\frac{S_n^\gamma}{\ell \log k}$ and (right) scaled genotype entropy $\frac{S_n^\gamma}{\ell \log k}$ versus robustness $\rho_n$ for RNA folding PrGP maps with $k = 4, \ell = 12$.

## III. FREQUENCY OF UNFOLDED PHENOTYPES FOR ALL RNA STUDIES

| **System** | **Details** | **DGP Map Frequency** $f_n$ | **PrGP Map Frequency** $f_n$ |
|---|---|---|---|
| RNA GC20 | $20\,^{\circ}$C | 0.000347 | 0.000650 |
| RNA GC20 | $37\,^{\circ}$C | 0.00147 | 0.00187 |
| RNA GC20 | $70\,^{\circ}$C | 0.0294 | 0.0265 |
| RNA12 | $37\,^{\circ}$C | 0.872 | 0.812 |

TABLE S3. Frequency of unfolded phenotypes for each RNA case, both for the DGP map and PrGP map.

We note that in all of the GC20 cases, the unfolded phenotype has very low frequency, meaning that the vast majority of phenotypes fold, even in the deterministic cases. In the RNA12 case, which has fewer phenotypes, the vast majority of sequences do not fold. However, the robustness curves remain qualitatively similar to the GC20 cases, and the topological properties of genotype networks formed in the RNA12 DGP map [5] are consistent with those of other DGP maps [4].

## IV. EXTENDED DATA FOR MAIN TEXT SPIN GLASS PrGP MAP

In the main text, we compared a spin glass DGP map with a fixed random external field $\{h_{0,i}\}$ with our spin glass PrGP map, which introduces a Gaussian distribution to the external field whose means are fixed at $\{h_{0,i}\}$ and whose variance $\sigma_h^2$ is varied as an independent variable. Figure S8 shows the topology of the graph $\mathcal{G}(V, E)$ (with $|V| = 9$, $|E| = 15$) that corresponds to the spin glass PrGP map data presented in the main text.

In Table S4, we include the Pearson correlation coefficient $r$ and Spearman rank correlation coefficient $\rho$ for each map (PrGP, DGP), external field variance ($\sigma_h^2 = 0.001$, $\sigma_h^2 = 0.01$, $\sigma_h^2 = 0.1$), and axis transformation presented in Figure 2. The primary feature we point out is the relative decrease of the PrGP Pearson $r$ coefficients in robustness versus $\log_{10}$(frequency) plots as compared to the DGP (deterministic) plot; this suggests a deviation from the empirical $\rho_n \propto \log f_n$ trend observed in the spin glass DGP study [6].

In Figure S9, we present plots of $\log_{10}$(frequency) versus normalized rank and $\log_{10}$(frequency) versus $\log_{10}$(normalized rank) for each external field variance and the deterministic case. Notably, the $\log_{10}$(frequency) versus $\log_{10}$(normalized rank) plot suggests a deviation from Zipf's law.

Figure S10 presents transition probabilities $\phi_{mn}$ for the most frequently occurring ground state $n$ to the other ground states $m$ due to a single bond perturbation. For each setting of external random field variance, a plot including and excluding the most robust transition is shown for added clarity. This figure demonstrates that the off-diagonal transition probabilities for PrGP maps maintained an approximate relationship $\phi_{mn} \propto f_m$ for $m \neq n$ in concordance with DGP maps, and in concordance with the random null expectation for PrGP maps (see main text). A proportionality constant not equal to 1 for $\phi_{mn} \propto f_m$ with $m \neq n$ is likely due to transition probability mass that is acquired by the diagonal element $\phi_{nn}$. It is also apparent that the most robust transition is much more likely than the transition to any other phenotype, in support of our claim that PrGP maps, like DGP maps, exhibit enhanced robustness.

G(V,E) for Main Text Data

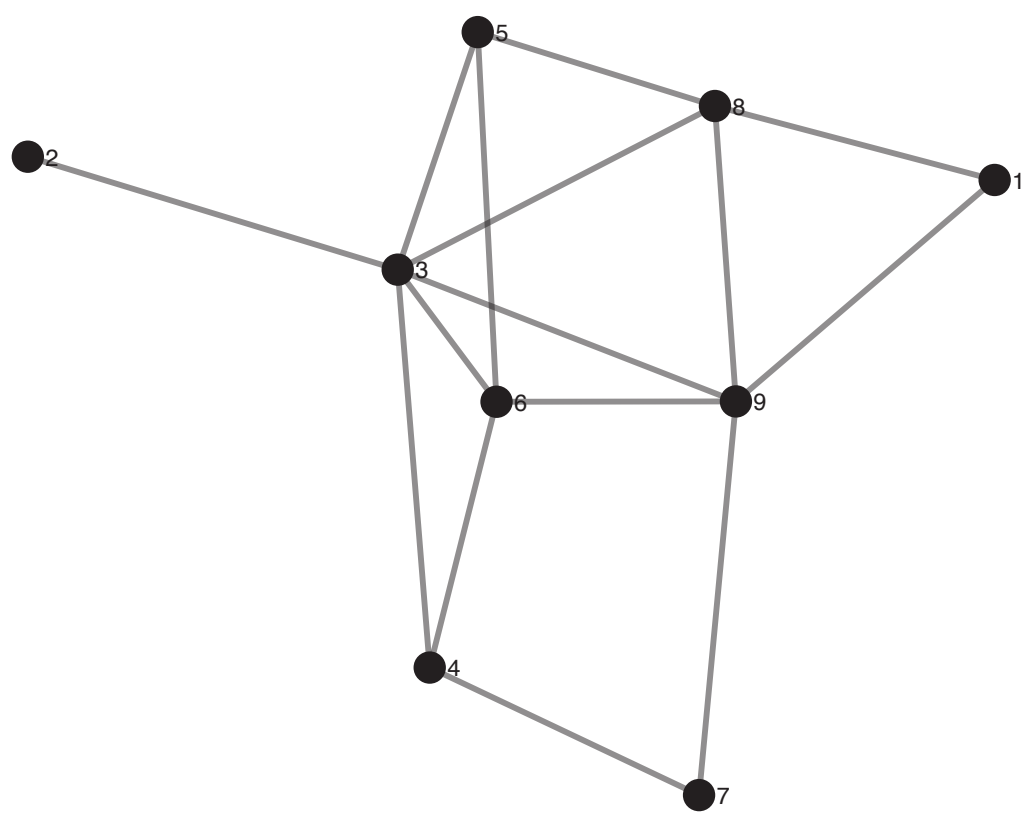

FIG. S8. Graph $\mathcal{G}(V, E)$ corresponding to the spin glass PrGP map data presented in the main text.

| System | Map | $\sigma_h^2$ | Axes | Pearson $r$ | Spearman $\rho$ |
|---|---|---|---|---|---|
| Spin glass | PrGP | 0.001 | Robust v. Freq | 0.766 | 0.962 |
| Spin glass | PrGP | 0.001 | Robust v. $\log_{10}$(Freq) | 0.940 | 0.962 |
| Spin glass | PrGP | 0.001 | $\log_{10}$(Robust) v. $\log_{10}$(Freq) | 0.920 | 0.962 |
| Spin glass | PrGP | 0.01 | Robust v. Freq | 0.874 | 0.985 |
| Spin glass | PrGP | 0.01 | Robust v. $\log_{10}$(Freq) | 0.924 | 0.985 |
| Spin glass | PrGP | 0.01 | $\log_{10}$(Robust) v. $\log_{10}$(Freq) | 0.986 | 0.985 |
| Spin glass | PrGP | 0.1 | Robust v. Freq | 0.976 | 0.987 |
| Spin glass | PrGP | 0.1 | Robust v. $\log_{10}$(Freq) | 0.954 | 0.987 |
| Spin glass | PrGP | 0.1 | $\log_{10}$(Robust) v. $\log_{10}$(Freq) | 0.989 | 0.987 |
| Spin glass | DGP | Deterministic | Robust v. Freq | 0.930 | 0.962 |
| Spin glass | DGP | Deterministic | Robust v. $\log_{10}$(Freq) | 0.964 | 0.962 |
| Spin glass | DGP | Deterministic | $\log_{10}$(Robust) v. $\log_{10}$(Freq) | 0.962 | 0.962 |

TABLE S4. Pearson and Spearman correlation coefficients for all robustness versus frequency plots for the spin glass PrGP map with $|V| = 9$ and $|E| = 15$ whose data are shown in the main text and here in the Supplemental Material.

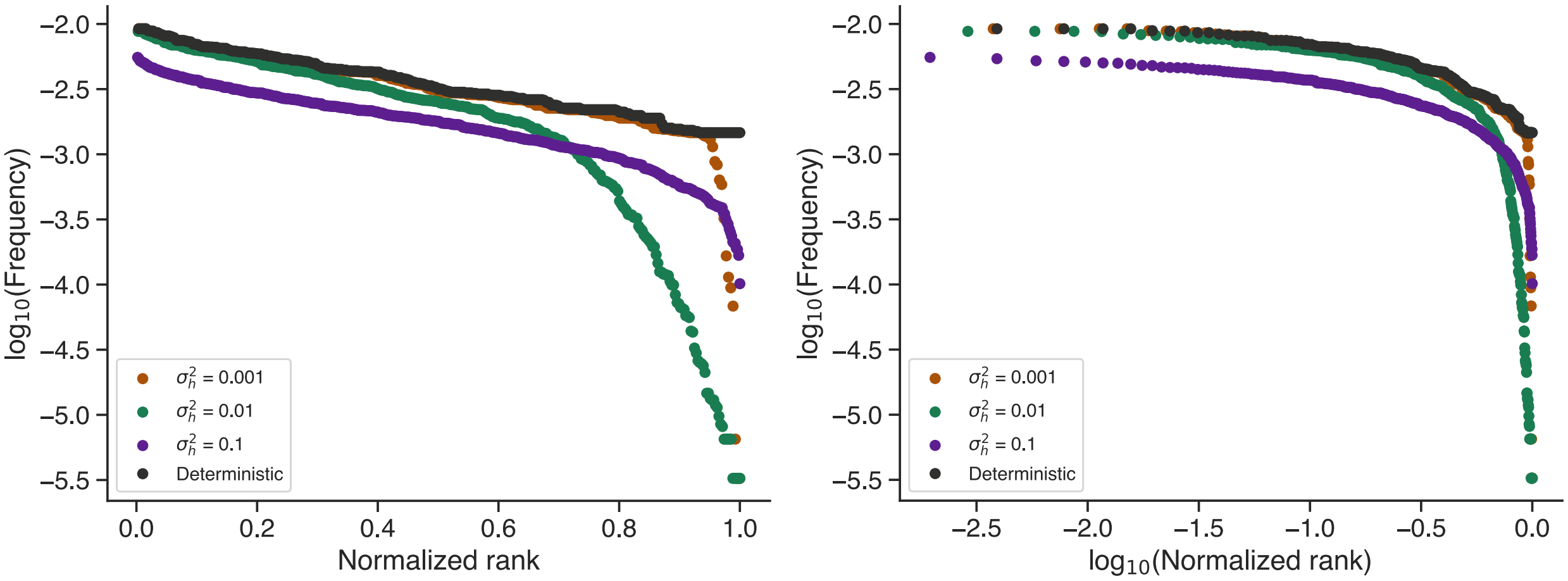

FIG. S9. Plots of (left) $\log_{10}$(frequency) versus normalized rank and (right) $\log_{10}$(frequency) versus $\log_{10}$(normalized frequency) for spin glass ground states for PrGP maps at three external field variances and DGP maps the deterministic case. When computing ranks, ties were broken arbitrarily.

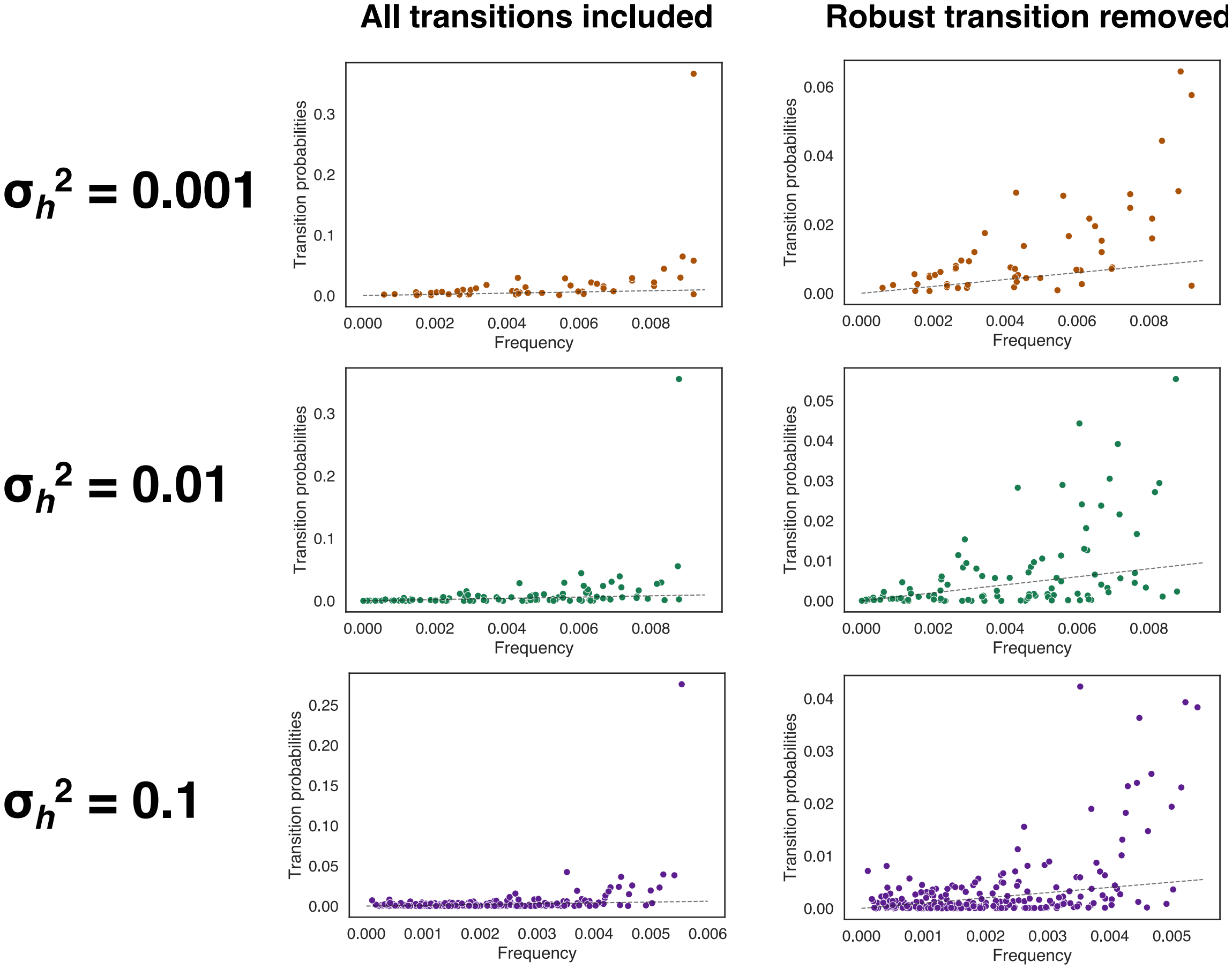

FIG. S10. Plots of transition probabilities versus frequency for spin glass ground states for PrGP maps at three external field variances, (top) $\sigma_h^2 = 0.001$, (middle) $\sigma_h^2 = 0.01$, and (bottom) $\sigma_h^2 = 0.1$. For each, plots include either (left) all transitions or (right) have the most robust transition removed. The dashed line is the random null expectation for both PrGP and DGP maps given by $\phi_{mn} = f_m$ for all $m$ and $n$.

## V. VALIDATION TRIAL FOR SPIN GLASS PrGP MAP

We provide a second spin glass PrGP map trial here in the Supplemental Material to illustrate that the spin glass trends described above and in the main text hold across multiple random graph instances. We generate a new $\mathcal{G}(V, E)$, once again with $|V| = 9$, $|E| = 15$ with topology shown in Figure S11. Figure S12 presents robustness versus frequency, robustness versus $\log_{10}$(frequency), and $\log_{10}$(robustness) versus $\log_{10}$(frequency) for spin glass PrGP maps at three different external field variances and for the deterministic case for DGP maps. The results from this validation trial exhibit the same behavior as observed in the trial presented in the main text. In particular, we see that as the disorder parameter increases the uncertainty in the genotype-phenotype pairing, the robustness versus frequency relationship in PrGP maps becomes suppressed relative to the DGP map limit. Again, these spin glass results are highly suggestive of a biphasic robustness relationship where at high frequencies, $\rho_n$ is substantially enhanced above the random null expectation and behavior close to the deterministic limit is observed. However, as is clear from the corresponding $\log_{10}$(frequency) versus $\log_{10}$(robustness) plots, nearly linear behavior is observed for the smallest frequencies with the empirical robustness nearly parallel to the random expectation, signaling $\rho_n \propto f_n$. See main text for discussion of these features. Additionally, Figure S12 plots the distribution of phenotype entropy $S(g)$ across all genotypes $g$ for PrGP maps at each external field variance experimental value. As is the case in the main text simulation, we observe that the entropy distributions shift rightward as the disorder parameter increases.

In Table S5, we include the Pearson correlation coefficient $r$ and Spearman rank correlation coefficient $\rho$ for each map (PrGP, DGP), external field variance ($\sigma_h^2 = 0.001$, $\sigma_h^2 = 0.00$, $\sigma_h^2 = 0.1$), and axis transformation presented in Figure S12. The primary feature we point out is the relative decrease of the PrGP Pearson $r$ coefficients in robustness versus $\log_{10}$(frequency) plots as compared to the DGP (deterministic) plot; this suggests a deviation from the empirical $\rho_n \propto \log f_n$ trend observed in the spin glass DGP study [6].

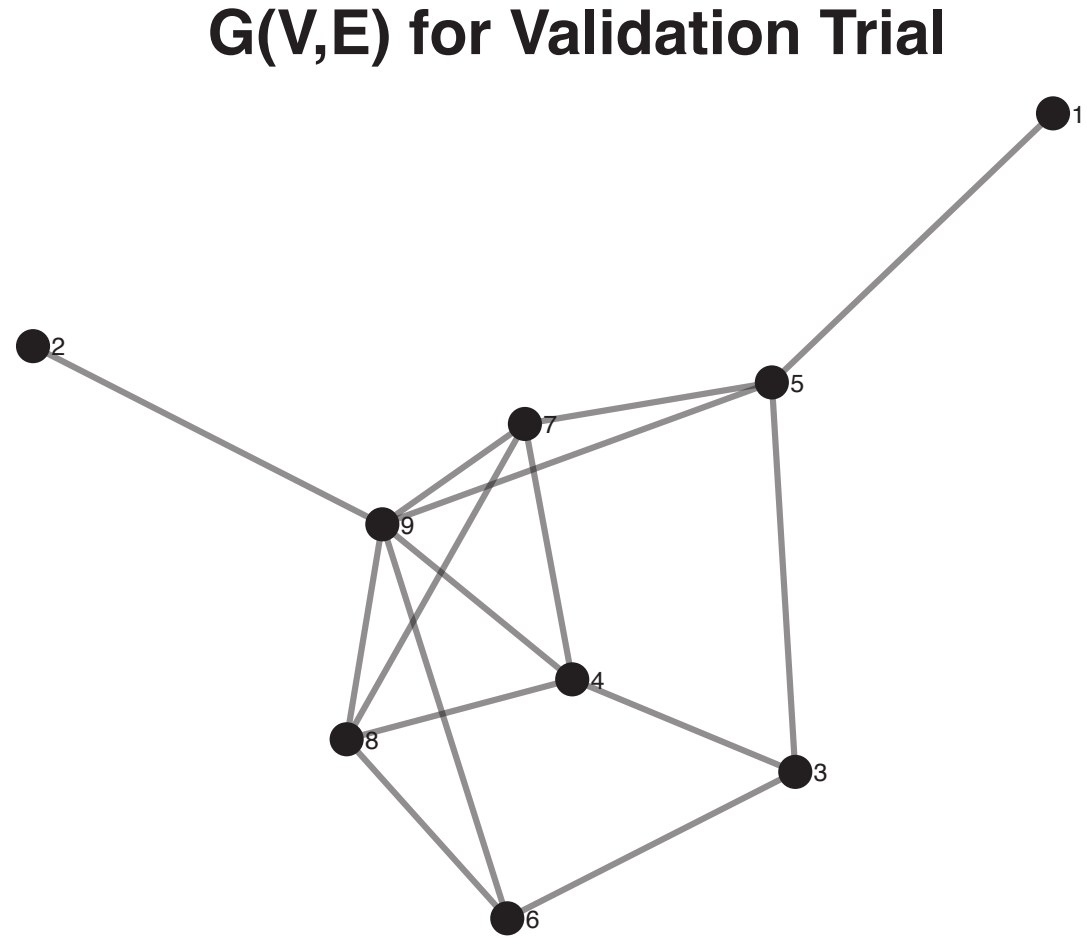

FIG. S11. Graph $\mathcal{G}(V, E)$ corresponding to the spin glass PrGP map validation trial data shown here in the Supplemental Material.

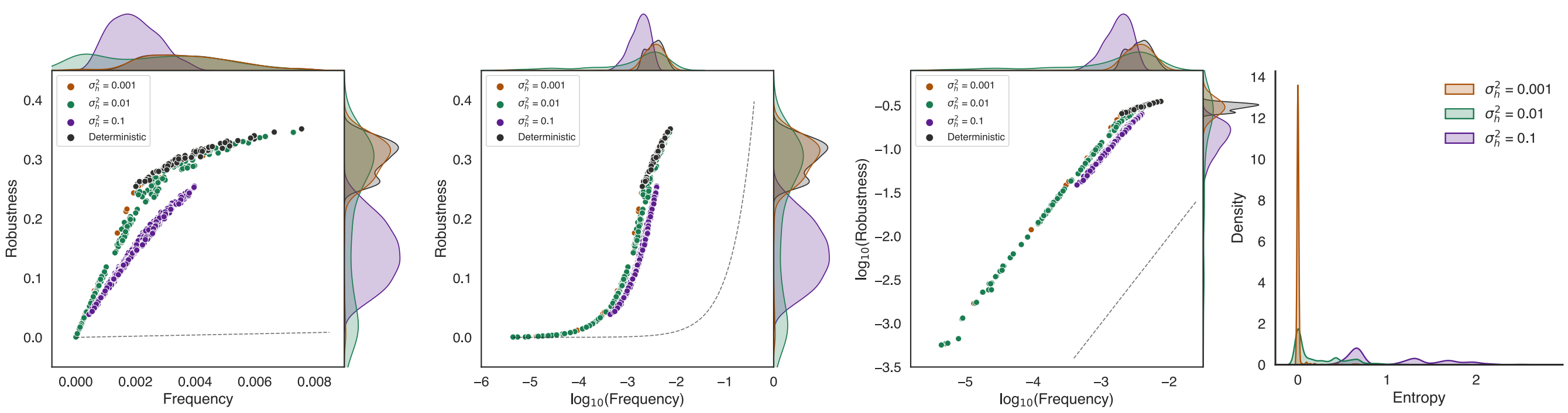

FIG. S12. Plots of (leftmost) robustness versus frequency, (middle left) robustness versus $\log_{10}$(frequency), and (middle right) $\log_{10}$(robustness) versus $\log_{10}$(frequency) for spin glass ground states for PrGP maps at three different external field variances and for the deterministic case for DGP maps. Additionally, the (rightmost) density versus phenotype entropy for the spin glass ground states at three difference external field variances is plotted. The dashed line is the random null expectation for both PrGP and DGP maps given by $\phi_{mn} = f_m$ for all $m$ and $n$.

| **System** | **Map** | $\sigma_h^2$ | **Axes** | **Pearson** $r$ | **Spearman** $\rho$ |
|---|---|---|---|---|---|
| Spin glass | PrGP | 0.001 | Robust v. Freq | 0.806 | 0.994 |
| Spin glass | PrGP | 0.001 | Robust v. $\log_{10}$(Freq) | 0.940 | 0.994 |
| Spin glass | PrGP | 0.001 | $\log_{10}$(Robust) v. $\log_{10}$(Freq) | 0.950 | 0.994 |
| Spin glass | PrGP | 0.01 | Robust v. Freq | 0.932 | 0.996 |
| Spin glass | PrGP | 0.01 | Robust v. $\log_{10}$(Freq) | 0.916 | 0.996 |
| Spin glass | PrGP | 0.01 | $\log_{10}$(Robust) v. $\log_{10}$(Freq) | 0.993 | 0.996 |
| Spin glass | PrGP | 0.1 | Robust v. Freq | 0.993 | 0.997 |
| Spin glass | PrGP | 0.1 | Robust v. $\log_{10}$(Freq) | 0.981 | 0.997 |
| Spin glass | PrGP | 0.1 | $\log_{10}$(Robust) v. $\log_{10}$(Freq) | 0.997 | 0.997 |
| Spin glass | DGP | Deterministic | Robust v. Freq | 0.962 | 0.995 |
| Spin glass | DGP | Deterministic | Robust v. $\log_{10}$(Freq) | 0.993 | 0.995 |
| Spin glass | DGP | Deterministic | $\log_{10}$(Robust) v. $\log_{10}$(Freq) | 0.990 | 0.995 |

TABLE S5. Pearson and Spearman correlation coefficients for all robustness versus frequency plots for the spin glass PrGP map validation trial with $|V| = 9$ and $|E| = 15$ whose data are shown above in this section.

## VI. QUANTUM CIRCUIT GENERATION ALGORITHM

In this study, we generated quantum circuits with 7 qubits and 4 layers. We take the genotype of the quantum circuit PrGP map to be a subset of single qubit gates (which are varied to reflect each genotype). We first start by seeding the circuit randomly with *CNOT* gates which cannot participate in the genotype gate list. Only certain pairs of qubits which are physically connected in the 7-qubit *ibm_lagos* v1.2.0 quantum computer can participate in the same *CNOT* gate. The remaining open places are seeded with single qubit gates, and we choose $\ell = 4$ of these gates to be the variable gates for the genotype. The alphabet chosen is of size $k = 8$: $\{Z, X, Y, H, S, S^{\dagger}, T, T^{\dagger}\}$. Circuit diagrams used in our experimental trials are shown in the subsequent sections.

## VII. EXTENDED DATA FOR MAIN TEXT QUANTUM CIRCUIT PrGP MAP

To our knowledge, this work is the first to analyze the structural properties of quantum circuit GP maps. We generate random quantum circuits as described in the main text and in the previous section with 7 qubits and 4 layers of gates. Figure S13 shows a schematic representation of the random quantum circuit generated for the quantum circuit PrGP map data presented in main text Figure 2 and in this Supplemental Material section.

In Table S6, we include the Pearson correlation coefficient $r$ and Spearman rank correlation coefficient $\rho$ for both exact and experimental quantum circuit PrGP results for each axis transformation presented in main text Figure 2. The primary features we point out are the high Perason correlation $r = 0.998$ of the robustness versus $\log_{10}$(frequency) relationship for the exact phenotype probability vectors, and the relative decrease of the experimental Pearson $r$ coefficients in robustness versus $\log_{10}$(frequency) plot as compared to the exact plot. This suggests that the exact relationship exhibits behavior similar to the empirical $\rho_n \propto \log f_n$ trend observed in DGP studies, and that the experimental trials introduce measurement noise which induces a deviation from the exact results.

In Figure S14, we present plots of $\log_{10}$(frequency) versus normalized rank and of $\log_{10}$(frequency) versus $\log_{10}$(normalized rank) for experimental and exact quantum circuit PrGP map results. Notably, the plot showing $\log_{10}$(frequency) versus $\log_{10}$(normalized rank) suggests a deviation from Zipf's law.

Figure S15 presents transition probabilities $\phi_{mn}$ for the most frequently occurring circuit output state $n$ to the other circuit output states $m$ due to a single variable gate perturbation. For both experimental and exact phenotype probability vectors, a plot including and excluding the most robust transition is shown for added clarity. This figure demonstrates that the off-diagonal transition probabilities for quantum circuit PrGP maps are positively correlated with the frequency $f_m$, though there appears to be some additional nonrandom relationship which is not predicted from standard DGP or PrGP theory. It is also apparent that the most robust transition is much more likely than the transition to any other phenotype, in support of our claim that PrGP maps, like DGP maps, exhibit enhanced robustness.

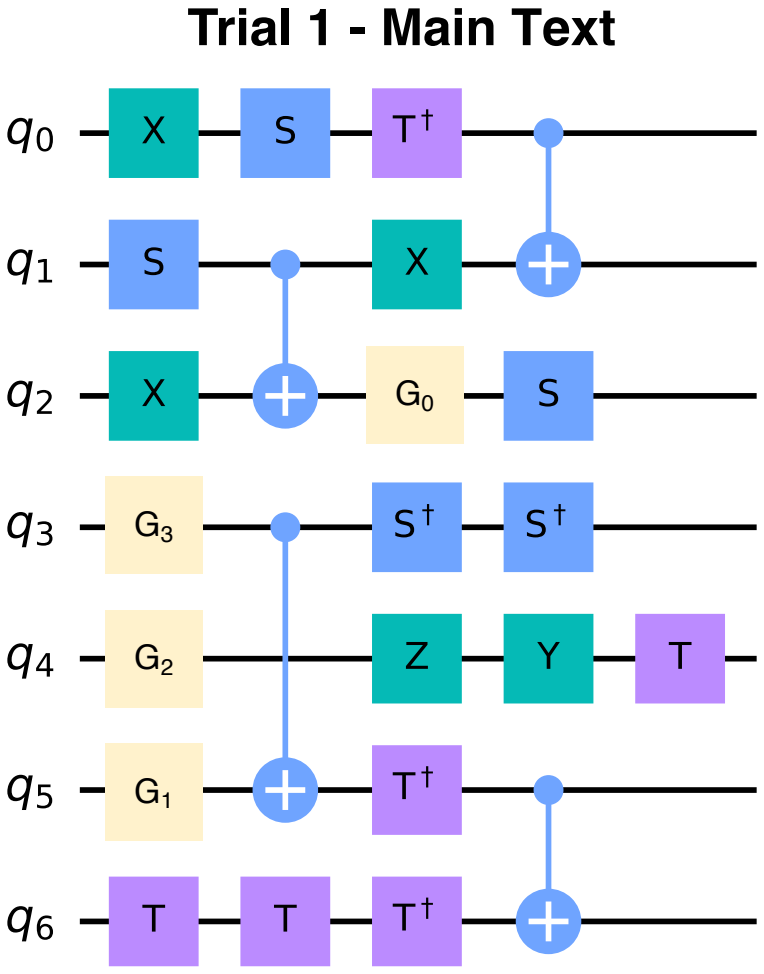

FIG. S13. Random circuit generated for quantum circuit trial 1, whose robustness data is plotted in the main text and below in the remainder of this section. The genotype is the set of variable gates $g = (G_0, G_1, G_2, G_3)$, so the length of the input sequence is $\ell = 4$ drawn from an alphabet of $k = 8$ single qubit gates: $\{Z, X, Y, H, S, S^\dagger, T, T^\dagger\}$.

| System | Map | Trial | Exact or Exp | Axes | Pearson $r$ | Spearman $\rho$ |
|---|---|---|---|---|---|---|
| Quantum circuit | PrGP | 1 | Exact | Robust v. Freq | 0.926 | 0.996 |
| Quantum circuit | PrGP | 1 | Exact | Robust v. $\log_{10}$(Freq) | 0.998 | 0.996 |
| Quantum circuit | PrGP | 1 | Exact | $\log_{10}$(Robust) v. $\log_{10}$(Freq) | 0.993 | 0.996 |
| Quantum circuit | PrGP | 1 | Experimental | Robust v. Freq | 0.912 | 0.987 |
| Quantum circuit | PrGP | 1 | Experimental | Robust v. $\log_{10}$(Freq) | 0.712 | 0.987 |
| Quantum circuit | PrGP | 1 | Experimental | $\log_{10}$(Robust) v. $\log_{10}$(Freq) | 0.983 | 0.987 |

TABLE S6. Pearson and Spearman correlation coefficients for all robustness versus frequency plots quantum circuit PrGP map whose robustness data was presented in the main text and in the above log-log plot. This includes both exact results as well as experimental results for realization/Trial 1, whose circuit is also printed earlier in this section.

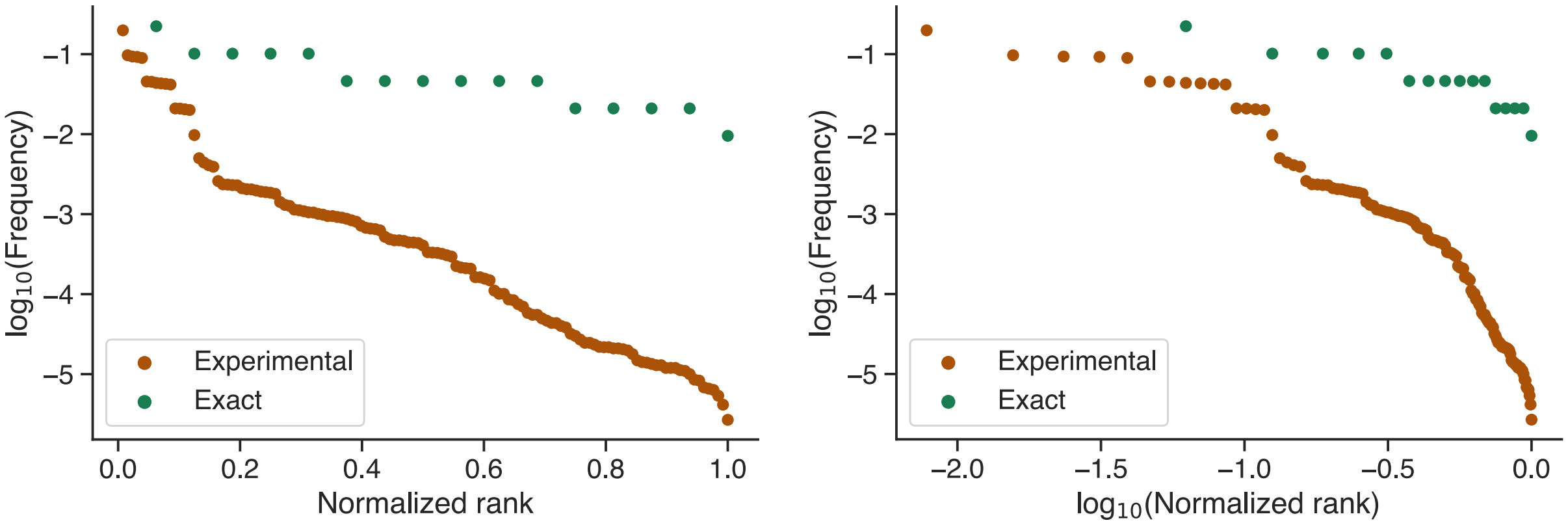

FIG. S14. Plots of (left) $\log_{10}$(frequency) versus normalized rank and (right) $\log_{10}$(frequency) versus $\log_{10}$(normalized frequency) for the quantum circuit trial 1 for experimental and exact data. The dashed line is the random null expectation for both PrGP and DGP maps given by $\phi_{mn} = f_m$ for all $m$ and $n$.

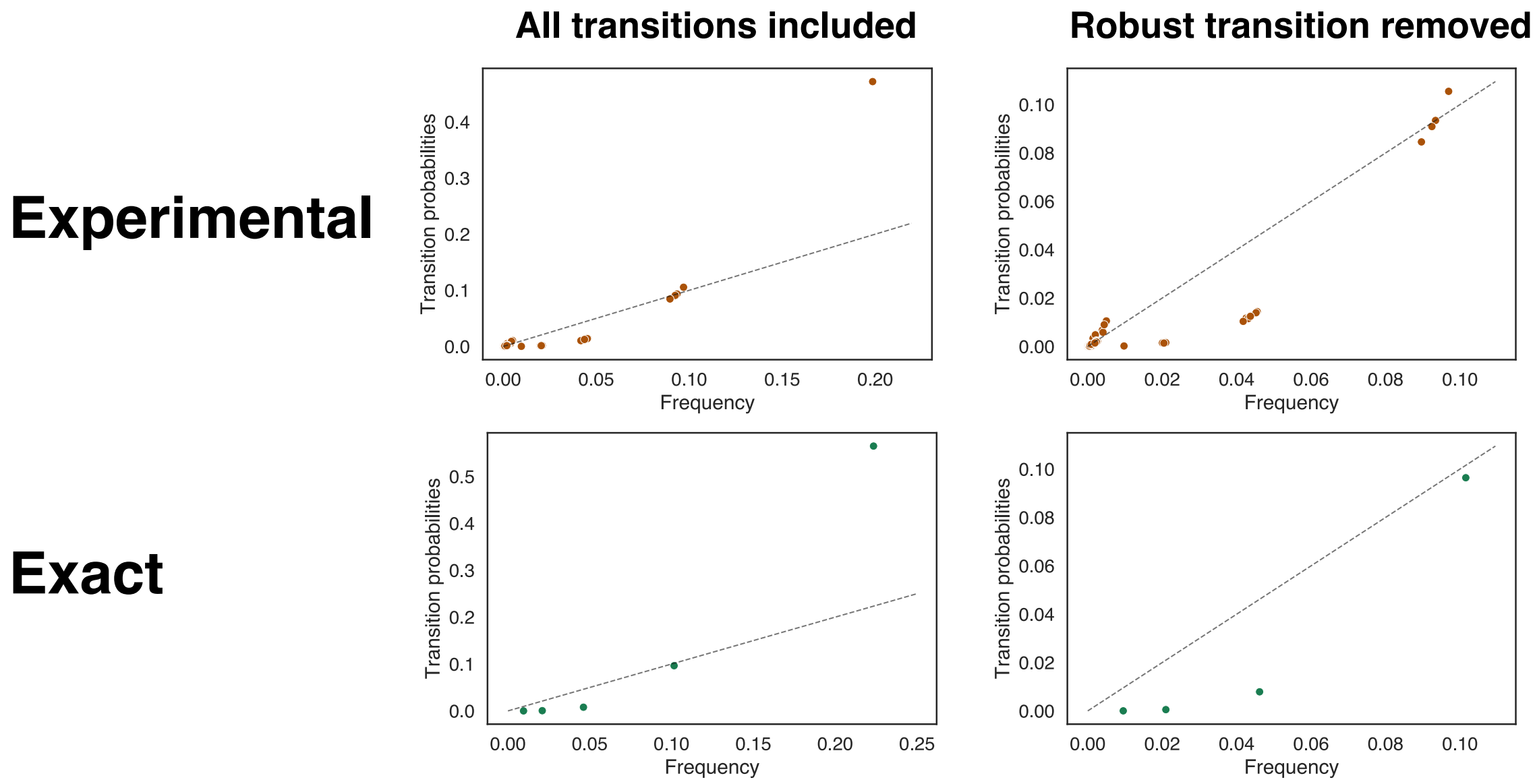

FIG. S15. Plots of (top) transition probabilities versus frequency for quantum circuit trial 1 for experimental and (bottom) exact data. For each model framework, plots include either (left) all transitions or (right) have the most robust transition removed. The dashed line is the random null expectation for both PrGP and DGP maps given by $\phi_{mn} = f_m$ for all $m$ and $n$.

## VIII. 7 QUBIT VALIDATION TRIALS FOR QUANTUM CIRCUIT PrGP MAP

To validate the quantum circuit PrGP map results presented in the main text and Supplemental Material, six additional trials were conducted. A schematic of the random quantum circuit generated for the first of these validation trials is shown in Figure S16. Figure S17 presents robustness versus frequency, robustness versus $\log_{10}$(frequency), and $\log_{10}$(robustness) versus $\log_{10}$(frequency) for this quantum circuit PrGP map validation trial. As with the first quantum circuit PrGP map trial, these data strongly support the enhanced $\rho_n \propto \log f_n$ scaling. Again, we see the spread of phenotypes observed in the frequency domain due to superposition and/or entanglement and that many of the phentoypes are degenerate with identical frequency and robustness. This degeneracy is broken in our experimental measurements, which exhibit measurement noise. Once again, the frequency and robustness of these logarithmically scaling phenotypes is suppressed relative to the exact case as probability mass is drawn towards additional phenotypes which are observed experimentally which were not observed in the exact case. These results illustrate our suggested biphasic robustness scaling in which the low frequency phenotypes, which are introduced due to measurement noise in the experimental trials, lie much closer to the random null expectation than the higher frequency phenotypes observed in the exact calculations, which rather scale with enhanced robustness similar to what is seen in standard DGP maps. Figure S17 also presents a plot of the distribution of phenotype entropy $S(g)$ across all genotypes $g$ for exact and experimental quantum circuit PrGP maps. Notably, the experimental entropy distribution is shifted rightward relative to the exact result due to measurement noise as well as a finite number of experimental trials.

In Table S7, we include the Pearson correlation coefficient $r$ and Spearman rank correlation coefficient $\rho$ for both exact and experimental quantum circuit PrGP results for each axis transformation presented in Figure S17. The primary features we point out are the high Perason correlation $r = 0.950$ of the robustness versus $\log_{10}$(frequency) relationship for the exact phenotype probability vectors, and the relative decrease of the experimental Pearson $r$ coefficients in robustness versus $\log_{10}$(frequency) plot as compared to the exact plot. This suggests that the exact relationship exhibits behavior similar to the empirical $\rho_n \propto \log f_n$ trend observed in DGP studies, and that the experimental trials introduce measurement noise which induces a deviation from the exact results.

Figure S18 presents robustness versus frequency and robustness versus $\log_{10}$(frequency) plots as well as schematics of the corresponding random quantum circuits for validation trials 3-7. In each trial, the suggested biphasic robustness scaling is clear. Additionally, these trials support the enhanced $\rho_n \propto \log f_n$ scaling.

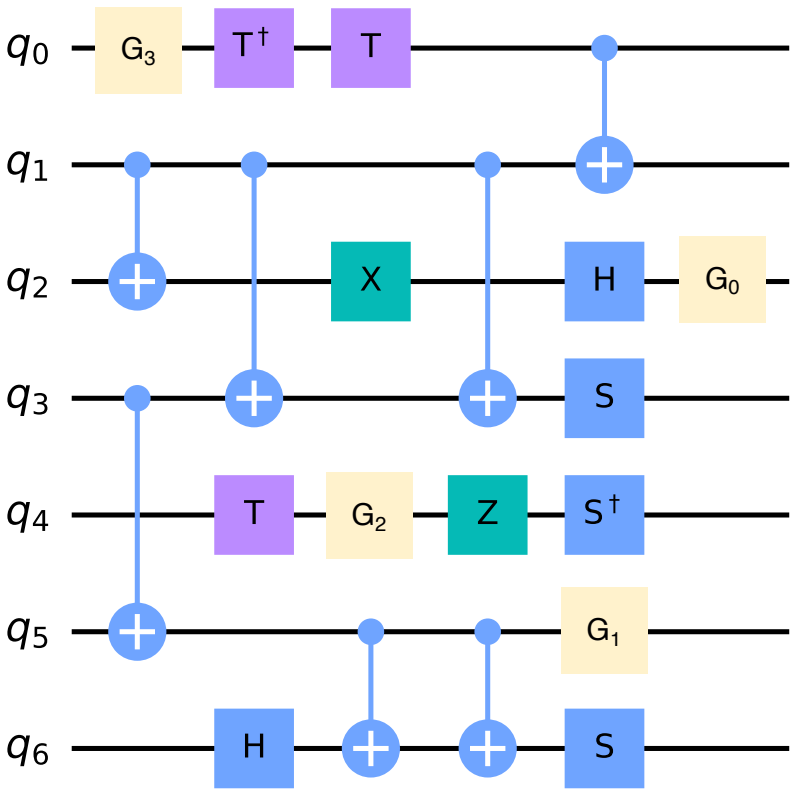

FIG. S16. Random circuit generated for quantum circuit trial 2, whose robustness and entropy data are plotted below as a validation trial. The genotype is the set of variable gates $g = (G_0, G_1, G_2, G_3)$, so the length of the input sequence is $\ell = 4$ drawn from an alphabet of $k = 8$ single qubit gates: $\{Z, X, Y, H, S, S^\dagger, T, T^\dagger\}$.

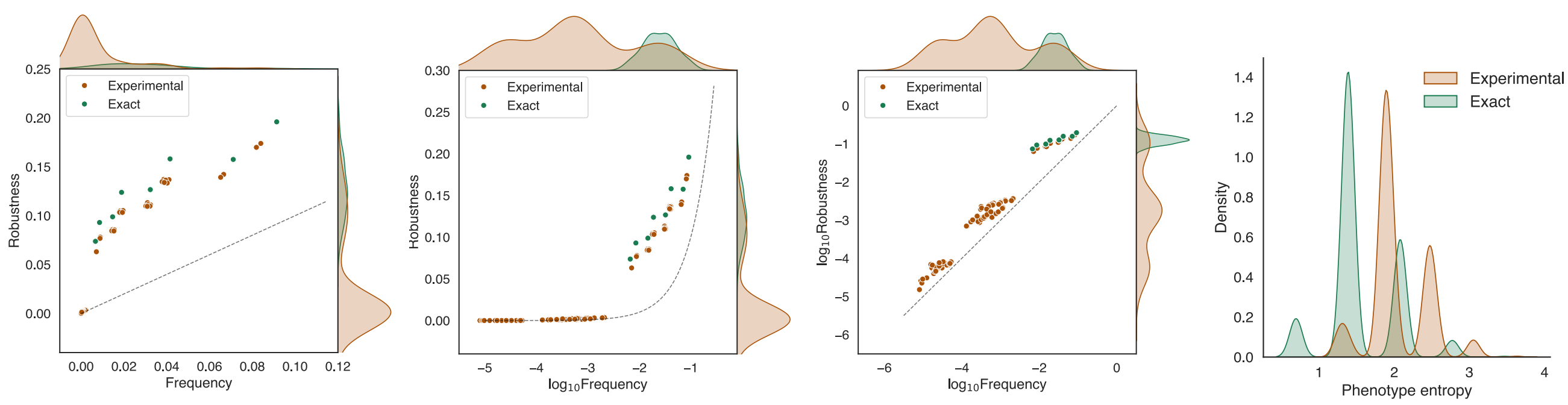

FIG. S17. Plot of (leftmost) robustness versus frequency, (left middle) robustness versus $\log_{10}$(frequency), and (middle right) $\log_{10}$(robustness) versus $\log_{10}$(frequency) for quantum circuit trial 2 for experimental and exact data. Additionally, the (rightmost) density versus phenotype entropy for quantum circuit trial 2 is plotted. The dashed line is the random null expectation for both PrGP and DGP maps given by $\phi_{mn} = f_m$ for all $m$ and $n$.

| **System** | **Map** | **Trial** | **Exact or Exp** | **Axes** | **Pearson $r$** | **Spearman $\rho$** |
|---|---|---|---|---|---|---|
| Quantum circuit | PrGP | 2 | Exact | Robust v. Freq | 0.910 | 0.973 |
| Quantum circuit | PrGP | 2 | Exact | Robust v. $\log_{10}$(Freq) | 0.950 | 0.973 |
| Quantum circuit | PrGP | 2 | Exact | $\log_{10}$(Robust) v. $\log_{10}$(Freq) | 0.954 | 0.973 |
| Quantum circuit | PrGP | 2 | Experimental | Robust v. Freq | 0.916 | 0.979 |
| Quantum circuit | PrGP | 2 | Experimental | Robust v. $\log_{10}$(Freq) | 0.837 | 0.979 |
| Quantum circuit | PrGP | 2 | Experimental | $\log_{10}$(Robust) v. $\log_{10}$(Freq) | 0.989 | 0.979 |

TABLE S7. Pearson and Spearman correlation coefficients for all robustness versus frequency plots quantum circuit PrGP map whose robustness data is shown above as a Validation trial (i.e. Trial 2). This includes both exact results as well as experimental results for Trial 2, whose circuit is also printed earlier in this section.

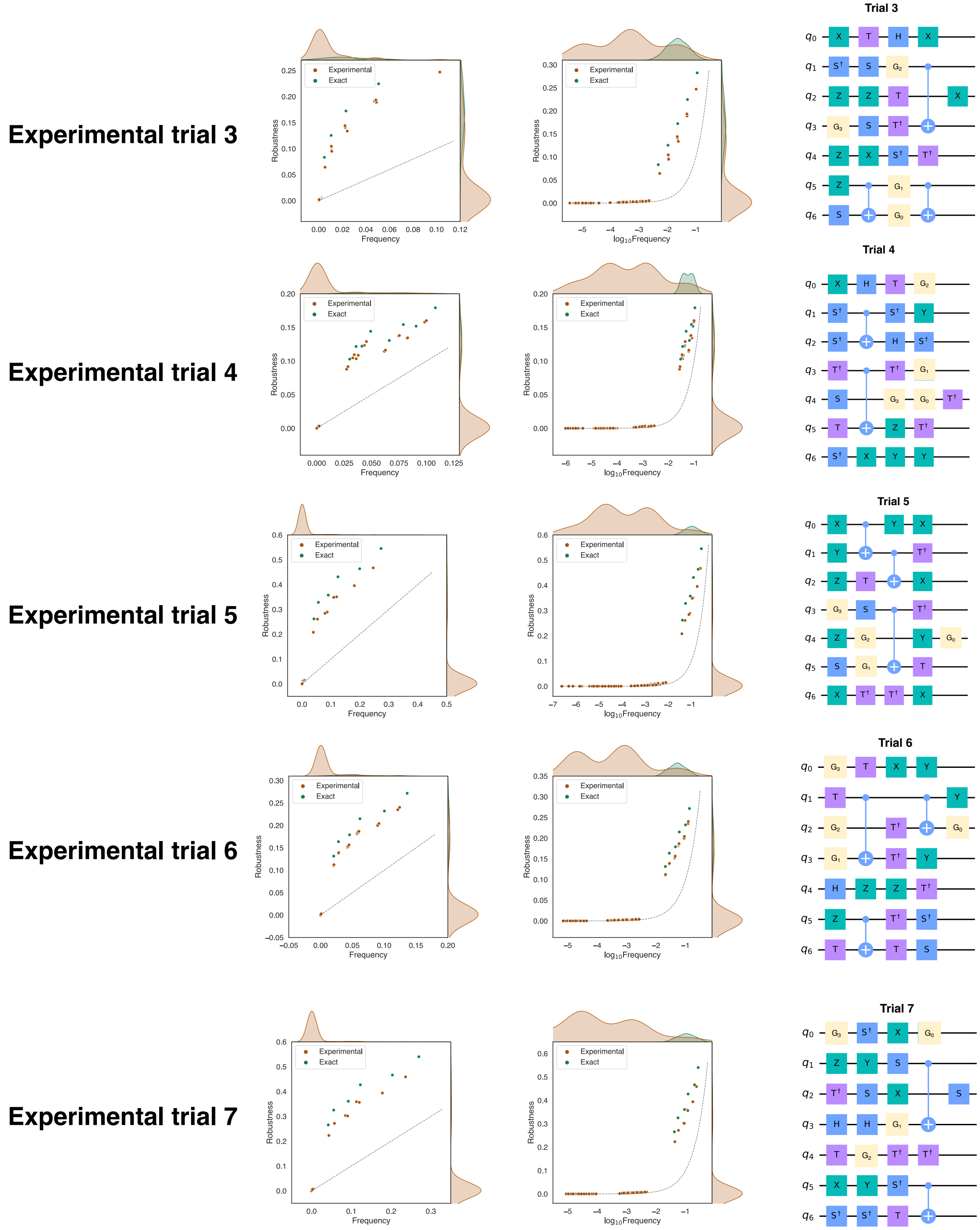

FIG. S18. Plots of (left) robustness versus frequency and (middle) robustness versus $\log_{10}$(frequency) for quantum circuit trials 3-7, as well as (right) corresponding random quantum circuits for Trials 3-7 (validation trials). The genotype is the set of variable gates $g = (G_0, G_1, G_2, G_3)$, so the length of the input sequence is $\ell = 4$ drawn from an alphabet of $k = 8$ single qubit gates: $\{Z, X, Y, H, S, S^\dagger, T, T^\dagger\}$. The dashed line is the random null expectation for both PrGP and DGP maps given by $\phi_{mn} = f_m$ for all $m$ and $n$.

## IX. 11 QUBIT VALIDATION TRIAL FOR QUANTUM CIRCUIT PrGP MAP

To further validate the quantum circuit PrGP map results presented in the main text and Supplemental Material, an additional validation trial with a larger circuit was conducted. In this trial, a noisy simulated quantum circuit was run with 11-qubits and 5 variable gates. Note that unlike the 7-qubit trials (which were run on the *ibm_lagos* machine), this simulation was conducted using the Qiskit Aer backend simulator with noise profile from *ibm_brisbane*.

A schematic of the random quantum circuit generated for this trial is shown in Figure S19. Figure S20 presents robustness versus frequency, robustness versus $\log_{10}$(frequency), and $\log_{10}$(robustness) versus $\log_{10}$(frequency) for this quantum circuit PrGP map validation trial. As with the 7-qubit quantum circuit PrGP map trials, these data strongly support the enhanced $\rho_n \propto \log f_n$ scaling. Again, we see the spread of phenotypes observed in the frequency domain due to superposition and/or entanglement and that many of the phentoypes are degenerate with identical frequency and robustness. This degeneracy is broken in our experimental measurements, which exhibit measurement noise. Once again, the frequency and robustness of these logarithmically scaling phenotypes is suppressed relative to the exact case as probability mass is drawn towards additional phenotypes which are observed experimentally which were not observed in the exact case. These results illustrate our suggested biphasic robustness scaling in which the low frequency phenotypes, which are introduced due to measurement noise in the experimental trials, lie much closer to the random null expectation than the higher frequency phenotypes observed in the exact calculations, which rather scale with enhanced robustness similar to what is seen in standard DGP maps.

An interesting feature of the robustness versus log frequency plots for the 11 qubit simulations is that the two phases of the robustness curve overlap, with some range of frequencies having phenotypes with both elevated log-scaling robustness, while other phenotypes have low, linear-scaling robustness. As we showed in previous sections, knowledge of $\xi_n$ allows one to obtain even this behavior from theory.

FIG. S19. Random circuit generated for 11-qubit quantum circuit trial, whose robustness data is plotted in below as a validation trial. The genotype is the set of variable gates $g = (G_0, G_1, G_2, G_3, G_4)$, so the length of the input sequence is $\ell = 5$ drawn from an alphabet of $k = 8$ single qubit gates: $\{Z, X, Y, H, S, S^\dagger, T, T^\dagger\}$.

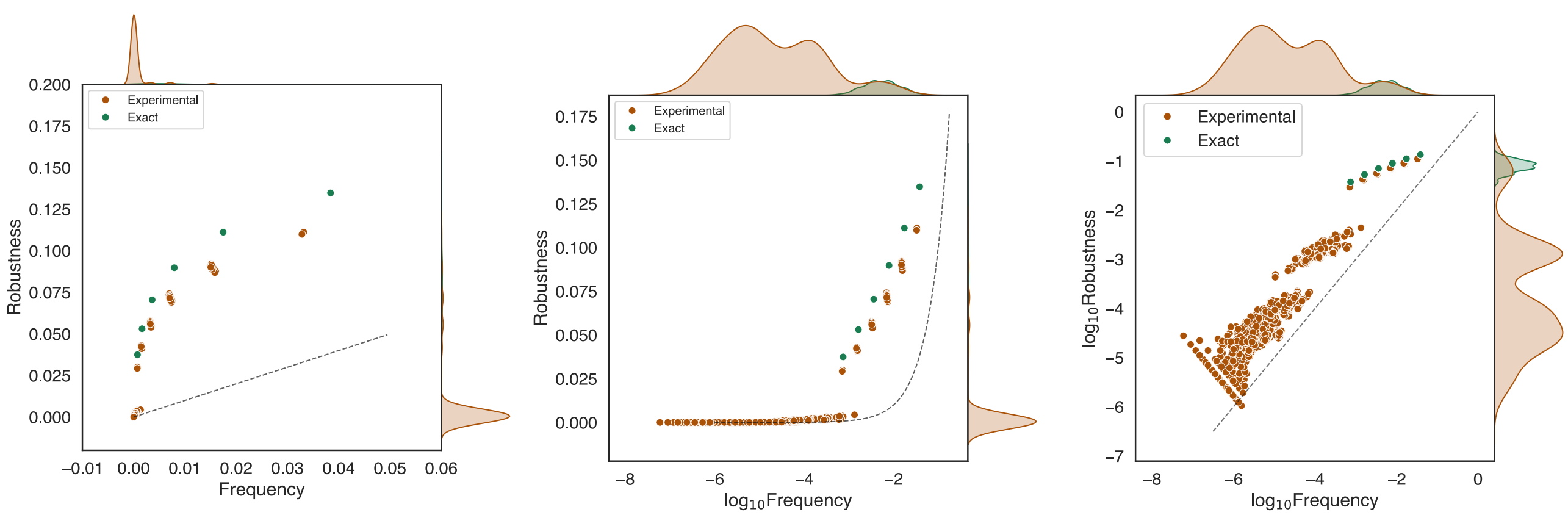

FIG. S20. Plot of (left) robustness versus frequency, (middle) robustness versus $\log_{10}$(frequency), and (right) $\log_{10}$(robustness) versus $\log_{10}$(frequency) for quantum circuit 11-qubit simulated validation trial for experimental and exact data. The dashed line is the random null expectation for both PrGP and DGP maps given by $\phi_{mn} = f_m$ for all $m$ and $n$.

---

[1] T. Jörg, O. C. Martin, and A. Wagner, Neutral network sizes of biological RNA molecules can be computed and are not atypically small, BMC Bioinformatics **9**, 464 (2008).
[2] S. E. Ahnert, Structural properties of genotype–phenotype maps, Journal of The Royal Society Interface **14**, 20170275 (2017).
[3] S. Manrubia, J. A. Cuesta, J. Aguirre, S. E. Ahnert, L. Altenberg, A. V. Cano, P. Catalán, R. Diaz-Uriarte, S. F. Elena, J. A. García-Martín, P. Hogeweg, B. S. Khatri, J. Krug, A. A. Louis, N. S. Martin, J. L. Payne, M. J. Tarnowski, and M. Weiß, From genotypes to organisms: State-of-the-art and perspectives of a cornerstone in evolutionary dynamics, Physics of Life Reviews **38**, 55 (2021).
[4] S. F. Greenbury, S. Schaper, S. E. Ahnert, and A. A. Louis, Genetic Correlations Greatly Increase Mutational Robustness and Can Both Reduce and Enhance Evolvability, PLOS Computational Biology **12**, e1004773 (2016).
[5] J. Aguirre, J. M. Buldú, M. Stich, and S. C. Manrubia, Topological Structure of the Space of Phenotypes: The Case of RNA Neutral Networks, PLoS ONE **6**, e26324 (2011).
[6] V. Mohanty and A. A. Louis, Robustness and stability of spin-glass ground states to perturbed interactions, Physical Review E **107**, 014126 (2023), publisher: American Physical Society.